\documentclass[letterpaper,12pt]{article}
\pdfoutput=1

\usepackage{jheppub}
\usepackage{hyperref}
\usepackage{amsmath}
\usepackage{amsfonts}
\usepackage{graphicx}
\usepackage{caption}
\usepackage{subcaption}
\usepackage{tikz}
\usepackage[english]{babel}

\usepackage{array}
\usepackage{mathtools}
\def\be{\begin{equation}}
\def\ee{\end{equation}}
\def\ba{\begin{eqnarray}}
\def\ea{\end{eqnarray}}

\newcommand{\MCG}{\ensuremath{\mathrm{MCG}}}
\newcommand{\Trop}{\ensuremath{\mathrm{Trop}}}

\usepackage{tcolorbox}
\title{
All Loop Scattering for All Multiplicity
} 

\author[a]{N.~Arkani-Hamed,}\author[b]{H.~Frost,}\author[c]{G.~Salvatori,}\author[d]{P-G.~Plamondon}\author[e]{H.~Thomas}

\affiliation[a]{School of Natural Sciences, Institute for Advanced Study, Princeton, NJ, 08540, USA}

\affiliation[b]{Mathematical Institute, Andrew Wiles Building, Woodstock Rd, Oxford, UK}

\affiliation[c]{Max-Plank-Instit\"ut fur Physik, Werner-Heisenberg-Institut, D-80805 M\"unchen, Germany}

\affiliation[d]{Laboratoire de Math\'ematiques de Versailles, UVSQ, CNRS, Universit\'e Paris-Saclay, IUF, France}

\affiliation[e]{LaCIM, D\'epartement de Math\'ematiques, Universit\'e du Qu\'ebec \`a Montr\'eal, Montr\'eal, QC, Canada}

\emailAdd{arkani@ias.edu}
\emailAdd{frost@maths.ox.ac.uk}
\emailAdd{giulios@mpp.mpg.de}
\emailAdd{pierre-guy.plamondon@uvsq.fr}
\emailAdd{thomas.hugh\_r@uqam.ca}

\date{\today}

\abstract{
This is part of a series of papers describing the new curve integral formalism for scattering amplitudes of the colored scalar tr$\phi^3$ theory. We show that the curve integral manifests a very surprising fact about these amplitudes: the dependence on the number of particles, $n$, and the loop order, $L$, is effectively decoupled. We derive the curve integrals at tree-level for all $n$. We then show that, for higher loop-order, it suffices to study the curve integrals for $L$-loop tadpole-like amplitudes, which have just one particle per color trace-factor. By combining these tadpole-like formulas with the the tree-level result, we find formulas for the all $n$ amplitudes at $L$ loops. We illustrate this result by giving explicit curve integrals for all the amplitudes in the theory, including the non-planar amplitudes, through to two loops, for all $n$.}

\begin{document}
  \maketitle

\section{Introduction and Summary}
This is the second in a series of papers laying out a new point of view on scattering amplitudes---the \emph{curve integral} formalism---which is based on combinatorial structures in the space of scattering data. \cite{us_qft} The curve integral exposes a number of hidden features of amplitudes. In this paper we describe a striking example of one such hidden structure. Ordinarily, the complexity of scattering amplitudes is thought to grow both with increasing the number of particles $n$, and also with increasing loop order $L$. The case of amplitudes at both large $n$ and large $L$ is thought to be vastly more complicated still. In this paper we will see that this expectation is misleading. First, we find that the tree-level calculation for all $n$ is simplified by the curve integral formula. Second, we find that the curve integral naturally decouples the dependence of amplitudes on $n$ and $L$. The calculation of an $L$-loop amplitude for small $n$ is all that is required to find a closed formula for all $n$.

The $n$-point tree amplitude in $\mathrm{tr}\phi^3$ theory is conventionally given by a sum of diagrams. There are
\begin{equation}
C_{n-2} = \frac{1}{n-1} {2n-4 \choose n-2}   
\end{equation}
such diagrams, and this number grows exponentially as $\sim 4^n/(\sqrt{\pi}n^{3/2})$ for large $n$. By contrast, the curve integral formula gives the amplitude as an integral over an $(n-3)$-dimensional space,
\begin{equation}
    {\cal A}_n^{\rm tree} = \int\limits_0^\infty d^{n-3} t\, Z.
\end{equation}
The integrand, $Z$, is given by computing $n(n-3)/2$ simple \emph{headlight functions}, $\alpha_{ij}$, each with no more than $O(n)$ terms:
\begin{equation}
    \log Z = - \sum_{1\leq i <j-1} \alpha_{ij} \left((p_i + \cdots + p_{j-1})^2+m^2\right).
\end{equation}
This simple function is all that is needed to obtain ${\cal A}_n^{\rm tree}$, which is naively a sum over exponentially many Feynman diagrams. We give explicit formulas for $\alpha_{ij}$ valid at all $n$.

We can give closed formulas for the headlight functions $\alpha_{ij}$ because they follow from an inductive calculation using two-by-two matrices. These \emph{curve matrices}, $M^{\rm tree}_C$, have simple polynomial entries. The curve matrices for the $n$-point amplitude are sufficient to determine the matrices for the $(n+1)$-point amplitude inductively, by matrix multiplication. This immediately leads to closed formulas for these matrices for all $n$.

The simplification of the curve integral at tree level becomes even more striking at loop level. We first study the very simplest \emph{tadpole-like} loop amplitudes, which have just one particle in every color trace-factor. These \emph{tadpole amplitudes} are determined by a basic set of $2 \times 2$ curve matrices, $M_C^{\rm tadpole}$. The surprise is that, having computed both the tadpole matrices $M_C^{\rm tadpole}$ and the tree matrices $M_C^{\rm tree}$, we find that this is everything needed to compute the curve integral for amplitudes at \emph{any} $n$, for fixed $L$. In other words, deriving the curve integral for an amplitude ${\cal A}_{n,L}$ is only as difficult as finding the curve integral for an $L$-loop tadpole.

Moreover, the dependence on $n$ and $L$ is completely factorized in the curve integrals. For any $n$ and $L$, the curve integral amplitude formula takes the form
\begin{equation}
    {\cal A}_{n,L} = \int d^E t ~ {\cal K} \left(\frac{\pi^L}{\cal U}\right)^{\frac{D}{2}}\, \times \exp\left(-\frac{{\cal F}_0}{\cal U} - {\cal Z}\right),
\end{equation}
where $E=n+3L-3$. In this formula, the first factor in the integrand,
\begin{equation}
    {\cal K} \left(\frac{\pi^L}{\cal U}\right)^{\frac{D}{2}}
\end{equation}
does not depend on $n$. It can be computed once and for all for a tadpole-like graph. The dependence on $n$ enters only via the exponential, whose exponent is a sum of terms which grow only polynomially with $n$.

The simplicity of these factorized formulas is illustrated by Table \ref{tab:counts}, which shows the number of curve matrices $M_C$ required in order to compute, first, the $n$-independent prefactor, and, second, the exponent of the exponential.

For example, this factorization can be seen in the 1-loop planar amplitudes. In appropriate coordinates, the curve integrals for these amplitudes are
\begin{equation}
    {\cal A}_{\text{1-loop}} = \int\limits_{-\infty}^0 dt_0 \int \limits_{-\infty}^\infty dt_1\cdots dt_{n-1} \, \left(\frac{\pi}{t_0}\right)^{\frac{D}{2}} \, \exp \left( -\frac{1}{t_0}{\cal F}_0 - {\cal Z} \right).
\end{equation}
The 1-loop planar tadpole has only two propagators, and defines two associated curve matrices, $M_C^{\rm tadpole}$. Just one of these matrices is required to compute the $n$-independent pre-factor, which in this case is
\begin{equation}
    \left(\frac{\pi}{t_0}\right)^{\frac{D}{2}}.
\end{equation}
Whereas, both of the two tadpole curve matrices are needed to compute the exponent. Given these two tadpole curve matrices, and the tree-level matrices $M_C^{\rm tree}$, both ${\cal F}_0$ and ${\cal Z}$ can be computed directly.

\begin{table}
\begin{center}
\begin{tabular}{ c | c | c}
Order & \# of Matrices for the prefactor & Total \# of Matrices\\
\hline
1-loop Planar & 1 & 2\\
1-loop Double-trace & 3 & 5\\
2-loop Planar & 9 & 17\\
2-loop Genus-one & 9 & 10\\
2-loop Double-trace & 14 & 23\\
2-loop Triple-trace & 24 & 33
\end{tabular}
\caption{The minimal number of curve matrices needed to derive the curve integral for all-multiplicity amplitudes (right column), and the number of matrices needed to derive just the pre-factor (middle column).}\label{tab:counts}
\end{center}
\end{table}

Needless to say, this Tree-Loop factorization is not at all manifest in the conventional diagrammatic approach to perturbation theory. Moreover, this factorization of the tree- and loop-like parts of the calculation plays a practical role when computing the amplitudes. In this paper, we illustrate this by giving concrete, all-$n$ formulas for all amplitudes in the tr$\phi^3$ theory through to two loops: tree-level, 1-loop planar, 1-loop double-trace, two-loop planar, two-loop triple trace, and the genus-one single-trace two-loop corrections. We focus on the practical methods for determining the curve integral formulas, while explaining the Tree-Loop factorization along the way. The mathematical and conceptual foundations for the results will be described in greater detail elsewhere.

\section{Review}
In \cite{us_qft} we gave a formula for amplitudes in a coloured cubic scalar theory at any order in perturbation theory. Each partial amplitude can be computed, in $D$ spacetime dimensions, using a \emph{curve integral} of the form
\begin{equation}\label{eq:1}
\mathcal{A} =    \int d^D \ell \int \frac{d^Et}{\MCG} \exp\left( -\sum_C\alpha_C(t) X_C \right).
\end{equation}
with $E = n-3 + 2L + 2g$, if ${\cal A}$ is a partial amplitude with $L$ loops and genus $g$. This formula resembles a standard Feynman integral with Schwinger parameters. However, here the Feynman parameters are replaced by piecewise linear functions, $\alpha_C$, of the $E$ integration variables $t_i$. Instead of integrating one Feynman diagram at a time, this formula computes the full amplitude.

Using the curve integral formula, we can integrate over the loop variables, $\ell_a$, once and for all (rather than separately for each diagram). The result is (in some $D=2d-\epsilon$ dimension)
\begin{equation}\label{eq:2}
\mathcal{A} =    \int \frac{d^Et }{\MCG} \left(\frac{\pi^L}{\mathcal{U}}\right)^{\frac{D}{2}}\, \exp\left({-\frac{\mathcal{F}_0}{\mathcal{U}}-\mathcal{Z}}\right),
\end{equation}
where $\mathcal{U},\mathcal{F}_0,\mathcal{Z}$ are homogeneous polynomials in the $\alpha_C$.

Equations \eqref{eq:1} and \eqref{eq:2} are formulas for the full amplitude, automatically summing over all Feynman diagrams. However, counter-intuitively, the starting point for these formulas is to choose a distinguished Feynman diagram, $\Gamma$. The Feynman diagrams for the theory are cubic fatgraphs. The main point in \cite{us_qft} is that a single given fatgraph, $\Gamma$, contains all the information we need to compute the sum over \emph{all} diagrams. The variables $t_i$ are each assigned to each of the $E$ edges of the diagram $\Gamma$. The propagator factors, $X_C$, correspond to paths, $C$, along the diagram $\Gamma$. And the functions $\alpha_C(t)$ are computed directly from the data of the path $C$, which can be described as series of left and right turns on $\Gamma$.

To be concrete, consider some curve, $C$, with path
\begin{equation}
C = e_0 L e_1 R e_2 \cdots.
\end{equation}
In other words, $C$ enters via external edge $e_0$, takes a left turn, goes down edge $e_1$, takes a right turn, and so on. To find $\alpha_C$, we first replace each turn and edge with a matrix:
\begin{equation}\label{eq:goncharov}
e \mapsto \begin{bmatrix} 1 & 0 \\ 0 & y_e \end{bmatrix}, ~ L \mapsto \begin{bmatrix} 1 & 0 \\ 1 & 1 \end{bmatrix},~ R \mapsto \begin{bmatrix} 1 & 1 \\ 0 & 1 \end{bmatrix}.
\end{equation}
In this way, each curve defines a matrix, given by the product of these matrices:
\begin{equation}
M_C = \begin{bmatrix} 1 & 0 \\ 1 & 1 \end{bmatrix}  \begin{bmatrix} 1 & 0 \\ 0 & y_1 \end{bmatrix} \begin{bmatrix} 1 & 1 \\ 0 & 1 \end{bmatrix}  \begin{bmatrix} 1 & 0 \\ 0 & y_2 \end{bmatrix} \cdots.
\end{equation}
Note that external edges have $y_e=1$. Each entry of $M_C$ is a polynomial in the variables $y_e$. Given $M_C$, the \emph{headlight function} is
\begin{equation}
\alpha_C = - \Trop M_C^{12} - \Trop M_C^{21}  + \Trop M_C^{11} + \Trop M_C^{22},
\end{equation}
where $\Trop$ denotes the \emph{tropicalization} of a polynomial. Note that if the curve $C$ ends in an infinite spiral around a closed loop path, $\Delta = e_1Le_2L\cdots e_mL$, this infinite path is replaced by the matrix
\begin{equation}\label{eq:MDelta}
\widetilde{M}_\Delta = \begin{bmatrix} 1 - y_1y_2\cdots y_m & 0 \\ y_1+y_1y_2+\ldots+y_1y_2\cdots y_m & 1 \end{bmatrix}.
\end{equation}
(See Section 5.5 of \cite{us_qft} for more details.)

The distinguished fatgraph, $\Gamma$, also defines a consistent momentum assignment to every curve, $C$. We again read this directly from the fatgraph. Momenta, $p_e^\mu$, are assigned to the edges of $\Gamma$ in the usual way. If $\Gamma$ has $L$ loops, the momenta are parameterized by a choice of $L$ loop momentum variables $\ell_a^\mu$.\footnote{This is equivalent to choosing a labeled basis of the first homology of the graph.} Then a curve, $C$, that enters via edge $e_1$, say, is assigned momentum
\begin{equation}\label{eq:momdef}
P_C^\mu = p_{e_1}^\mu + \sum_{\text{right turn}} p_{\text{left}}^\mu,
\end{equation}
where, at each right turn, we add the momentum incident on the vertex from the left.

Given this momentum assignment, each $P_C^\mu$ can be written as a sum of a loop-dependent and non-loop-dependent part,
\begin{equation}
P_C^\mu = K_C^\mu + \sum_{a=1}^L \ell_C^a \ell_a^\mu,
\end{equation}
for some integers $\ell_C^a$. In terms of this momentum assignment, the polynomials $\mathcal{U},\mathcal{F}_0,\mathcal{Z}$ are given by
\begin{align}\label{eq:UF0}
\mathcal{U} = \det \Lambda,\qquad {\cal F}_0 = {\cal J}^T \tilde{\Lambda}{\cal J},\qquad {\cal Z} = \sum_C (K_C^2+m^2) \alpha_C,
\end{align}
where the $L\times L$ matrix $\Lambda$ and vector $B$ are given by
\begin{equation}
    \Lambda^{ab} = \sum_C \ell_C^a \ell_C^b \alpha_C,\qquad {\cal J}^{a,\mu} = \sum_C \ell_C^a K_C^\mu \alpha_C.
\end{equation}
We write $\tilde{\Lambda} = (\det \Lambda)\, \Lambda^{-1}$ for the adjugate of $\Lambda$. For practical purposes, it is convenient to leave $\mathcal{U}$ and $\mathcal{F}_0$ expressed in terms of $\Lambda^{ab}$ and ${\cal J}^a$. However, these polynomials also have a beautiful combinatorial definition. Expanding the determinant, ${\cal U}$ becomes a sum of monomials, each corresponding to a cut of the diagram $\Gamma$ that reduces it to a tree graph. ${\cal U}$ is then the sum over all such \emph{maximal cuts}. Moreover, the monomial terms of $\mathcal{F}_0$ correspond to \emph{factorizing} cuts of $\Gamma$ into two trees.

The amplitude formulas, \eqref{eq:1} and \eqref{eq:2}, still look a little formal because the integral requires us to mod out by the Mapping Class Group, $\MCG$. $\MCG$ is a discrete finitely generated group of symmetries of the fatgraph, $\Gamma$. Rather than finding a fundamental domain for this group action (which is tricky in general), we can evaluate the integrals by introducing an integration kernel. The final formula for the amplitude then reads,
\begin{equation}\label{eq:3}
\mathcal{A} =    \int d^Et\, \mathcal{K}\, \left(\frac{\pi^L}{\mathcal{U}}\right)^{D/2}\, \exp\left({-\frac{\mathcal{F}_0}{\mathcal{U}}-\mathcal{Z}}\right),
\end{equation}
for some kernel $\mathcal{K}(t)$, that we call the Mirzakhani kernel.\footnote{${\cal K}$ is a tropical version of the kernel introduced by Mirzakhani to compute Weil-Petterson volumes, \cite{mirzakhani2007} and which has been applied to physical calculations in, e.g. \cite{Saad:2019lba}.} General methods for finding a kernel are given in \cite{us_qft}. In this article, we will see that the computation of ${\cal K}$ can be simplified further. Amazingly, both ${\cal U}$ and ${\cal K}$ (and also the matrix $\Lambda$) are independent of the number of external particles, $n$. The number of external particles, and their momenta, only enter the formula through the exponent in the formula, which is linear in the momentum invariants.

\section{Tree-level Amplitudes}
The curve integral formula for amplitudes is already interesting at tree level. Naively, a tree-level $n$-point amplitude is a sum over $\sim 4^n$ different Feynman diagrams. By contrast, the tree-level curve integral is an integral over $n-3$ dimensions of an integrand formed by $\sim n^2$ headlight functions, $\alpha_C$, each of which can be computed by multiplying $\sim n$ matrices. The resulting formula for the tree level amplitude is
\begin{align}
{\cal A}_\mathrm{tree}(123...n) = \int d^{n-3} t\, Z,
\end{align}
where
\begin{align}
-\log Z= \sum_C \alpha_C X_C.
\end{align}
This sum is over all $n(n-3)/2$ possible propagators, $C_{ij}$. If $k_i^\mu$ are the external momenta, the propagator variables, $X_C$, are given by
\begin{equation}
X_{ij} = P_{ij}^2 + m^2,
\end{equation}
where $P_{ij} = k_i + \ldots + k_{j-1}$ is the momentum of the $C_{ij}$ propagator. We stress again that computing $Z$ is much easier than enumerating Feynman diagrams.

\begin{figure}
\begin{center}
\includegraphics[scale=0.35]{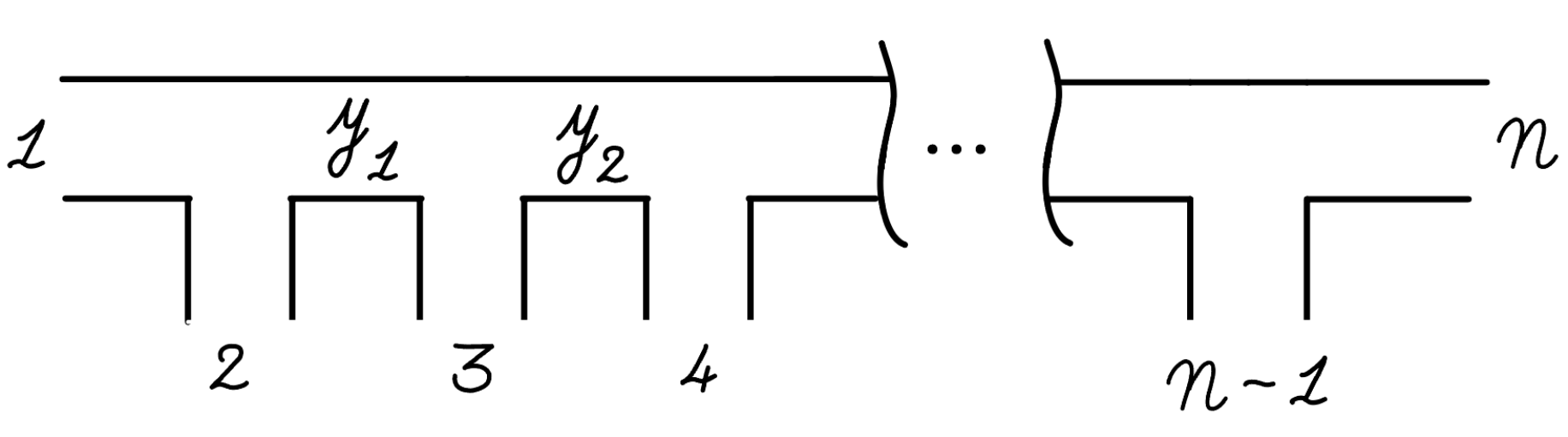}
\caption{The $n$-point tree-level comb graph.}
\label{fig:tree-fan}
\end{center}
\end{figure}

In fact, this section finds a simple all-$n$ expression for $Z$. The headlight functions, $\alpha_C$, depend on our choice of distinguished fatgraph, $\Gamma$. If we choose $\Gamma$ to be the comb graph, as in Figure \ref{fig:tree-fan}, then we obtain particularly simple formulas for the $\alpha_C$. This leads to the following compact all-$n$ formula for the amplitude,
\begin{align}\label{eq:treeampfin}
{\cal A}_\mathrm{tree}(123...n) = \int d^{n-3} t\, \exp\left( \sum_{i=1}^{n-3} t_{i} X_{i+1,n}  - \sum_{1\leq i\leq j}^{j=n-3} f_{ij} c_{ij} \right).
\end{align}
Here we write $f_{ij}$ for the tropical functions
\begin{align}
f_{ij}(\mathbf{t}) = \max\{0, t_i, t_i+t_{i+1},...,t_i+t_{i+1}+...+t_{j}\},
\end{align}
and $c_{ij}$ are the Mandelstam-like kinematic variables,
\begin{align}\label{eqn:XXXX}
c_{ij} = X_{i,j+3}+ X_{i+1,j+2} - X_{i,j+2}-X_{i+1,j+3}.
\end{align}
Not only is \eqref{eq:treeampfin} an all-$n$ formula for the tree amplitude, it also contains the ingredients we need to compute amplitudes for all-$n$ at higher orders in perturbation theory. Because of this, we will derive \eqref{eq:treeampfin} in detail in this section. The formulas in this section will be used repeatedly in subsequent calculations in this paper. 

\subsection{Tree-level Headlight Functions}\label{sec:tree:head}

Take $\Gamma$ to be the comb graph with end points labelled $1$ and $n$, and with the ordering $1234...n$. For each $i,j$, there is a unique curve, $C_{ij}$, that connects external line $i$ to external line $j$. The paths of these curves are
\begin{align}
C_{ij} &= R e_{i-1} L e_i L e_{i+1} ... L e_{j-2} R,\\
C_{1j}&= L e_1 L e_2 ... L e_{j-2} R,\\
C_{in}&= R e_{i-1} L e_i L e_{i+1} ... L e_{n-3} L.
\end{align}
where $1<i<j-1$ and $j<n$. Using the replacements \eqref{eq:goncharov}, we find that these curves have curve matrices
\begin{gather}\label{eq:treeW}
M_{ij} = \begin{bmatrix} F_{i-1,j-3} & F_{i-1,j-2} \\ y_{i-1}F_{i,j-3} & y_{i-1} F_{i,j-2} \end{bmatrix}, \\
M_{1j}= \begin{bmatrix} 1 & 1 \\ F_{1,j-3} & F_{1,j-2} \end{bmatrix}, \qquad
M_{in}= \begin{bmatrix} F_{i-1,n-3} & y_{i-1,n-3} \\ y_{i-1} F_{i,n-3} & y_{i-1,n-3} \end{bmatrix},
\end{gather}
where $y_{i,j} = y_i y_{i+1}\ldots y_j$ and the $F_{ij}$ are the polynomials\footnote{The polynomials $F_{ij}(y)$ are the \emph{F-polynomials} associated to this A-type cluster algebra.}
\begin{align}\label{eq:treeF}
F_{ij} = 1+ y_i + y_iy_{i+1} + ...  + y_iy_{i+1}...y_j,
\end{align}
These matrices will be used throughout our calculations in this paper.

Write $f_{ij} = \text{Trop} F_{ij}$ for the tropicalization of the F-polynomials, \eqref{eq:treeF}. It follows from \eqref{eq:treeW} that the headlight functions are
\begin{align}\label{eq:Atropf}
\alpha_{ij}(\mathbf{t}) &=  f_{i-1,j-3} + f_{i,j-2} - f_{i-1,j-2} - f_{i,j-3}\\
\alpha_{1j}(\mathbf{t}) &= f_{1,j-2} - f_{1,j-3},\\
\alpha_{in}(\mathbf{t}) &= - t_{i-1} + f_{i-1,n-3} - f_{i,n-3},\label{eq:Atropf3}
\end{align}
with the tropical functions $f_{ij}$ given by
\begin{align}
f_{ij}(\mathbf{t}) = \max\{0, t_i, t_i+t_{i+1},...,t_i+t_{i+1}+...+t_{j}\}.
\end{align}

\subsection{The All-multiplicity Amplitude}\label{sec:tree:amp}
To obtain the amplitude, we first consider the assignment of momenta to curves. Using the momentum assignment rule, \eqref{eq:momdef}, the momentum of the curve $C_{ij}$ is (with $i<j-1$)
\begin{align}\label{eq:treeP}
P_{ij} & = p_i + p_{i+1} + ... + p_{j-1},
\end{align}
where $p_a^\mu$ are the external momenta. Because $\Gamma$ is planar, we can alternatively assign these momenta using \emph{dual momentum variables}, $z_a^\mu$, so that
\begin{equation}
P_{ij}^\mu = z_j^\mu - z_i^\mu,
\end{equation}
where, say, $z_i = p_1+\cdots + p_{i-1}$. In either case, the propagator factors are
\begin{align}\label{eq:treeX}
X_{ij} = m^2 + \left( \sum_{a=i}^{j-1} p_a\right)^2,\qquad \text{with}~X_{ii+1} = 0,~ X_{1n} = 0. 
\end{align}

With these momentum assignments understood, the amplitude is
\begin{align}
{\cal A}_n = \int d^{n-3} t\, Z,
\end{align}
where
\begin{align}\label{eq:treeI}
-\log Z = \sum_{1\leq i < j-1}^n \alpha_{ij} X_{ij}.
\end{align}
The $\alpha_{ij}(t)$ are given by (\ref{eq:Atropf}--\ref{eq:Atropf3}).

An alternative form of $\log Z$ is obtained by expanding the $\alpha_{ij}(t)$. This gives
\begin{align}
-\log Z = - \sum_{i=1}^{n-3} t_{i} X_{i+1,n} + \sum_{1\leq i<j-1}^{j=n} (f_{i,j-2} + f_{i-1,j-3} - f_{i-1,j-2} - f_{i,j-3} ) X_{ij},
\end{align}
where, in the second sum, we set $f_{kl}\equiv 0$ if $k=0$, or $l=n-2$. The second sum can be rewritten to find
\begin{align}\label{eq:treeI2}
\log Z =  \sum_{i=1}^{n-3} t_{i} X_{i+1,n}  + \sum_{1\leq i\leq j}^{j=n-3} f_{ij} c_{ij},
\end{align}
where the Mandelstam-like variables $c_{ij}$ are
\begin{equation}
c_{ij} = X_{i,j+3} + X_{i+1,j+2} - X_{ij+2} - X_{i+1,j+3}.
\end{equation}
Evaluating these using \eqref{eq:treeX}, we find
\begin{align}\label{eqn:XXXX}
c_{ij} = \left\{ \begin{matrix} s_{i,j+3}-m^2 & j=i \\ s_{1n-3} - m^2 &~~~ (i,j) =(1,n-3)\\ s_{i,j+3} & \text{otherwise}, \end{matrix} \right.
\end{align}
where the $s_{ij} = (k_i+ k_j)^2$ are the Mandelstam variables.

\subsection{Grafting Trees}\label{sec:tree:graft}
Throughout must of this paper, we will join comb tree graphs to other, higher-loop fatgraphs. By convention, we will take a $n+1$-point comb graph, $\Gamma$, and join its $(n+1)$st leg to an external leg of some other graph, $\Gamma'$. The result is a larger graph $\Gamma \sqcup \Gamma'$. Consider a curve, $C$, that enters the graph through external line $i$ of $\Gamma$, and then proceeds into $\Gamma'$. It has a path
\begin{equation}
    C = R e_{i-1} L e_i L \cdots L e_{n-1} D,
\end{equation}
for some path $D$. Now that $e_{n-1}$ is an internal edge of the larger graph $\Gamma\sqcup\Gamma'$, it is assigned its own variable, $y_{n-1}$. The curve matrix $M_C$ then factors into a product
\begin{equation}
    M_C = M_{i,n+1} \begin{bmatrix} 1 & 0 \\ 0 & y_{n-1} \end{bmatrix} M_{D}.
\end{equation}
Given this, it is convenient to define the matrix
\begin{equation}
    W_i = M_{i,n+1} \begin{bmatrix} 1 & 0 \\ 0 & y_{n-1} \end{bmatrix},
\end{equation}
so that $M_C = W_i M_D$. For $i>1$ we have
\begin{equation}
   W_i = \begin{bmatrix} F_{i-1,n-2} & y_{i-1,n-1} \\ y_{i-1} F_{i,n-2} & y_{i-1,n-1} \end{bmatrix},
\end{equation}
and for $i=1$ we have
\begin{equation}
    W_1 = \begin{bmatrix} 1 & 0 \\ F_{1,n-2} & y_{1,n-1} \end{bmatrix}.
\end{equation}

When multiple tree graphs $\Gamma^1,\ldots,\Gamma^h$ are joined to different trace-factors of a graph $\Gamma$, it is helpful to record the trace-factor by adding an additional index and writing $W_i^A$ (for $A=1,\ldots,h$) for the tree-level matrix starting from $i$ in $\Gamma^A$.

\section{The Planar 1-loop Amplitudes}\label{sec:Dn}
Curve integrals allow us to compute the 1-loop planar amplitudes for all multiplicity. In this section, we illustrate the curve integral method by deriving an all-$n$ formula using the \emph{wheel} diagram. Later, in Section \ref{sec:Dnrev}, we revisit this amplitude with new tools, that allow us to disentangle the loop-like and tree-like parts of this formula.

\subsection{1-Loop Headlight Functions}
\begin{figure}
\begin{center}
\includegraphics[scale=0.35]{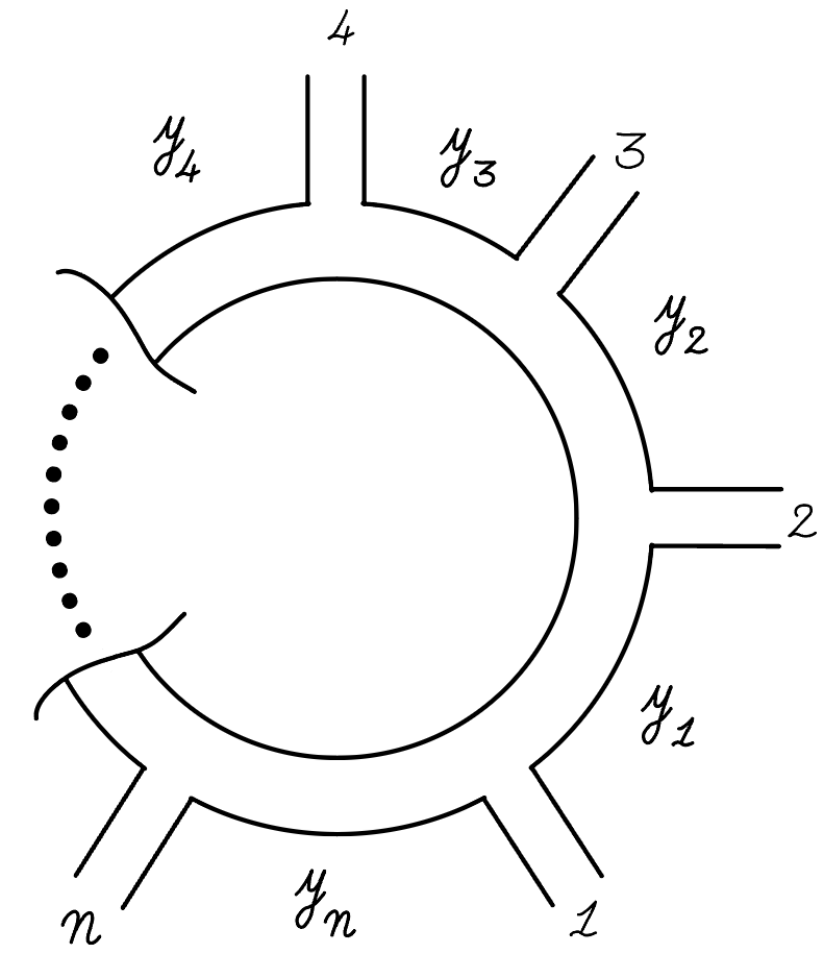}
\caption{An $n$-point 1-loop planar graph.}
\label{fig:wheel}
\end{center}
\end{figure}

Take $\Gamma$ to be the 1-loop \emph{wheel diagram}, in Figure \ref{fig:wheel}, with $n$ external legs labeled $1,2,...,n$, cyclically. This fatgraph has finitely many curves. There are $n(n+1)/2$ curves connecting external legs, that we call $C_{ij}$ with $1\leq i \leq n$ and $i<j$ (cyclically). There are also $n$ spiraling curves $C_{i0}$ that start at an external leg, $i$, and end in a spiral around $\Gamma$'s internal boundary.

Examining $\Gamma$, in Figure \ref{fig:wheel}, the curves $C_{ij}$ ($i\leq j$) and $C_{i0}$ have paths
\begin{gather}
C_{ij} = R e_i L e_{i+1} L \cdots L e_{j-1} R,\\
C_{i0} = R (e_i L e_{i+1} \cdots L e_{i-1} L)^\infty.
\end{gather}
Substituting the matrices in equation \eqref{eq:goncharov}, these paths determine the curve matrices of these curves\footnote{For the infinitely spiraling path the limit has to be defined carefully and $M_{i0}$ is defined by replacing the infinite spiral with the matrix $\widetilde{M}_\Delta$ given by equation \eqref{eq:MDelta}.}
\begin{align}
M_{ij} = \begin{bmatrix} F_{i,j-1} & F_{ij} \\ y_i F_{i+1,j-1} & y_i F_{i+1,j} \end{bmatrix},\qquad 
M_{i0}  = \begin{bmatrix} F_{i,i-2} & 1\\ y_i F_{i+1,i-1} & 1 \end{bmatrix},
\end{align}
where
\begin{align}
{F}_{ij} = 1+y_i+y_{i}y_{i+1}+\ldots +y_iy_{i+1}...y_j.
\end{align}
These curve matrices define headlight functions. Write
\begin{align}
\mathrm{Trop} {F}_{ij} = {f}_{ij} = \max(0,t_i,t_{i}+t_{i+1},...,t_i+t_{i+1}+...+t_j),
\end{align}
for the tropicalization of the $F$-polynomials, where the indices are understood cyclically (mod $n$). Then (in the region $\sum_{i=1}^n t_i <0$)
\begin{align}
\alpha_{{ij}} &=  f_{i,j-1}+f_{i+1,j} - f_{ij} - f_{i+1,j-1}\\
\alpha_{i0} &= - t_i - f_{i+1,i-1} + f_{i,i-2}.
\end{align}
Note that the $\alpha_{i0}$ satisfy the identity
\begin{align}
\sum_{i=1}^n \alpha_{i0} = - \sum_{i=1}^n t_i.
\end{align}
This identity will be useful for simplifying the formula for the amplitude.

\subsection{The Loop Integrand}
Since $\Gamma$ is a planar diagram, it is convenient to use \emph{dual momentum variables} to record momentum assignments. With dual variables $x_i^\mu$ ($i=0,1,\ldots,n$), the propagator factors are
\begin{equation}
    X_{ij} = (x_j-x_i)^2 + m^2.
\end{equation}
If $p_i^\mu$ ($i=1,\ldots,n$) are the external momenta, take
\begin{align}
x_i^\mu = p_1^\mu + \ldots + p_{i-1}^\mu.
\end{align}
It is convenient to use $x_0^\mu$ as the loop momentum variable.

With these momentum assignments, the integrand (pre-loop integration) is given by the curve integral
\begin{align}
{\cal I}_\mathrm{1-loop} = \int\limits_{\sum t_i \leq 0} d^n t\, Z,
\end{align}
where
\begin{equation}
    -\log Z = \sum_{i<j}^\text{cyclic} X_{ij} \alpha_{ij} + \sum_{i=1}^n X_{i0} \alpha_{i0}.
\end{equation}

A more explicit formula for $Z$ follows from substituting the expressions for the headlight functions. This gives
\begin{align}
\log Z =  \sum_{i=1}^n t_i X_{i,0} + \sum_{i=1}^n f_{i,i-2} (X_{i-1,0}-X_{i,0}) + \sum_{i\leq j}^\text{cyclic} f_{ij} c_{ij},
\end{align}
where $c_{ij} = X_{ij}+X_{i-1,j+1} - X_{i,j+1}-X_{i-1,j}$.

\subsection{The All-multiplicity Amplitude}
The integrand, ${\cal I}$, depends on the loop momentum variable $x_0^\mu$. But the integral over $x_0^\mu$ is Gaussian. Performing this integration in $D=2d-\epsilon$ dimensions gives the curve integral formula for the amplitude,
\begin{align}\label{eq:1looppost}
  {\cal A}_\text{1-loop} = \int\limits_{\sum_i t_i \leq 0} d^n\mathbf{t} \left( \frac{\pi}{\mathcal{U}} \right)^{\frac{D}{2}} \exp\left(-\frac{\mathcal{F}_0}{\mathcal{U}} - \mathcal{Z} \right),
  \end{align}
where $\mathcal{U},\mathcal{F}_0,\mathcal{Z}$ are the \emph{surface Symanzik polynomials} for $\Gamma$. In terms of the headlight functions,
\begin{equation}
    \mathcal{U} = \sum_{i=1}^n \alpha_{i0},\qquad
\mathcal{F}_0 = \sum_{i,j=1}^n \alpha_{i0}\alpha_{j0} z_i\cdot z_j,
\end{equation}
\begin{equation}
    \mathcal{Z} = \sum_{i<j-1}^\text{cyclic} \alpha_{ij} X_{ij} + \sum_{i=1}^n \alpha_{i0} (z_i^2+m^2).
\end{equation}
We find even more explicit formulas for ${\cal U},{\cal F}_0$ by substituting the formulas for the headlight functions. This gives:
\begin{align}
\mathcal{U} = - \sum_{i=1}^nt_i,\qquad
\mathcal{F}_0 = \sum_{i,j=1}^n t_i t_j z_i\cdot z_j + f_{i,i-2}f_{j,j-2} \, p_i\cdot p_j.
\end{align}

\section{The Telescopic Property}

The computations of the headlight functions, $\alpha_C$, are not very sensitive to the number of external particles, $n$. This is because the transfer matrices, $M_C$, are computed multiplicatively. This is related to a fundamental \emph{telescopic property} of the headlight functions, which we explain in this section. The telescopic property leads to simplifications of the curve integral formulas.

Consider a path, $C$, that has not yet reached an external line of some fatgraph, $\Gamma$. The path has an associated curve matrix,
\begin{equation}
    M_C = \begin{bmatrix} a & b \\ c & d \end{bmatrix},
\end{equation}
for some $a,b,c,d$, which are polynomials for the $y$-variables. The associated headlight function for this matrix is
\begin{equation}
\alpha_C = \text{Trop}(a)+\text{Trop}(d) - \text{Trop}(b)-\text{Trop}(c).
\end{equation}

\begin{figure}
\begin{center}
\includegraphics[scale=0.35]{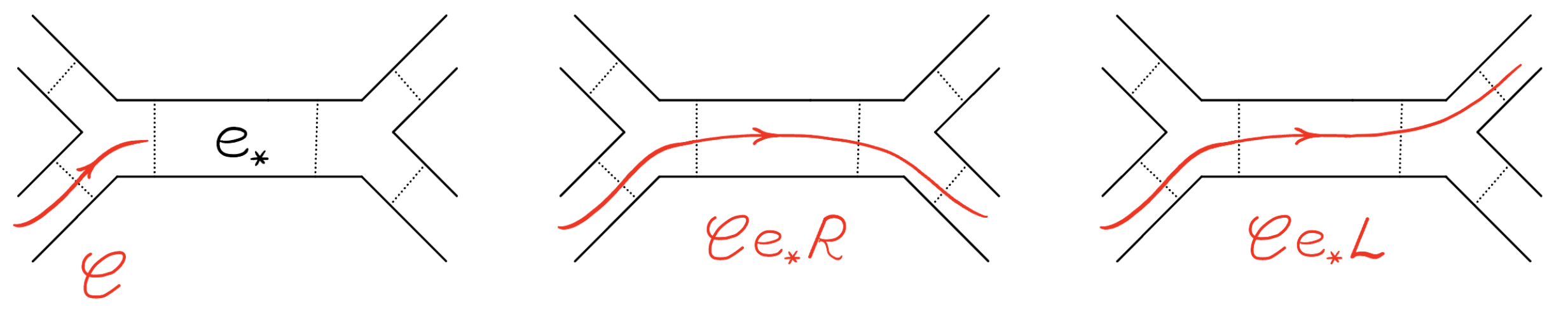}
\caption{An incomplete curve can be extended in one of two ways: to the left, or to the right.}
\label{fig:telescope}
\end{center}
\end{figure}

Suppose that $C$ ends just before traversing an edge, $e_*$, with variable $y_*$. Then $C$ can be extended in one of two ways, as in Figure \ref{fig:telescope}. It can traverse $e_*$ and turn left, to get a longer path: $Ce_*L$. Or it can turn right instead: $Ce_*R$. The associated transfer matrices are
\begin{equation}
    M_{Cy_*L} = \begin{bmatrix} a + b y_*& b y_* \\ c + d y_* & d y_* \end{bmatrix}, \qquad M_{Cy_*R} = \begin{bmatrix}a & a + b y_* \\ c & c + d y_* \end{bmatrix}.
\end{equation}
If we stop here to compute the new headlight functions of these paths, we find
\begin{align}
    \alpha_{Ce_*L} &= \text{Trop}(b) + \text{Trop}(c+dy_*) - \text{Trop}(d) - \text{Trop}(a+by_*)\\
    \alpha_{Ce_*R} &= \text{Trop}(a+by_*) + \text{Trop}(c) - \text{Trop}(a) - \text{Trop}(c+dy_*).
\end{align}
But note that the sum of these two functions is equal to $\alpha_C$, corresponding to the matrix $M_C$:
\begin{equation}\label{eq:tele1}
    \alpha_{Ce_*L}+    \alpha_{Ce_*R} = \alpha_C.
\end{equation}
In other words, the sum $\alpha_{Ce_*L}+\alpha_{Ce_*R}$ can be computed without finding $M_{Cy_*L}$ and $M_{Cy_*R}$ separately. It is only necessary to compute the matrix $M_C$.

This relation can be telescopically nested to compute larger sums of headlight functions. Suppose $\Gamma$ can be decomposed into a join of two graphs, $\Gamma = \Gamma^1 \sqcup \Gamma^2$, where $\Gamma^1$ is a tree graph. And suppose that $C$ is a path in $\Gamma$ with an endpoint at the join of the two graphs. We can sum over all ways to extend $C$ to an external line of the tree graph $\Gamma^1$. If $\Gamma^1$ has $k$ external lines (excluding the join), we obtain $k$ curves $C_1,...,C_k$. Then applying \eqref{eq:tele1} gives
\begin{equation}\label{eq:tele2}
    \sum_{i=1}^k \alpha_{C_k} = \alpha_C.
\end{equation}
Similarly, if $C$ is a path in $\Gamma$ with \emph{both} of its endpoints at the join of the two graphs, we can sum over all ways to extend $C$ to a complete path. Suppose $C_{ij}$ is the curve obtained by extending $C$ to have endpoints $i$ and $j$ in $\Gamma_0$ (with, say, $i\leq j$). Then
\begin{equation}
    \sum_{\substack{i=1\\ i\leq j}}^k \alpha_{C_{ij}} = \alpha_C.
\end{equation}
In other words, the telescopic property dramatically simplifies any sum over headlight functions $\alpha_C$ for a class of curves, $C$, that differ only by their endpoint on a tree subgraph.

\section{Decoupling Trees and Loops}\label{sec:treeloop}
One drastic consequence of the telescopic property is that the curve integral integrands \emph{factorize} into the product of a loop-like factor, which is independent of $n$, and a second $n$-dependent factor which contains the kinematics. 

\begin{figure}
\begin{center}
\includegraphics[scale=0.3]{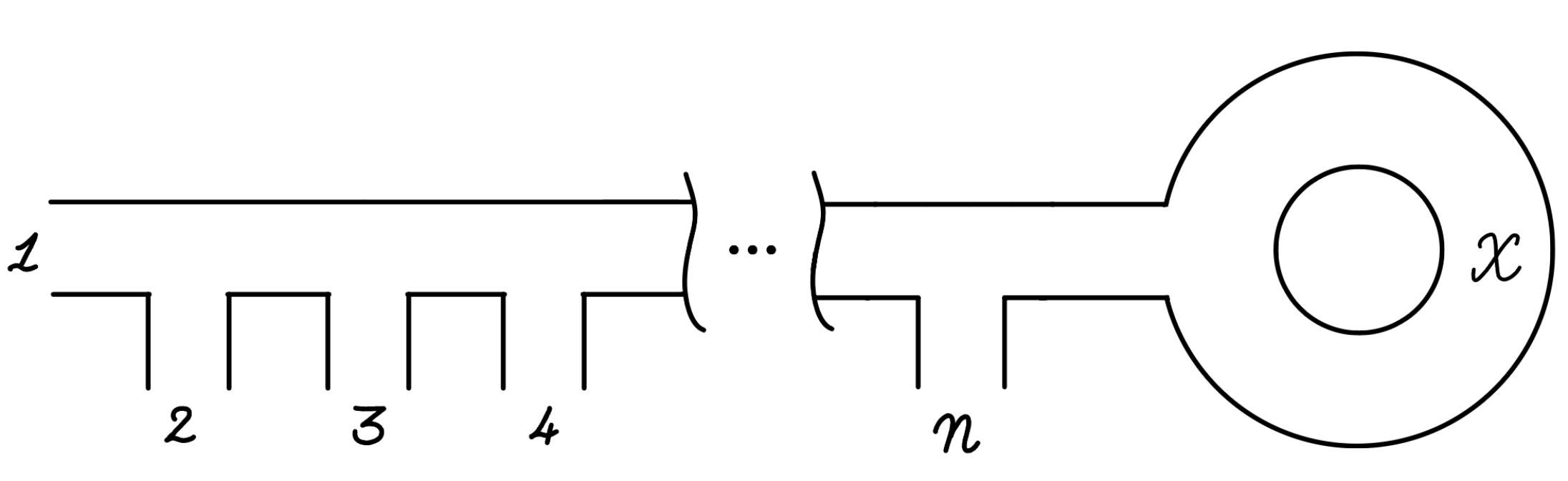}
\caption{A Tree-Loop fatgraph for the planar 1-loop amplitudes.}
\label{fig:Dtadpole}
\end{center}
\end{figure}

To make this factorization manifest, we define a class of \emph{Tree-Loop graphs}. We say that an $L$-loop fatgraph is a \emph{tadpole graph} if it has exactly one particle per trace-factor. This generalises the usual notion of a tadpole (i.e. a graph with just one external leg). A Tree-Loop graph is then formed by joining tree graphs to the legs of a tadpole graph. For example, Figure \ref{fig:Dtadpole} is a Tree-Loop graph formed by joining a 1-loop tadpole and an $(n+1)$-point tree graph. 

Take some $L$-loop tadpole fatgraph, $\Gamma_\ell$, with $h$ external legs (one for each trace-factor). Then, for some comb tree graphs $\Gamma^A$ ($A=1,\ldots,h$) we can form a Tree-Loop graph
\begin{equation}
    \Gamma = \Gamma_\ell \sqcup \bigsqcup_{A=1}^h \Gamma_A,
\end{equation}
where $\Gamma^A$ is joined to leg $A$ of $\Gamma_\ell$. Now consider the curves on this Tree-Loop graph, $\Gamma$. A curve that connects particle $i$ on trace-factor $A$ to particle $j$ on trace-factor $B$ can be decomposed into three parts:
\begin{equation}
    C = C_{i}^AC_{AB}C_{j}^B,
\end{equation}
where $C_{i}^A$ and $C_{j}^B$ are paths in the tree graphs, and $C_{AB}$ is a path in the tadpole, $\Gamma_\ell$. The curve matrix for $C$ is then the product
\begin{equation}
    M_C = W_{i}^A M_{C_{AB}} \left( W_{j}^B\right)^T,
\end{equation}
where the $W_{i}^A$ are the tree-level matrices we have already computed for all $n$ in Section \ref{sec:tree:graft}. In other words, to find the curve matrices for $\Gamma$, it suffices to compute only the the matrices $M_{C_{AB}}$ for the curves on the tadpole graph, $\Gamma_\ell$.

Note that the momentum assignments also become easier to compute. For the curve $C = C_{i}^AC_{AB}C_{j}^B$, the momentum can be written as a sum
\begin{equation}\label{eq:momgen}
    P_C = {P}_{C_{AB}} +z_j^B-z_i^A,
\end{equation}
where $P_{C_{AB}}$ is the momentum of $C_{AB}$ on the tadpole graph, $\Gamma_{\ell}$. Here, $z_i^A$ is the momentum
\begin{equation}
    z_i^A = p_{i+1} + \cdots + p_{n_A}
\end{equation}
formed by summing momenta on trace-factor $A$. $n_A$ is the number of particles incident on trace-factor $A$, and the particles are ordered $(12\cdots n_A)$ by convention. So the momenta of curves on $\Gamma$ follow  from the momentum assignments for the tadpole graph, $\Gamma_\ell$. 

The upshot is that $n$-point amplitudes at $L$ loops with $h$ trace-factors, can be computed by studying an $h$-point $L$-loop fatgraph.

\section{Planar One-Loop Amplitudes Revisited}\label{sec:Dnrev}
In Section \ref{sec:Dn} we computed the planar one-loop amplitudes. We now revisit these amplitudes, by deriving them from a study of the 1-loop planar tadpole.

\begin{figure}
\begin{center}
\includegraphics[scale=0.37]{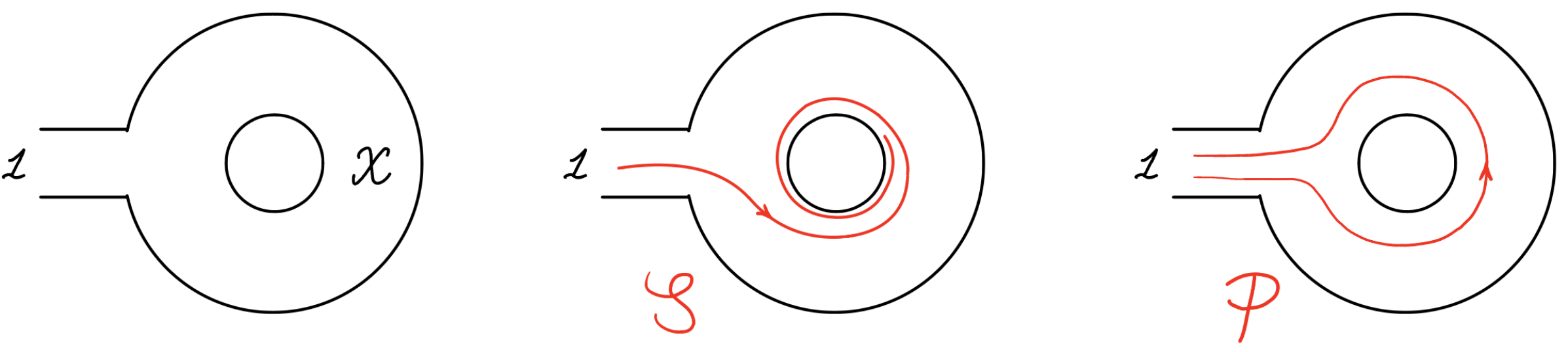}
\caption{The Planar One-loop tadpole (left), and its two curves: $S$ and $P$.}
\label{fig:tadpole}
\end{center}
\end{figure}

The 1-loop planar tadpole graph, $\Gamma_\ell$, is shown in Figure \ref{fig:tadpole}. The tadpole has two curves: a single loop, $P$, and a spiral, $S$. We discard the clockwise spiral.\footnote{In general, only one helicity of spirals enter into the formulas for amplitudes. As explained in \cite{us_qft}, including both spirals leads to overcounting of Feynman diagrams.} The paths of these curves are
\begin{equation}
   P=RxR,\qquad\text{and}\qquad S=R(xL)^\infty.
\end{equation}
So the two curve matrices are
\begin{align}\label{eq:PS}
    P = \begin{bmatrix} 1 & 1+x \\ 0 & x \end{bmatrix},\qquad\text{and}\qquad S = \begin{bmatrix} 1 & 1 \\ x & 1 \end{bmatrix}.
\end{align}
The headlight functions are
\begin{equation}
    \alpha_P = t_x - \max(0,t_x),\qquad \alpha_S = -t_x.
\end{equation}
Note that $P$ has zero momentum, whereas $S$ has momentum $x_0^\mu$, if $x_0^\mu$ is the loop momentum variable running through the one edge of the graph.

As an aside, consider evaluating the tadpole amplitude. The curve integral formula for the loop integrand gives
\begin{equation}
    {\cal I}_{\rm tadpole} = \int\limits_{-\infty}^0 dt_x\, \exp\left({-\alpha_P m^2 - \alpha_S (m^2+\ell^2)}\right) = \int\limits_{-\infty}^0 dt_x\, e^{t_x \ell^2}.
\end{equation}
After integrating out the loop momentum, the curve integral for the amplitude is
\begin{equation}
    {\cal A}_{\rm tadpole} = \int\limits_{-\infty}^0 dt_x\,\left(\frac{\pi}{-t_x}\right)^{\frac{D}{2}},
\end{equation}
as expected.

To compute the $n$-point amplitude, join a comb tree graph $\Gamma^1$ to the one leg of the tadpole. The result is the \emph{Tree-Loop} graph, $\Gamma$, in Figure \ref{fig:Dtadpole}. The curves on $\Gamma$ can be written in Tree-Loop factorised form. In addition to the curves on $\Gamma^1$, $C_{ij}$ (with $i<j$), that do not intersect the tadpole, the other curves are (with $i\leq j$)
\begin{align}
    C_{ji} = W_{i}^1 \, P\, \left({W}_{j}^1\right)^T,\qquad C_{i0} = W_{i}^1 S,
\end{align}
where $W_i^1$ is the curve that connects edge $i$ on the tree, to the edge that joins $\Gamma^1$ to the the tadpole. $(~)^T$ denotes word reversal. Note that $M_{C^T}$ is given by the matrix transpose, $M_C^T$. The curve matrices for these paths are given in Section \ref{sec:tree:graft}.

Using \eqref{eq:PS}, the curve matrix for $C_{i0}$, for example, is
\begin{equation}
    M_{i0} = W_i^1 S = \begin{bmatrix} 1 & 0 \\ F_{1,n-2}&y_{1,n-1}\end{bmatrix} \begin{bmatrix} 1 & 1 \\ x & 1 \end{bmatrix} = \begin{bmatrix} 1 & 0 \\ F_{1,n-2}+xy_{1,n-1} & F_{1,n-2}+y_{1,n-1} \end{bmatrix},
\end{equation}
so that the headlight function for this curve is
\begin{align}
    \alpha_{i0} &= \max(0~,~t_1~,~\ldots~,~t_1+\cdots+t_{n-1}) \\ &~~~~~ - \max(0~,~t_1~,~\ldots~,~t_1+\cdots+t_{n-2}~,~t_1+\cdots+t_{n-1}+t_x),\notag
\end{align}
where $t_x$ is the tropical variable for $x$. The other headlight functions can be similarly computed.

Substituting the headlight functions into our curve integral formula gives a new formula for the planar one-loop amplitudes. In fact, an interesting simplification arises. Recall that ${\cal U}$ is given by
\begin{equation}
    {\cal U} = \sum_{i=1}^n \alpha_{i0}.
\end{equation}
Applying the telescopic property, this simplifies to
\begin{equation}
    {\cal U} = \alpha_S = - t_x.
\end{equation}
The amplitude then takes the form
\begin{equation}
    {\cal A}_{\text{1-loop}} = \int\limits_{t_x<0} d^n {\bf t} \, \left(\frac{\pi}{-t_x}\right)^{\frac{D}{2}} \, \exp \left( \frac{{\cal F}_0}{t_x} - {\cal Z} \right),
\end{equation}
where ${\bf t}=(t_1,\ldots,t_{n-1},t_x)$. In this integral, the first factor,
\begin{equation}
    \left(\frac{\pi}{-t_x}\right)^{\frac{D}{2}},
\end{equation}
is independent of $n$, and is precisely what is obtained when we evaluate the (divergent) tadpole amplitude.

This is a first sign that Tree-Loop factorization simplifies the final amplitude formulas. As we will see in the following sections, Tree-Loop factorization becomes even more useful at higher orders in the perturbation series.

\section{The Non-planar One-loop Amplitudes}\label{sec:apq}
As a more complicated example, consider the one-loop double-trace amplitudes, for any number of particles, $n=n_1+n_2$. The all-multiplicity case follows from an analysis of the 1-loop double-trace tadpole graph. The final formula we obtain for the amplitude takes the form (with $E=n_1+n_2$)
\begin{equation}
    {\cal A}_{n_1,n_2} = \int d^E t\, {\cal K}\, \left(\frac{\pi}{\cal U}\right)^{\frac{D}{2}} \, \exp \left( -\frac{{\cal F}_0}{\cal U} - {\cal Z}\right).
\end{equation}
Here, ${\cal U}$ and ${\cal K}$ have simple formulas that follow from studying the tadpole graph. Whereas ${\cal F}_0$ and ${\cal Z}$ have all-$n$ formulas involving the kinematics, that we derive by adding trees to the tadpole fatgraph. The number of terms in ${\cal F}_0$ and ${\cal Z}$ grows polynomially with $n$.

\subsection{The Tadpole}\label{sec:apq:tadpole}
\begin{figure}
\begin{center}
\includegraphics[scale=0.37]{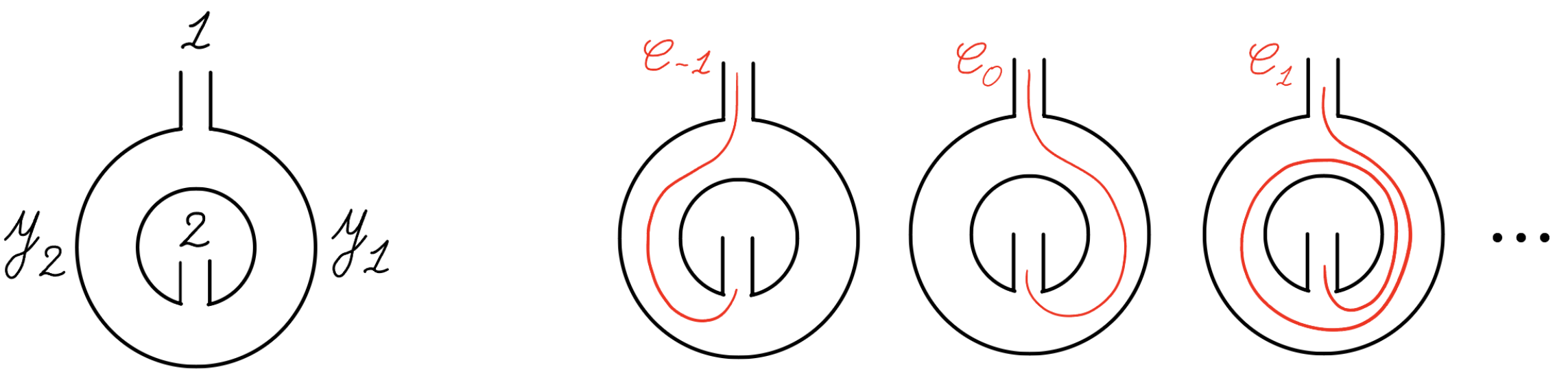}
\caption{The non-planar one-loop tadpole-like graph (left), and examples of curves on the graph (right).}
\label{fig:A11}
\end{center}
\end{figure}

The tadpole fatgraph, $\Gamma_\ell$, is given in Figure \ref{fig:A11}. It has two trace-factors, each with one particle, labelled $1$ and $2$. There are two curves on $\Gamma_\ell$ that begin and end at the same point, and carry zero momentum:
\begin{equation}\label{eq:a11:cur}
C_{11} = L e_1 L e_2 L, \qquad C_{22} = L e_1 L e_2 L.
\end{equation}
There are also curves that connect $1$ and $2$. For example, reading off from the fatgraph, the first few of these curves are
\begin{equation}\label{eq:a11:curves}
     C_{12}^{-1} = R e_2 L,\qquad
    C_{12}^0 = L e_1 R,\qquad
    C_{12}^1 = L e_1 (L e_2 R e_1) R.
\end{equation}
These three curves differ from each other by Dehn twists around the closed path
\begin{equation}
    \Delta = L e_2 R e_1.
\end{equation}
Dehn twists around $\Delta$ generate the MCG for this diagram.

Let $p_1^\mu = - p^\mu$ and $p_2^\mu =  p^\mu$ be the momentum assigned to the external legs. Then the momentum associated to $\Delta$ is
\begin{equation}
P_\Delta = - p^\mu.
\end{equation}
Assign a momentum routing to the internal edges of the fatgraph ($e_1$ and $e_2$) by setting, say, $p_{e_2} = \ell - p$, and $p_{e_1} = \ell - 2p$. Then the three curves have momenta
\begin{equation}\label{eq:a11:mom}
    P_{12}^{-1} = \ell - p,\qquad P_{12}^0 = \ell,\qquad P_{12}^1 = \ell+p.
\end{equation}

\subsection{Aside on the Role of the Mirzakhani Kernel}\label{sec:apq:aside}
$\Gamma_\ell$ is sufficiently simple that it is a feasible exercise to write down paths for \emph{all} curves on the fatgraph. The complete set of curves can be labelled as
\begin{equation}
    C_{11},\qquad C_{22},\qquad C_{12}^w,
\end{equation}
for $w\in\mathbb{Z}$. Here, $C_{12}^{w+1}$ is obtained from $C_{12}^w$ by inserting a twist around the closed curve $\Delta$. The curves $C^{-1}_{12},C^0_{12},C^1_{12}$ are as above. Moreover, the momentum assignments in \eqref{eq:a11:mom} generalise to
\begin{equation}
    P_{12}^m = \ell + m p,\qquad P_{11} = 0,\qquad P_{22} = 0.
\end{equation}
With these momentum assignments, the tadpole amplitude can be given by the curve integral
\begin{equation}
    {\cal A}_{\text{tadpole}} = \int \frac{d^2 t}{\MCG}\,\left(\frac{\pi}{\cal U}\right)^{\frac{D}{2}} \exp\left( -\frac{{\cal F}_0}{\cal U} - {\cal Z} \right),
\end{equation}
where it is still necessary to quotient by the action of MCG. In this formula, the surface Symanzik polynomials are formally given as infinite sums:
\begin{gather}
    {\cal U} = \sum_{w=-\infty}^\infty \alpha_{12}^w,\qquad{\cal F}_0 = {\cal J}^\mu {\cal J}_\mu,\\
{\cal Z} = m^2(\alpha_{11}+\alpha_{22}) + \sum_{w=-\infty}^\infty (m^2+w^2p^2) \alpha_{12}^w,
\end{gather}
where
\begin{equation}
    {\cal J}^\mu = p^\mu \sum_{w=-\infty}^\infty \,w \,\alpha_{12}^w.
\end{equation}

In practice, these infinite expressions are not useful for computations. This emphasizes the important role played by the \emph{Mirzakhani kernel}, ${\cal K}$. The integral becomes
\begin{equation}
        {\cal A}_{\rm tadpole} = \int {d^2 t}\,{\cal K}\,\left(\frac{\pi}{\cal U}\right)^{\frac{D}{2}} \exp\left( -\frac{{\cal F}_0}{\cal U} - {\cal Z} \right).
\end{equation}
In the region where ${\cal K}\neq 0$, most headlight functions, $\alpha_C$, are vanishing. In general, only finitely many headlight functions contribute to this formula. So ${\cal U}, {\cal F}_0$ and ${\cal Z}$ are all given instead by \emph{finite sums}. And it is only necessary to find the paths for a finite number of curves in order to use the formula. 

\subsection{MCG and Mirzakhani Kernel}\label{sec:apq:mcg}
The MCG is generated by Dehn twists around the closed curve $\Delta$. Write $\gamma_\Delta$ for the action of this Dehn twist on curves. $\Delta$ does not intersect $C_{11}$ or $C_{22}$, so these curves are MCG invariant. Whereas $\Delta$ intersects any curve, $C_{12}$, connecting $1$ and $2$. So these curves are acted on by $\gamma_\Delta$.

The curves $C_{12}$ form a single coset under the action of $\gamma_\Delta$. Take $C_{12}^0$ as a coset representative. Then a Mirzakhani kernel is given by
\begin{equation}
    {\cal K} = \frac{\alpha_{12}^0}{\rho},
\end{equation}
where $\rho$ is the sum
\begin{equation}
    \rho = \sum_{C_{12}} \alpha_{C_{12}}
\end{equation}
over all curves connecting $1$ and $2$.

However, here a simplification arises. Draw $C_{12}^0$ on the fatgraph $\Gamma_\ell$. There are only \emph{four} curves that do not intersect $C_{12}^0$. These are
\begin{equation}
    C_{11}, ~~C_{22},~~C_{12}^{-1},~~C_{12}^1,
\end{equation}
whose paths are given above (see \eqref{eq:a11:cur} and \eqref{eq:a11:curves}). So the Mirzakhani kernel is in fact
\begin{equation}
    {\cal K} = \frac{\alpha_{12}^0}{\alpha_{12}^{-1}+\alpha_{12}^0+\alpha_{12}^1}.
\end{equation}

Moreover, the surface Symanzik polynomials for the tadpole, which were given as formal expressions in Section \ref{sec:apq:aside}, also simplify in the region where ${\cal K}$ is non-vanishing. They become (if $p^\mu$ is on-shell, say)
\begin{gather}
{\cal U} = \alpha_{12}^{-1}+\alpha_{12}^0+\alpha_{12}^1,\qquad {\cal F}_0 = p^2 (\alpha_{12}^1-\alpha_{12}^{-1})^2\\
{\cal Z} = m^2(\alpha_{11}+\alpha_{22}+\alpha_{12}^0).
\end{gather}

In summary, the pre-factor appearing in the tadpole amplitude formula, 
\begin{equation}
    {\cal K}\,\left(\frac{\pi}{\cal U}\right)^{\frac{D}{2}},
\end{equation}
can be computed using the paths for 3 curves. While the whole integrand for the integral can be computed using the paths for a set, ${\cal S}$ of 5 curves. The curve matrices for these five curves is given in Appendix \ref{app:apq}.

As a final remark, using the Appendix, the headlight functions for the five curves are
\begin{gather}
    \alpha_{12}^{-1} = \max(0,t_2)-t_2,\qquad \alpha_{12}^0 = \max(0,t_1),~ \\ \alpha_{12}^1 = \max(0,2t_1,2t_1+t_2)-2\max(0,t_1),~\\
    \alpha_{11} = \alpha_{22} = t_1+t_2 - \max(0,t_1,t_1+t_2).
\end{gather}
Restricting to the region $t_1 > 0$, where ${\cal K}\neq 0$, they simplify to the headlight functions become
\begin{gather}
    \alpha_{12}^{-1} = \max(0,t_2)-t_2,\qquad \alpha_{12}^0 = t_1, \\ \alpha_{12}^1 = \max(0,t_2),\qquad
    \alpha_{11} = t_2 - \max(0,t_2).
\end{gather}
So, in fact, ${\cal K}$ and ${\cal U}$ take especially simple forms in coordinates:
\begin{equation}\label{eq:apq:KU}
    {\cal K} = \frac{t_1}{t_1+|t_2|},\qquad{\cal U} = t_1 + |t_2|.
\end{equation}
We will see below that these formulas continue to hold for the all-multiplicity formula.

\subsection{The Tree-Loop Fatgraph}
\begin{figure}
\begin{center}
\includegraphics[scale=0.25]{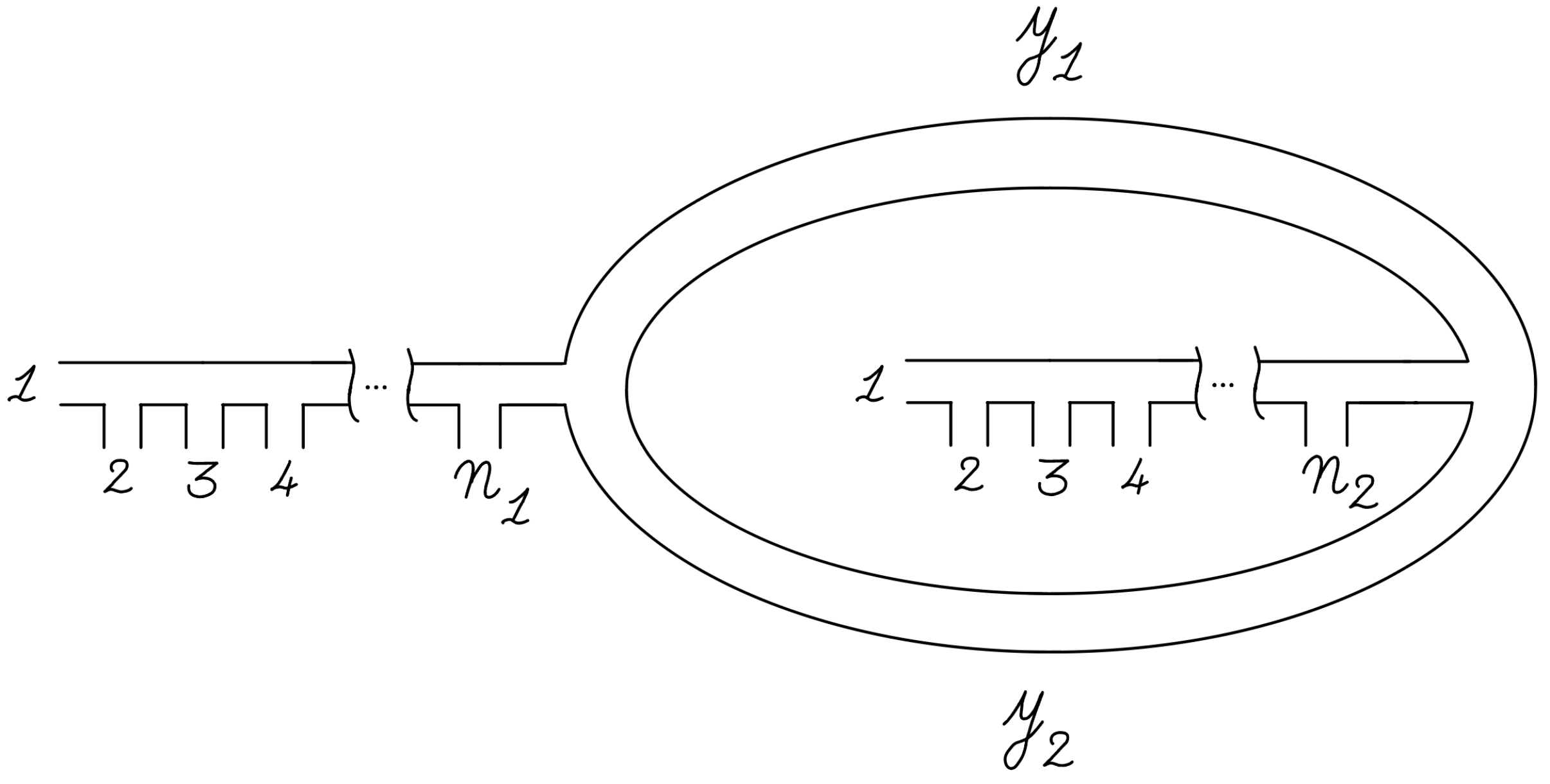}
\caption{The Tree-Loop graph for the 1-loop non-planar amplitude.}
\label{fig:Atildetadpole}
\end{center}
\end{figure}

To compute the $n$-point amplitude, consider joining the two ends of the tadpole $\Gamma_\ell$ to two comb tree graphs, $\Gamma^1$ and $\Gamma^2$, with $n_1$ and $n_2$ particles respectively. The result is the Tree-Loop graph for the problem, $\Gamma$, as in Figure \ref{fig:Atildetadpole}. In this section, we see how to find the curves on $\Gamma$ and their momenta explicitly. 

The curves on $\Gamma$ can be obtained from the curves on $\Gamma_\ell$. For example, the curve $C_{12}^0$ can be extended to a path
\begin{equation}\label{eq:Apq:Cijm}
    C_{ij}^0 = W_{i}^1 C_{12}^0 \left( W_{j}^2\right)^T,
\end{equation}
that connects particle $i$ on trace-factor $1$ and particle $j$ on trace-factor $2$. Here, $W_i^A$ is the path through the tree $\Gamma_A$ from particle $i$ to the edge joining the tree to $\Gamma_\ell$, and $(~)^T$ denotes path-reversal. Similarly, a curve that begins and ends on a single trace-factor, $A$, (with $A=1$ or $2$) has a path of the form (with $i<j$)
\begin{equation}\label{eq:Apq:Cij}
    C_{ij} = W_{ij}^A,\qquad\text{or}\qquad C_{ji} = W_{i}^A C_{AA} \left( W_{j}^A\right)^T,
\end{equation}
where $W_{ij}^A$ is the curve connecting particles $i,j$ in $\Gamma^A$ that does not pass through $\Gamma_\ell$.

The momenta of curves on $\Gamma$ also follow from the momentum assignment to curves on the tadpole, $\Gamma_\ell$. Let $p_{i}^1$ and $p_{j}^2$ ($i=1,\ldots,n_1$, $j=1,\ldots,n_2$) be the external momenta. By momentum conservation,
\begin{equation}
    \sum_{i=1}^{n_1} p_{i}^1 =  p^\mu, \qquad \sum_{j=1}^{n_2} p_{j}^2 = - p^\mu.
\end{equation}
Then the momentum of $C_{ij}^0$, for example, is given by
\begin{equation}
    P_{ij}^0 = \ell + z_j^2- z_i^1,
\end{equation}
where recall that the curve $P_{12}^0$ on the tadpole has momentum $\ell$. Here,
\begin{equation}
    z_i^A = p_{i+1}^A + \cdots + p_{n_A}^A
\end{equation}
are the dual-like variables on each trace-factor. (The general rule was given as equation \eqref{eq:momgen}.) Similarly, the curves in \eqref{eq:Apq:Cij} have momenta (with $A=1$ or $2$)
\begin{equation}
    P_{ij}^A = p^A_{i+1}+\cdots+p^A_j,\qquad P_{ji}^A = z_j^A - z_i^A.
\end{equation}

\subsection{The All-multiplicity Amplitude}\label{sec:apq:alln}
We now use the Tree-Loop graph, $\Gamma$, to find the curve integral for the all-multiplicity amplitude. We find that the end result is not that much more complicated than the tadpole case. This is true in general, for reasons explained further in Section \ref{sec:general}.

The MCG action on $\Gamma$ is induced by the MCG on the tadpole, $\Gamma_\ell$. It is generated by the Dehn twist $\gamma_\Delta$, which acts non-trivially only on the curves $C_{ij}^w$,
\begin{equation}
    C_{ij}^w = C_i^1 C_{12}^w C_j^2,
\end{equation}
that connect trace-factor 1 and trace-factor 2. Here, $C_{12}^w$ are the curves connecting 1 and 2 on the tadpole. $\gamma_\Delta$ does not alter the tree-parts of the curve. So this set of curves is decomposed into cosets: one coset for every pair $i,j$ of particles. Take $C_{ij}^0$ as coset representatives. Then a Mirzakhani kernel is
\begin{equation}
    {\cal K} = \sum_{i=1}^{n_1}\sum_{j=1}^{n_2} \frac{\alpha_{ij}^0}{\rho},
\end{equation}
where $\rho$ is the sum
\begin{equation}
    \rho = \sum_{C_{ij}^w} \alpha_{ij}^w
\end{equation}
over all such curves.

However, the telescopic property dramatically simplifies this expression. In particular,
\begin{equation}
    \sum_{i=1}^{n_1}\sum_{j=1}^{n_2} {\alpha_{ij}^0} = \alpha^0,
\end{equation}
where $\alpha^0$ is the headlight function of the curve $C_{12}^0$ on the tadpole graph. Likewise, $\rho$ simplifies to
\begin{equation}
    \rho = \alpha^{-1}+\alpha^0+\alpha^1,
\end{equation}
where we recall (from Section \ref{sec:apq:mcg}) that these are the only terms that contribute in the region $\alpha^0\neq 0$. So the all-multiplicity Mirzakhani kernel is
\begin{equation}
    {\cal K} = \frac{\alpha^0}{\alpha^{-1}+\alpha^0+\alpha^1},
\end{equation}
which is identical to the ${\cal K}$ found for the tadpole.

In the region $\alpha^0\neq 0$, $\alpha_C$ vanishes unless $C$ is a curve that does not intersect the curve $C_{12}^0$ on the fatgraph. In other words, the only curves we need to consider are those constructed from the special set of five curves
\begin{equation}
    C_{11}, ~~C_{22},~~C_{12}^{-1},~~C_{12}^0~~C_{12}^1,
\end{equation}
considered in Section \ref{sec:apq:mcg}. ${\cal U}$ is a sum over all curves that carry a loop momentum. These are the curves joining trace-factor 1 and trace-factor 2. So
\begin{equation}
    {\cal U} = \sum_{w=-1}^1 \sum_{i=1}^{n_1}
\sum_{j=1}^{n_2} \alpha_{ij}^w = \alpha^{-1}+\alpha^0+\alpha^1,
\end{equation}
which has again been simplified using the telescopic relation. Note that this ${\cal U}$ is identical to the one obtained for the tadpole case

Using the coordinate expressions for ${\cal K}$ and ${\cal U}$, equation \eqref{eq:apq:KU}, the all-multiplicity amplitude is then given by
\begin{equation}\label{eq:apq:final}
    {\cal A}_{n_1,n_2} = \int\limits_{t_1>0} d^E t\, {\frac{t_1}{t_1+|t_2|}}\, \left(\frac{\pi}{t_1+|t_2|}\right)^{\frac{D}{2}}\, \exp\left(-\frac{{\cal F}_0}{t_1+|t_2|} - {\cal Z}\right).
\end{equation}
The remaining surface Symanzik polynomials are given by finite sums of curves. In particular, ${\cal F}_0 = {\cal J}^\mu {\cal J}_\mu$ where
\begin{equation}
{\cal J}^\mu = \sum_{w=-1}^1 \sum_{i,j} \alpha_{ij}^m (z_j^2-z_i^1+wp).
\end{equation}
And ${\cal Z}$ is a sum over the finite set of all curves that do not intersect $C_{12}^0$:
\begin{equation}
    {\cal Z} = \sum_{A=1,2}\sum_{i\neq j} \alpha_{ij}^A \left( \left(P_{ij}^A\right)^2 +m^2\right)+ \sum_{w=-1}^1\sum_{i,j} \alpha_{ij}^w \left( (z_j^2-z_i^1 + wp)^2 + m^2\right).
\end{equation}

\subsection{Comment on Evaluating the Curve Integral}
The curve integral formula for the amplitude, \eqref{eq:apq:final}, has two parts. There is the pre-factor
\begin{equation}
    {\cal K}\,\left(\frac{\pi}{\cal U}\right)^{\frac{D}{2}},
\end{equation}
and the exponential part,
\begin{equation}\label{eq:apq:finalFZ}
    \int d^{E-2} t\,\exp\left(-\frac{{\cal F}_0}{\cal U} - {\cal Z}\right),
\end{equation}
which contains the kinematics. To evaluate the pre-factor, it was only necessary to know the curve matrices for 3 curves on the tadpole graph. (These matrices are given in appendix \ref{app:apq}.)

To evaluate the kinematic part, it was necessary to use the set of five curves on the tadpole graph that are non-intersecting with $C_{12}^0$:
\begin{equation}\label{eq:apq:CCC}
C_{11},~~C_{22},~~C_{12}^{-1},~~C_{12}^0,~~C_{12}^1.
\end{equation}
Out of these five curves, we can build the relevant set of curves on $\Gamma$:
\begin{equation}
    C_{ij}^1,~ C_{kl}^2,~C_{ik}^{-1},~C_{ik}^0,~C_{ik}^1,
\end{equation}
for $i,j=1,\ldots,n_1$, $k,l=1,\ldots,n_2$ and $i\leq j$, $k\leq l$. There are 
\begin{equation}
3n_1n_2+n_1(n_1-2)+n_2(n_2-2)    
\end{equation}
such curves. The curve matrices for all these curves follow from taking products of the five tadpole curve matrices (for the curves \eqref{eq:apq:CCC}), as well as the tree matrices
\begin{equation}
W_{ij}^1, ~W_{kl}^2, ~W_{i}^1, ~W_{k}^2,
\end{equation}
which were computed in Section \ref{sec:tree:graft}. There are
\begin{equation}
\frac{1}{2}n_1(n_1-1)+\frac{1}{2}n_2(n_2-1)
\end{equation}
such matrices. So both the number of terms in the exponent of \eqref{eq:apq:finalFZ}, and the number of matrices needed to compute them, grow polynomially in $n_1,n_2$.

\section{Prefactors for All-multiplicity}\label{sec:general}
For the 1-loop double-trace amplitudes studied in Section \ref{sec:apq}, both the Mirzakhani kernel, ${\cal K}$, and the surface Symanzik polynomial, ${\cal U}$, did not depend on the number of particles. In this section, we explain that this \emph{always} happens, at all orders in perturbation theory, when using a Tree-Loop graph.

\subsection{Mirzakhani Kernels}

\begin{figure}
\begin{center}
\includegraphics[scale=0.45]{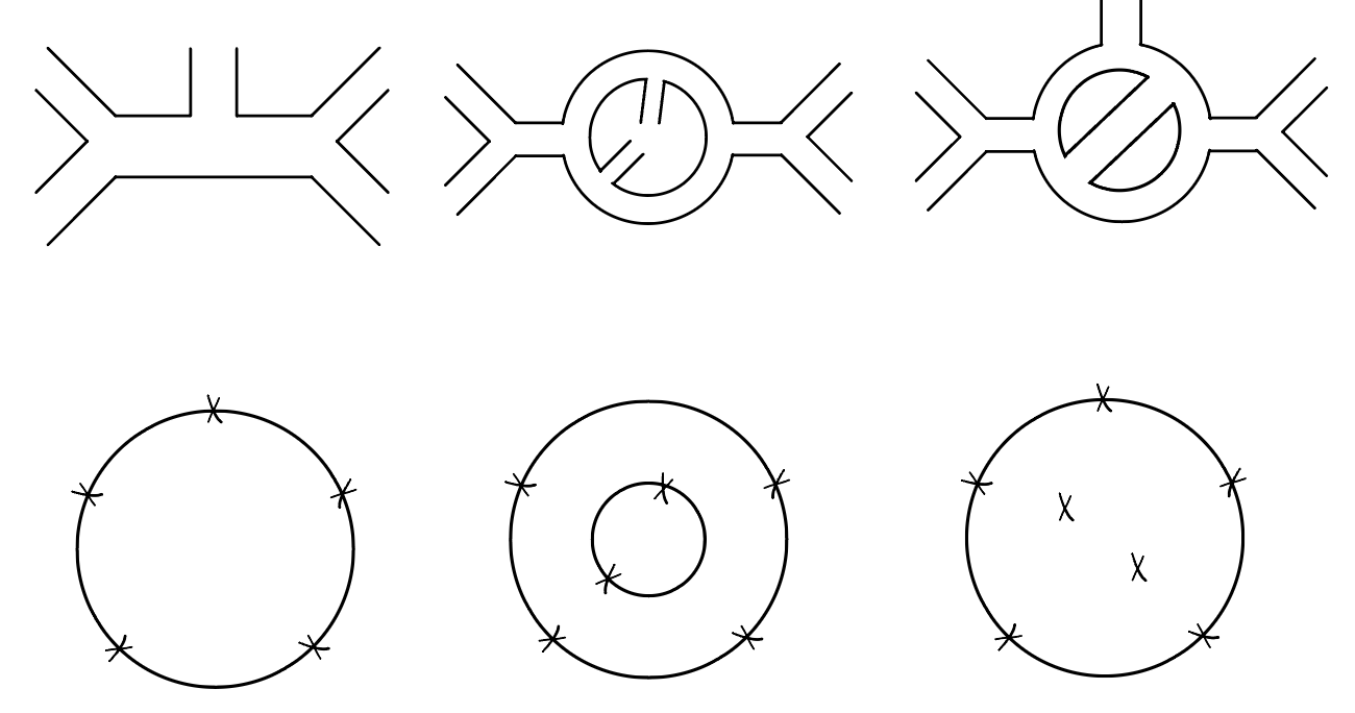}
\caption{Turning a fatgraph into a surface.}
\label{fig:fat}
\end{center}
\end{figure}

\begin{figure}
\begin{center}
\includegraphics[scale=.45]{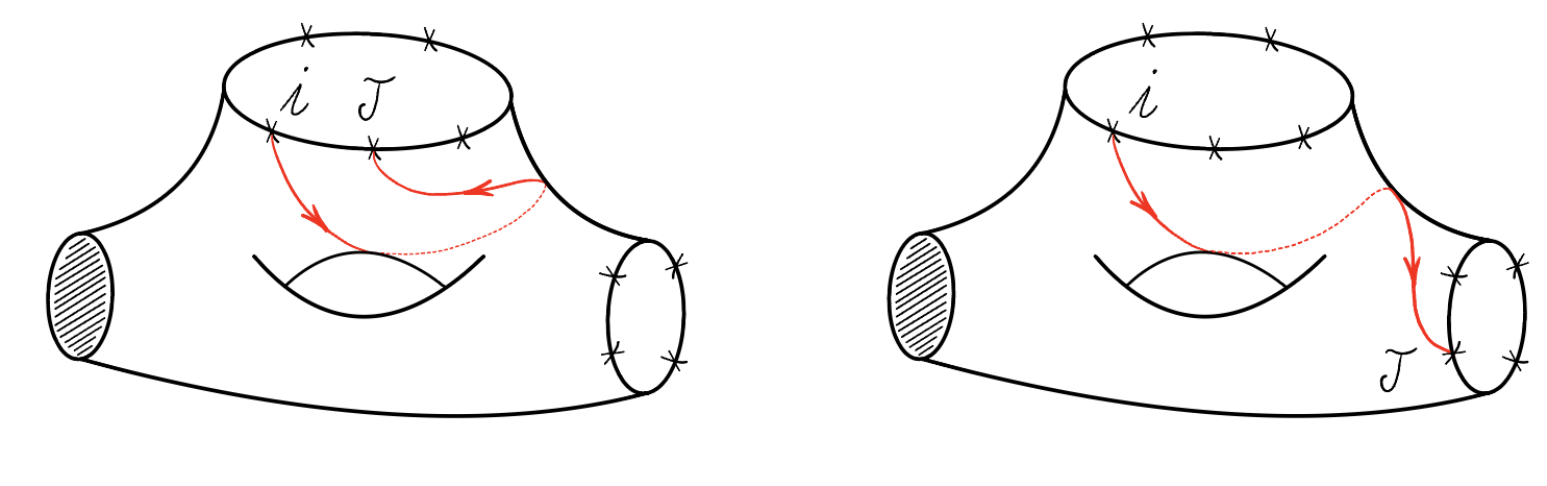}
\caption{The two types of cosets that arise when forming the Mirzakhani kernel: curves that lower the genus (left), and curves that decrease the number of trace-factors (right).}
\label{fig:cosets}
\end{center}
\end{figure}

Consider any Tree-Loop graph, $\Gamma$, with $h$ trace-factors labelled $A=1,2,\ldots,h$. It can be a useful book-keeping device when constructing the Mirzakhani kernel to also draw the surface, $S$, obtained by fattening $\Gamma$. Examples of surfaces obtained from fatgraph are in Figure \ref{fig:fat}. A path $C$ on $\Gamma$ lifts to a curve on $S$, and one can draw $C$ on $S$ without knowing its exact path. 

There are two types of curves that arise when constructing the Mirzakhani kernel, as shown in Figure \ref{fig:cosets}. Starting from a fixed trace-factor, $A$, we consider the set of curves, ${\cal S}_A$, that contain: all non-separating curves\footnote{A non-separating curve is one that does not divide the surface into two regions.} that return to trace-factor $A$; and all curves that connecting trace-factor $A$ to a second trace-factor $B$ ($B\neq A$). 

The first step in constructing the Mirzakhani kernel is to consider the coset decomposition of the set of curves ${\cal S}_A$ under the $\MCG$ action. The set can be written as the disjoint union
\begin{equation}
    {\cal S}_A = {\cal S}_{AA} \sqcup \bigsqcup_{B\neq A} {\cal S}_{AB},
\end{equation}
where ${\cal S}_{AB}$ are the curves connecting distinct trace-factors, $A$ and $B$, and ${\cal S}_{AA}$ are the set of non-separating curves that begin and end on trace-factor $A$.

Any curve in the set ${\cal S}_{AB}$ can be written in the factorized form
\begin{equation}
    C = W_{i}^A C_{AB} \left( W_{j}^B\right)^T,
\end{equation}
for some external lines $i$ and $j$, and some curve $C_{AB}$ connecting the two trace-factors on the tadpole sub-graph, $\Gamma_\ell$. The set ${\cal S}_{AB}$ decomposes into cosets, corresponding to each possible choice of start and end points, $i$ and $j$. Fix some choice of curve $C^0_{AB}$ connecting the two trace-factors on the tadpole sub-graph, $\Gamma_\ell$. Then the curves
\begin{equation} C_{ij}^0 = W_{i}^A C_{AB}^0 \left(W_{j}^B\right)^T,\end{equation}
for each $i$ and $j$ can be taken as coset representatives. In building a Mirzakhani kernel, we sum over each of these cosets, to obtain the following terms:
\begin{equation}
    {\cal K}_{AB} = \sum_{i=1}^{n_A}\sum_{j=1}^{n_B} \frac{\alpha_{ij}^0}{\rho}.
\end{equation}
where $n_A$ and $n_B$ are the number of particles on each trace-factor, and
\begin{align}
\rho = \sum_{C\in{\cal S}} \alpha_C.
\end{align}

The telescopic property gives a large simplification of the terms ${\cal K}_{AB}$. The property implies that
\begin{equation}
\sum_{i,j} \alpha_{C_{ij}^0} = \alpha_{C^0_{AB}}.
\end{equation}
So the terms in ${\cal K}_{AB}$ simplify to become
\begin{equation}
    {\cal K}_{AB} = \frac{\alpha_{AB}^0}{\rho}.
\end{equation}
Moreover, the telescopic property also simplifies the denominator, $\rho$, which can now be written as
\begin{equation}
    \rho = \sum_{B=1}^h \sum_{C_{AB}} \alpha_{C_{AB}},
\end{equation}
where the second sum is over curves $C_{AB}$ on the tadpole graph that connect $A$ and $B$. It follows that ${\cal K}_{AB}$ only depends on tadpole graph curves.

The analysis of the set ${\cal S}_{AA}$ of non-separating curves is similar. After applying the telescopic property, the sum over cosets of this set gives
\begin{equation}
    {\cal K}_{AA} = \frac{\alpha_{AA}^0}{\rho},
\end{equation}
for some choice of non-separating curve $C_{AA}^0$ on the tadpole graph. The Mirzakhani kernel for the set ${\cal S}$ is therefore
\begin{equation}
    {\cal K} = \sum_{B=1}^h \frac{\alpha_{AB}^0}{\rho}, 
\end{equation}
and this is precisely a Mirzakhani kernel for the tadpole graph, $\Gamma_\ell$. 

Inductively applying this result, to obtain a Mirzakhani kernel for the full $\MCG$, we can conclude that: \emph{A Mirzakhani kernel for a Tree-Loop graph $\Gamma$ is given by the Mirzakhani kernel ${\cal K}$ of its tadpole sub-graph, $\Gamma_\ell$.}

\subsection{The Surface Symanzik Polynomial}
The surface Symanzik function ${\cal U}$ for a Tree-Loop fatgraph $\Gamma$ also does not depend on $n$. This is because the loop momentum $\ell_C$ carried by a curve, $C = C_i^A C_{AB} C_j^B$, is just the loop momentum carried by the tadpole part of its path, $C_{AB}$. Similarly, the loop momentum $\ell_C$ carried by $C = C_i^A C_{AA} C_j^A$ is the loop momentum carried by the subpath $C_{AA}$. The entries of the $\Lambda$ matrix can therefore be written as
\begin{equation}
    \Lambda^{ab} = \sum_{A,B=1}^h \sum_{C_{AB}} \sum_{i,j}\, \ell_{C_{AB}}^a \ell_{C_{AB}}^b \alpha_{C_{iABj}},
\end{equation}
where $C_{iABj} = C_i^A C_{AB} C_j^B$. 
By the telescopic property,
\begin{equation}
    \sum_{i,j}\, \alpha_{C_{iABj}} = \alpha_{C_{AB}}.
\end{equation}
So it follows that the matrix entry $\Lambda^{ab}$ for the Tree-Loop graph $\Gamma$ is equal to the matrix entry for the tadpole subgraph $\Gamma_\ell$:
\begin{equation}
    \Lambda^{ab} = \sum_{A,B=1}^h \sum_{C_{AB}} \ell_{C_{AB}}^a \ell_{C_{AB}}^b \alpha_{C_{AB}} = \sum_{\substack{C \\ {\rm on}~\Gamma_\ell}} \ell_C^a\ell_C^b \alpha_C.
\end{equation}
We can conclude that: the surface Symanzik function for a Tree-Loop graph is equal to the surface Symanzik ${\cal U}$ of its tadpole sub-graph, $\Gamma_\ell$. 

\section{Summary of Tree-Loop factorization}
\label{sec:summary}
In this section we summarize how Tree-Loop fatgraphs, and the general results in Sections \ref{sec:treeloop} and \ref{sec:general}, can be used in practice to compute amplitudes at arbitrary multiplicity. In the subsequent sections, we then apply this method to all 2-loop amplitudes.

Let $\Gamma$ be any Tree-Loop graph with $h$ trace-factors and $L$ loops. Let $\Gamma_{\ell}$ be its tadpole subgraph. As a first step, we can compute a Mirzakhani kernel, ${\cal K}$, for the tadpole graph by analysing the MCG action on $\Gamma_\ell$.

Given this ${\cal K}$, recall that ${\cal K}$ restricts the domain of integration. We can restrict our attention to the finite set of curves, ${\cal S}$, on $\Gamma_\ell$, that have non-vanishing headlight functions on this domain. Using this set of curves, we can compute the surface Symanzik polynomial
\begin{equation}
   {\cal U} =  \det\Lambda,
\end{equation}
where $\Lambda$ is the $L\times L$ $\Lambda$-matrix of the tadpole graph. This is computed by working out the momentum assignments, $P_C$, to the curves $C\in{\cal S}$. Each momentum takes the form
\begin{equation}
P_C^\mu = K_C^\mu + \sum_{a=1}^L \ell_C^a \ell_a^\mu,
\end{equation}
for some choice of loop momentum variables $\ell_a^\mu$. Then the $\Lambda$-matrix is
\begin{equation}
    \Lambda^{ab} = \sum_{C\in{\cal S}} \ell_C^a\ell_C^b \alpha_C,
\end{equation}
which is always a finite sum.

After these preliminaries, we have almost all the data needed to write down the curve integral for all-multiplicity. The final result takes the form
\begin{align}
    {\cal A}_\Gamma = \int d\textbf{t} \mathcal{K}\left(\frac{\pi^L}{\mathcal{U}}\right)^{D/2} \exp\left(-\frac{\mathcal{F}_0}{\mathcal{U}}-\mathcal{Z}\right),
\end{align}
where ${\cal K}$ and ${\cal U}$ have already been computed. 

The remaining two functions, ${\cal F}_0$ and ${\cal Z}$ are given by sums over the finitely many curves on $\Gamma$ that can be produced by extending the curves ${\cal S}$ to the full graph. In particular, 
\begin{equation}
    {\cal F}_0 = {\cal J}_{\mu} \tilde{\Lambda} {\cal J}^\mu,
\end{equation}
where $\tilde{\Lambda} = \det(\Lambda) \Lambda^{-1}$ denotes the adjugate of the $\Lambda$-matrix, and ${\cal J}^\mu$ is an $L$-vector given by
\begin{equation}
    {\cal J}^{a,\mu} = \sum_{C_{AB}\in{\cal S}} \sum_{C = W_i^AC_{AB}W^{B}_j}\, \ell_C^a \,\alpha_{C}\,\left(K_{C_{AB}}^\mu + z_j^B - z_i^A\right).
\end{equation}
The sum is over all curves $C = W_i^A C_{AB} W^B_j$ that can be arrived at by extending a curve $C_{AB}$ in the set ${\cal S}$ to a curve on the full graph $\Gamma$. Likewise, ${\cal Z}$ is given by a similar sum
\begin{equation}
    {\cal Z} = \sum_{C_{AB}\in{\cal S}} \sum_{C = W_i^AC_{AB}W^B_j}\, \left(K_{C_{AB}}^\mu + z_j^B - z_i^A\right)^2 \alpha_C + \sum_{\substack{C\\{\rm tree}}} K_C^2\alpha_C.
\end{equation}
We also include in this sum curves $C=W_{ij}^A$ which are entirely in the tree-part of the tree-loop graph. If we have already computed the momentum assignments, $P_{C}$, for the tadpole graph curves in ${\cal S}$, these sums are easy to evaluate. In particular, the headlight function $\alpha_C$, for a curve $C = W_i^AC_{AB}W^{B}_j$, is given by computing the curve matrix
\begin{equation}
    M_C = W_i^A\,M_{AB}\,\left(W^{B}_j\right)^T.
\end{equation}
The $W_i^A$ matrices have been computed for all-multiplicity in Section \ref{sec:tree:graft}.

In summary, to find the all-multiplicity formula for the amplitude, it suffices to first study the tadpole graph. For the tadpole graph, one first derives a kernel, $\mathcal{K}$, and the corresponding finite set of curves, $\mathcal{S}$, that are compatible with ${\cal K}$. For each of these curves, $C \in \mathcal{S}$, one computes the corresponding curve matrix, $M_C$, and momentum, $P_C = \ell_C + K_C$.  These data depend only on the tadpole subgraph $\Gamma_\ell$, and they suffice to compute the all-multiplicity formula. The dependence on the particle data in $\mathcal{A}$ enters only through the vector ${\cal J}^\mu$ and the function $\mathcal{Z}$. But the computation of these functions explained above does not depend on the topology of the graph $\Gamma_\ell$. 


\section{The Planar 2-Loop Amplitudes}\label{sec:Dntilde}
\begin{figure}
\begin{center}
\includegraphics[scale=0.25]{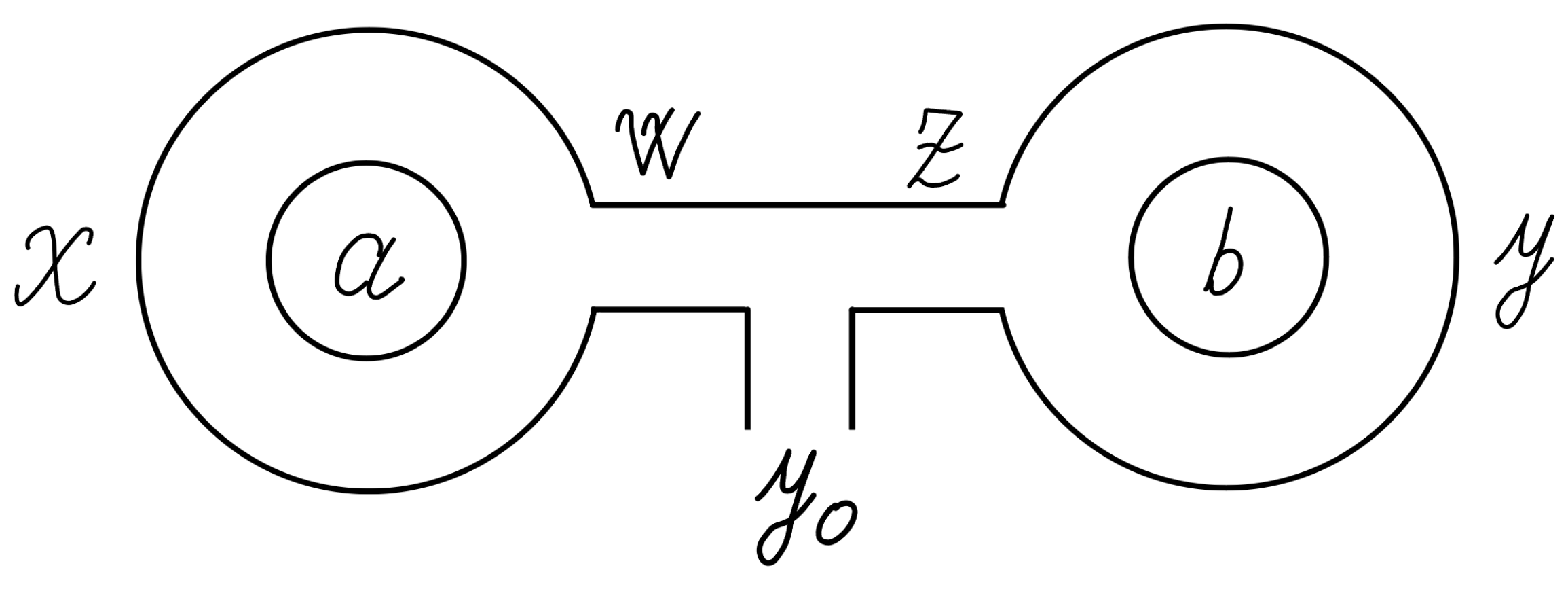}
\caption{The 2-loop planar tadpole.}
\label{fig:Dntadpole}
\end{center}
\end{figure}

This section uses the Tree-Loop method described in Section \ref{sec:summary} to compute the planar 2-loop amplitudes. Most of the computation is concerned with the planar 2-loop tadpole, and then the all-multiplicity result is derived at the end.

\subsection{MCG and Mirzakhani Kernel}
\begin{figure}
\begin{center}
\includegraphics[scale=0.3]{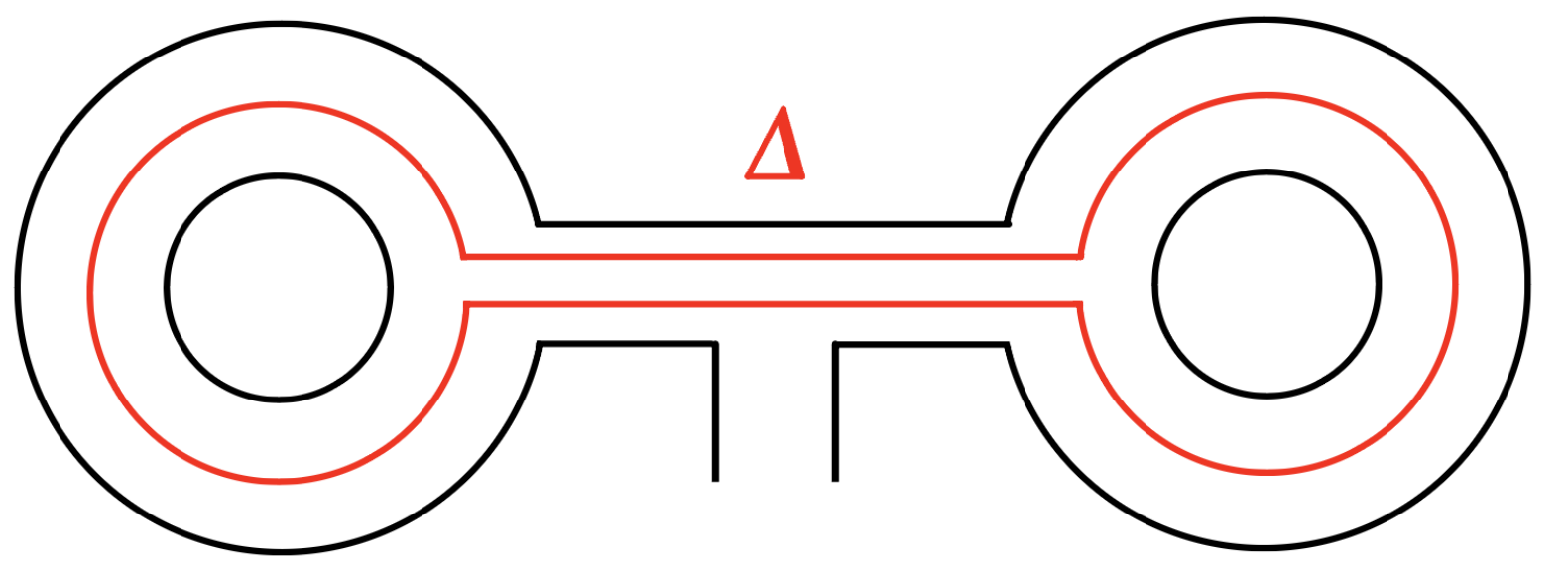}
\caption{The closed curve, $\Delta$, on the 2-loop planar tadpole graph.}
\label{fig:Dntilde:delta}
\end{center}
\end{figure}
\begin{figure}
\begin{center}
\includegraphics[scale=0.25]{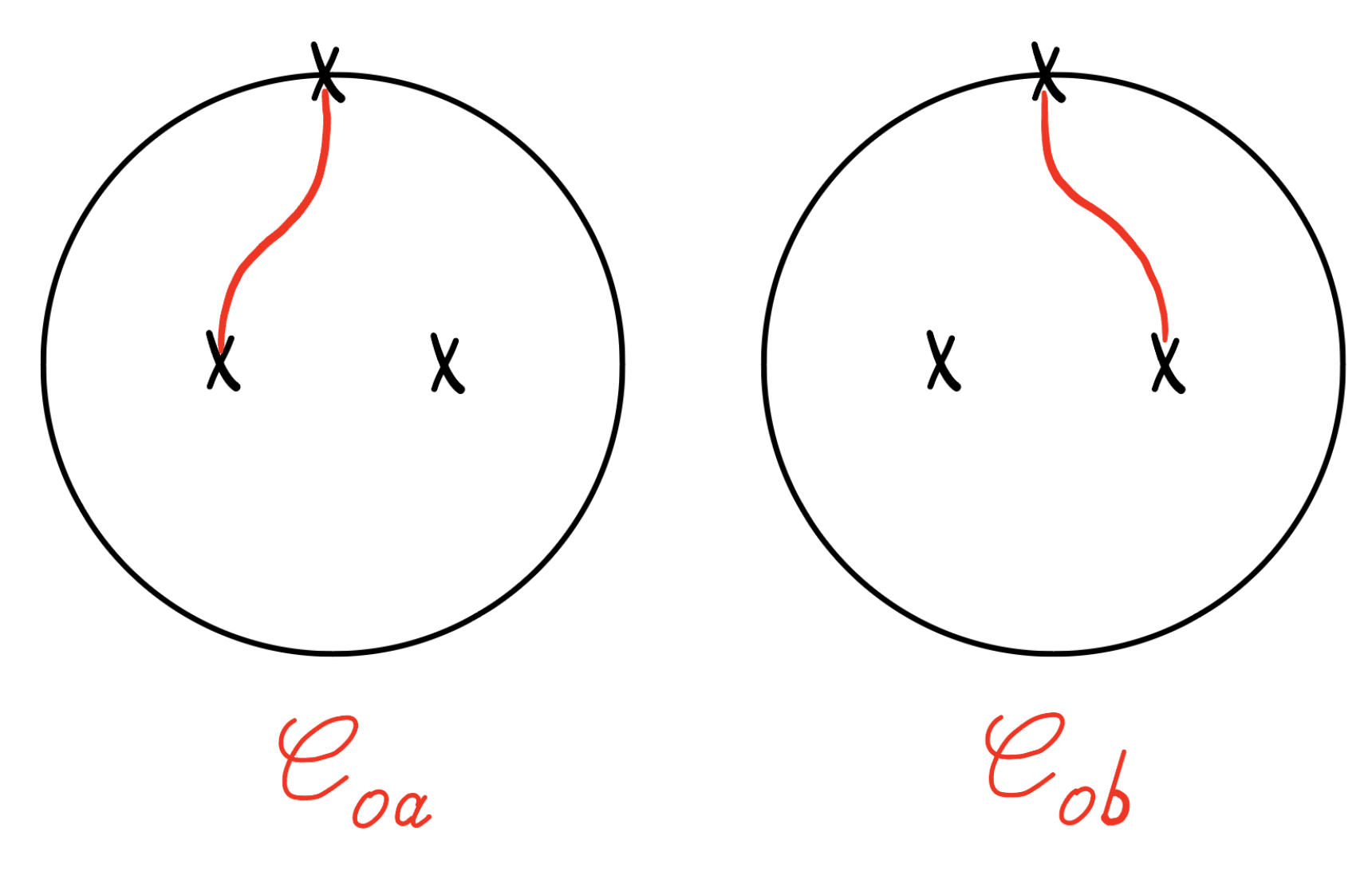}
\caption{The surface picture obtained by fattening the tadpole graph, with two MCG cosets: curves $C_{0a}$ that end in a spiral around $a$ (left) and curves $C_{0b}$ that end in a spiral around $b$ (right).}
\label{fig:Dntilde:surface}
\end{center}
\end{figure}

A 2-loop planar tadpole fatgraph, $\Gamma_\ell$, is given in Figure \ref{fig:Dntadpole}. The graph is planar, with two closed loops we label $a$ and $b$. There is one nontrivial closed curve on $\Gamma_\ell$, $\Delta$, shown in Figure \ref{fig:Dntilde:delta}. This closed curve has (anti-clockwise) path
\begin{equation}
    \Delta = z R y R z R w R x R w L.
\end{equation}
Dehn twists around $\Delta$, $\gamma_\Delta$, generate the MCG of the graph.

To find a Mirzakhani kernel, consider how Dehn twists around $\Delta$ act on the set of curves with one endpoint on the trace-factor. (To produce the Mirzakhani kernel, we ignore curves that factorize $\Gamma_\ell$ into two graphs.) Using the surface diagram, Figure \ref{fig:Dntilde:surface}, obtained by fattening $\Gamma_\ell$ to a surface, it is easy to identify that there are two cosets under MCG: the curves $C_{0a}$ that end in a spiral around $a$, and the curves $C_{0b}$ that end in a spiral around $b$.

Exploring possible paths on the fatgraph, $\Gamma_\ell$, we identify two possible coset representatives,
\begin{align}
    C_{0a}^0= L w R (xL)^\infty,\qquad \text{and}\qquad C_{0b}^0 = R z R (yL)^\infty.
\end{align}
These are the shortest paths connecting $0$ to a spiral around $a$ or $b$. Other paths can be obtained by acting with Dehn twists $\gamma_\Delta$. Acting once with $\gamma_\Delta$ gives paths
\begin{align}
    C_{0a}^1 = R z R y R z R w R (xL)^\infty, \qquad  C_{0b}^1 = R \Delta z R (yL)^\infty.
\end{align}
Whereas acting with $\gamma_\Delta$ (or twisting `clockwise' around $\Delta$) gives paths
\begin{align}
    C_{0a}^{-1} = L w L x L w L z L y L z R w R (xL)^\infty,\qquad
    C_{0b}^{-1} = L w L x L w L z R (yL)^\infty.
\end{align}

The Mirzakhani kernel is then
\begin{equation}\label{eq:DntildeK}
    {\cal K} = \frac{\alpha_{0a}^0}{\rho} + \frac{\alpha_{0b}^0}{\rho},
\end{equation}
where $\rho$ is a sum over all curves of the form $C_{0a}$ or $C_{0b}$ that are compatible with either $C_{0a}^0$ or $C_{0b}^0$. We write
\begin{equation}
     \rho = \sum_{C_{0a}\in{\cal S}} \alpha_{C_{0a}} + \sum_{C_{0b}\in{\cal S}} \alpha_{C_{0b}},
\end{equation}
where ${\cal S}$ be the set of \emph{all} curves on $\Gamma_\ell$ which are compatible with either $C_{0a}^0$ or $C_{0b}^0$.

We find that there are 17 curves in ${\cal S}$. All of these curves and their paths are listed in Appendix \ref{app:Dntilde}.

\subsection{Tadpole Curves and Momenta}
The tadpole fatgraph is planar, so the momentum assignments can be solved using dual momentum variables: $x_0^\mu, x_a^\mu, x_b^\mu$. Curves that begin at the external line, $0$, and end in a spiral, $a$ or $b$, have propagators
\begin{align}\label{eq:X0aX0b}
    X_{0a} = (x_a-x_0)^2+m^2,\qquad
    X_{0b} = (x_b-x_0)^2+m^2.
\end{align}
Likewise curves that begin spiraling around $a$ and end spiraling around $b$ have propagator
\begin{equation}\label{eq:Xab}
    X_{ab} = (x_a-x_b)^2+m^2.
\end{equation}
Finally, any curve that begins and ends at $0$ has zero momentum. It is convenient to set $x_0=0$ and use $x_a$ and $x_b$ as loop momentum variables.

The $\Lambda$-matrix is constructed from the headlight functions $\alpha_C$ for curves $C$ that carry loop momentum. But from this momentum assignment, it follows that the only curves that carry loop momentum, $x_a$ and $x_b$, are the curves $C_{0a}, C_{0b}$, that have one spiral, and any curve, $C_{ab}$, that spirals around $a$ and spirals around $b$. In fact, there is just one curve $C_{ab}$. In the set of curves ${\cal S}$, listed in Appendix \ref{app:Dntilde}, we see that of a total of 17 curves, only 9 of them carry loop momentum and appear in $\Lambda$.

Using \eqref{eq:X0aX0b} and \eqref{eq:Xab}, the $\Lambda$-matrix is
\begin{equation}\label{eq:A:Dtilde}
    \Lambda = \begin{bmatrix} \alpha_{ab}+\alpha_{0a} & -\alpha_{ab}\\ -\alpha_{ab}& \alpha_{ab}+ \alpha_{0b} \end{bmatrix}.
\end{equation}
Here, $\alpha_{0a},\alpha_{0b}$ is a convenient notation for the sums
\begin{equation}
    \alpha_{0a} = \sum_{C_{0a}\in{\cal S}}\alpha_{C_{0a}}, \qquad \alpha_{0b} = \sum_{C_{0b}\in{\cal S}}\alpha_{C_{0b}}.
\end{equation}
The surface Symanzik polynomial, ${\cal U}$, is given by ${\cal U} = \det \Lambda$. Expanding the determinant as a sum of monomials gives
\begin{equation}\label{eq:DntildeU}
    {\cal U} = \alpha_{ab}\alpha_{0a}+\alpha_{ab}\alpha_{0b} + \alpha_{0a}\alpha_{0b}.
\end{equation}

In summary, ${\cal K}$ and ${\cal U}$ can be computed from the headlight functions for a set of 9 curves, whose paths are listed in Appendix \ref{app:Dntilde}. In addition to these, ${\cal S}$ has an additional 8 curves which do not carry loop momentum. These 8 curves do not contribute to computing ${\cal K}$, ${\cal U}$, ${\cal F}_0$, but are needed in what follows to compute when computing ${\cal Z}$.

\subsection{The Tree-Loop Fatgraph}
\begin{figure}
\begin{center}
\includegraphics[scale=0.25]{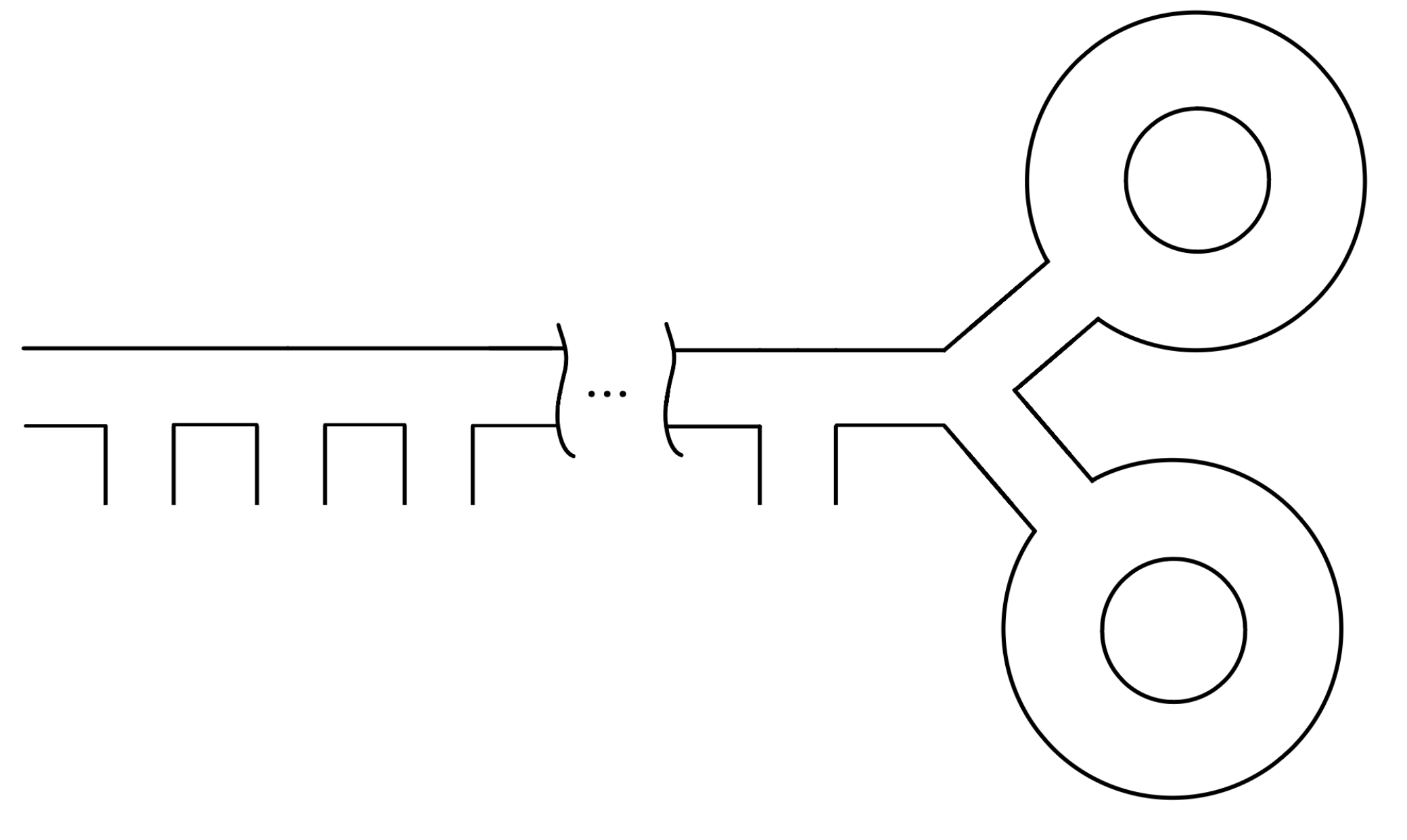}
\caption{The Tree-Loop fatgraph for the 2-loop planar amplitudes.}
\label{fig:Dntilde}
\end{center}
\end{figure}

To find the all-$n$ amplitude, attach a tree graph $\Gamma^0$ to the external leg of the tadpole, to obtain the Tree-Loop graph, $\Gamma$, in Figure \ref{fig:Dntilde}.

Since $\Gamma$ is still planar, the momenta of curves can still be assigned using dual momentum variables, $x_i^\mu$ (for $i=1,\ldots,n,a,b$). If the external momenta are $p_i^\mu$, we can take
\begin{equation}
    x_i^\mu = p_1^\mu + \cdots + p_{i-1}^\mu.
\end{equation}
The propagator of a curve that connects line $i$ to line $j$ is then
\begin{equation}
    X_{ij} = (x_j-x_i)^2+m^2.
\end{equation}
The propagators for a curve that begins at line $i$ and ends in a spiral ($a$ or $b$) are
\begin{equation}\label{eq:XiaXib}
    X_{ia} = (x_a-x_i)^2+m^2,\qquad X_{ib} = (x_b-x_i)^2+m^2.
\end{equation}
We again take $x_a,x_b$ as loop momentum variables.

To compute ${\cal F}_0$ and ${\cal Z}$, we need to enumerate the curves on $\Gamma$. But these are obtained directly from the curves on the tadpole graph, $\Gamma_\ell$, by adding tree paths in all possible ways. All the work was in enumerating the curves on $\Gamma_\ell$. We find that the complete list of loop-carrying curves on $\Gamma$ is 
\begin{equation}
     C_{ia} = W_i C_{0a}, \qquad C_{ib} = W_i C_{0b},\qquad C_{ab}
\end{equation}
for all curves $C_{0a},C_{0b} \in {\cal S}$ on the tadpole fatgraph. In addition, we have curves that do not carry loop momentum, (with $i\leq j$)
\begin{flalign}
    C_{ij} = W_i C_{00} \,W_j^T, \qquad C_{ij} = W_{ij}, \qquad C_{ji} = W_i C_{B} \,W_j^T,
\end{flalign}
for all tadpole curves $C_{00}\in{\cal S}$, and where $C_B$ is the boundary curve on the tadpole fatgraph.

The paths for all the tadpole curves in ${\cal S}$ are given in Appendix \ref{app:Dntilde} and can be used to find the curve matrices $M_C$ for all of these curves by taking matrix products. 

\subsection{The All-multiplicity Amplitude}
The resulting curve integral for the amplitude is (with $E=n+2$)
\begin{equation}
    {\cal A}_n = \int d^Et\,{\cal K}\, \left(\frac{\pi^2}{\cal U}\right)^{\frac{D}{2}}\,\exp\left(-\frac{{\cal F}_0}{\cal U} - {\cal Z}\right),
\end{equation}
where ${\cal U},{\cal K}$ are given by the tadpole calculation, equations \eqref{eq:DntildeU} and \eqref{eq:DntildeK}. It remains to find ${\cal F}_0$ and ${\cal Z}$.

The formulas for ${\cal F}_0$ and ${\cal Z}$ then follow from the momentum assignments to these curves. ${\cal F}_0$ is given by
\begin{equation}
    \mathcal{F}_0 = {\cal J}^T \tilde{\Lambda} {\cal J},
\end{equation}
where the vector $\mathcal{J}$ follows from the momentum assignments, \eqref{eq:XiaXib}, and is given by
{\everymath={\displaystyle}
\begin{equation}
    {\cal J}^\mu = \sum_{i=1}^n z_i^\mu \begin{bmatrix} \sum_{C_{ia}\in{\cal S}}\alpha_{C_{ia}} \\  \sum_{C_{ib}\in{\cal S}}\alpha_{C_{ib}} \end{bmatrix}.
\end{equation}
}
$\tilde{\Lambda}$ is the matrix adjugate of the matrix, $\Lambda$, computed for the tadpole, \eqref{eq:A:Dtilde}. The adjugate is
\begin{equation}
    \tilde{\Lambda} = \begin{bmatrix} \alpha_{ab} + \alpha_{0b} & \alpha_{ab} \\ \alpha_{ab} & \alpha_{ab} + \sum_k \alpha_{0a}^k \end{bmatrix}.
\end{equation}

Finally, ${\cal Z}$ is a sum over all curves on $\Gamma$ that can be obtained by from the set ${\cal S}$,
\begin{equation}
    {\cal Z} = \sum_C \alpha_C (K_C^2+m^2),
\end{equation}
where $K_C^\mu$ is the part of the momentum of $C$ that does not depend on $z_a^\mu,z_b^\mu$. Explicitly, this is
\begin{multline}
    {\cal Z} = \sum_{i<j}^n X_{ij} \left( \alpha_{ij} +\alpha_{ji}+ \sum_{C_{00}\in{\cal S}} \alpha_{W_iC_{00}W_j^T} \right) \\ + \sum_{i=1}^n (z_i^2+m^2) \left( \sum_{C_{ia}\in{\cal S}} \alpha_{C_{ia}} +  \sum_{C_{ib}\in{\cal S}} \alpha_{C_{ib}}\right).
\end{multline}

\section{The 2-Loop Triple-trace Amplitudes}\label{sec:Apqr}
In the previous example, the $\MCG$ had only one generator, just as in the 1-loop double-trace amplitudes. The triple-trace amplitudes are first example with a more interesting $\MCG$. The $\MCG$ in this case is still abelian, but has three generators.

\subsection{MCG and Mirzakhani Kernel}
\begin{figure}
        \begin{center}
        \includegraphics[scale=0.2]{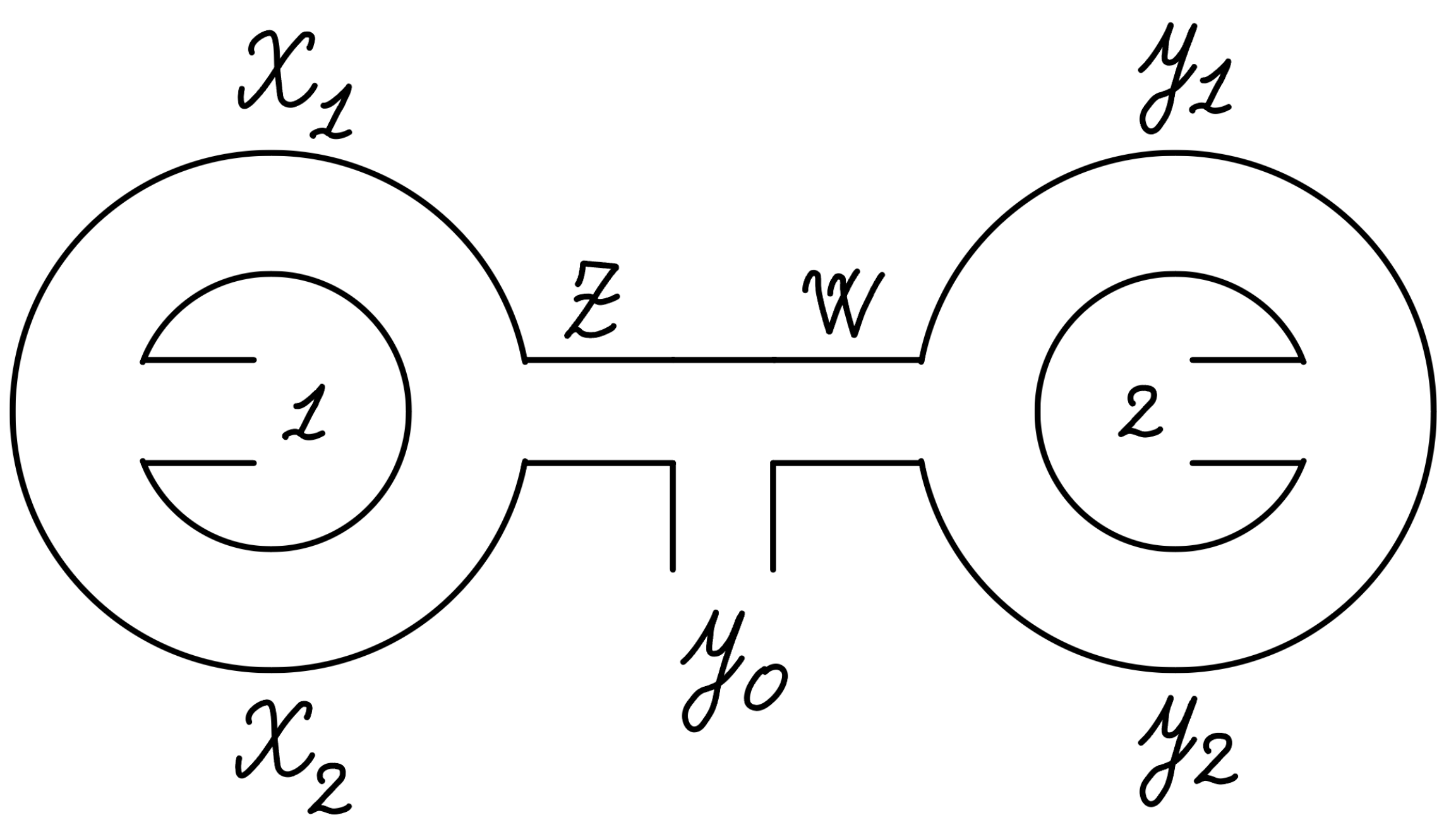}
       \caption{A 2-loop triple-trace tadpole graph.}
        \label{fig:A111}
        \end{center}
    \end{figure}

\begin{figure}
        \begin{center}
        \includegraphics[scale=0.15]{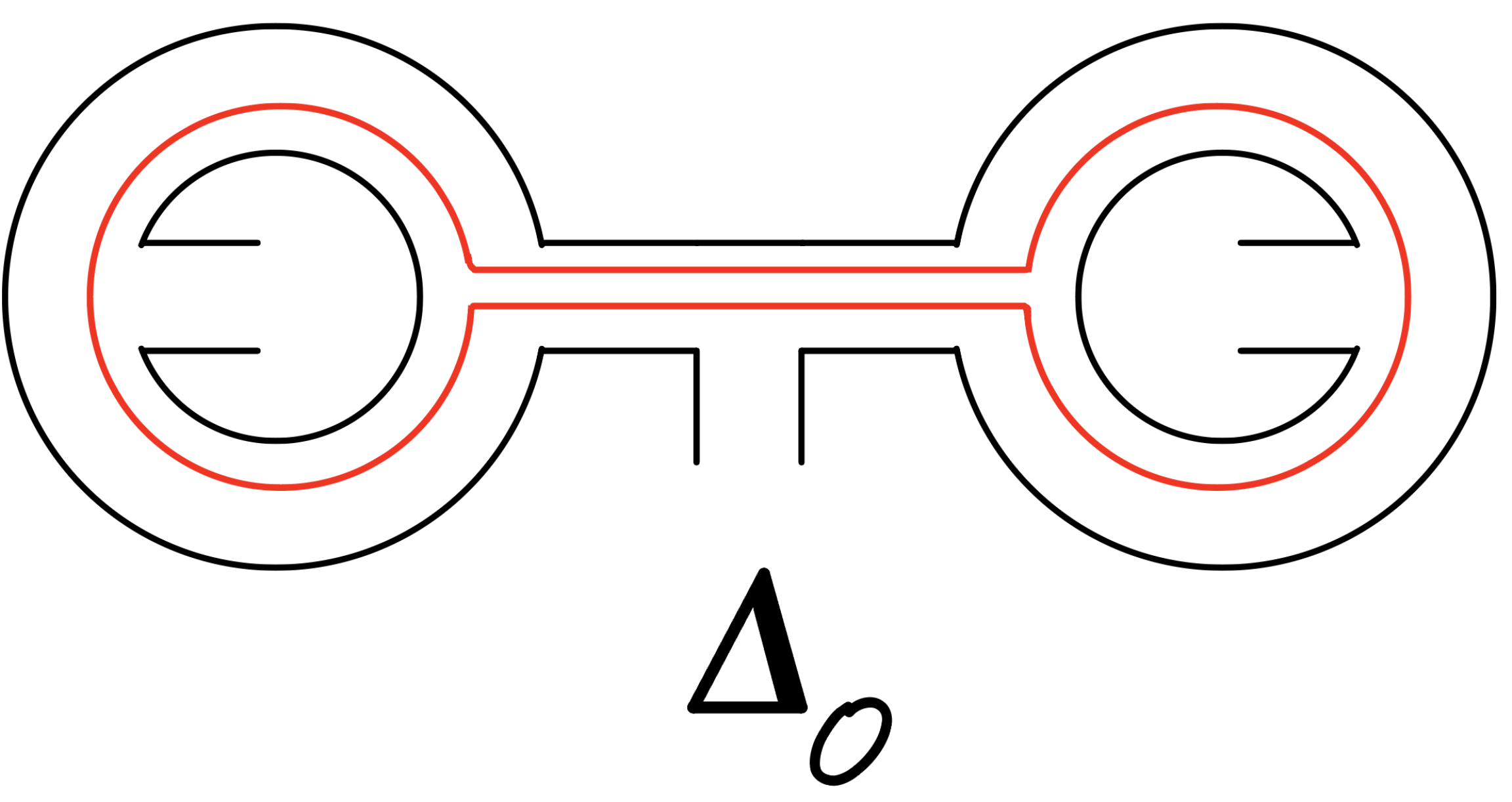}
        \includegraphics[scale=0.15]{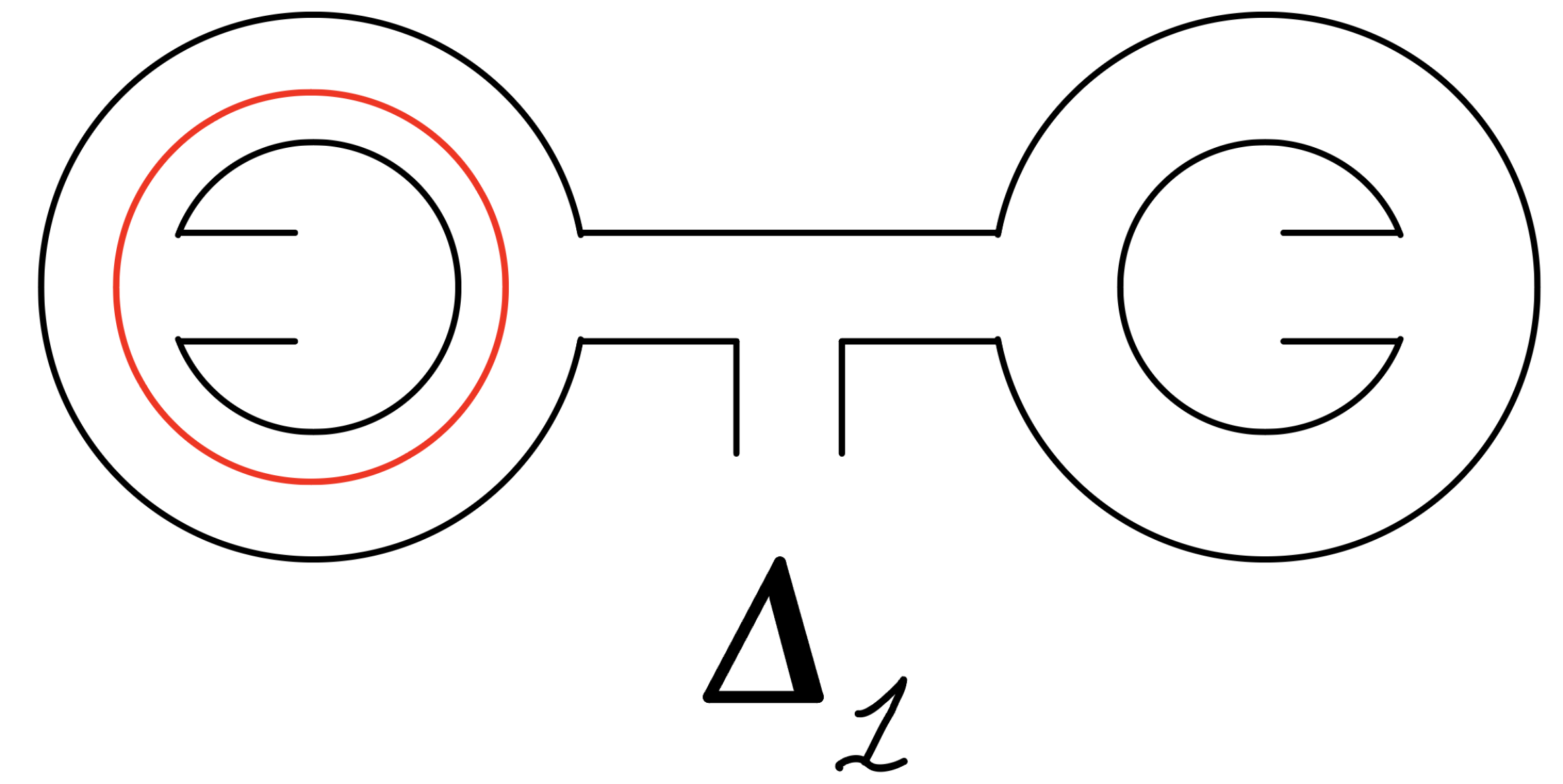}
        \includegraphics[scale=0.15]{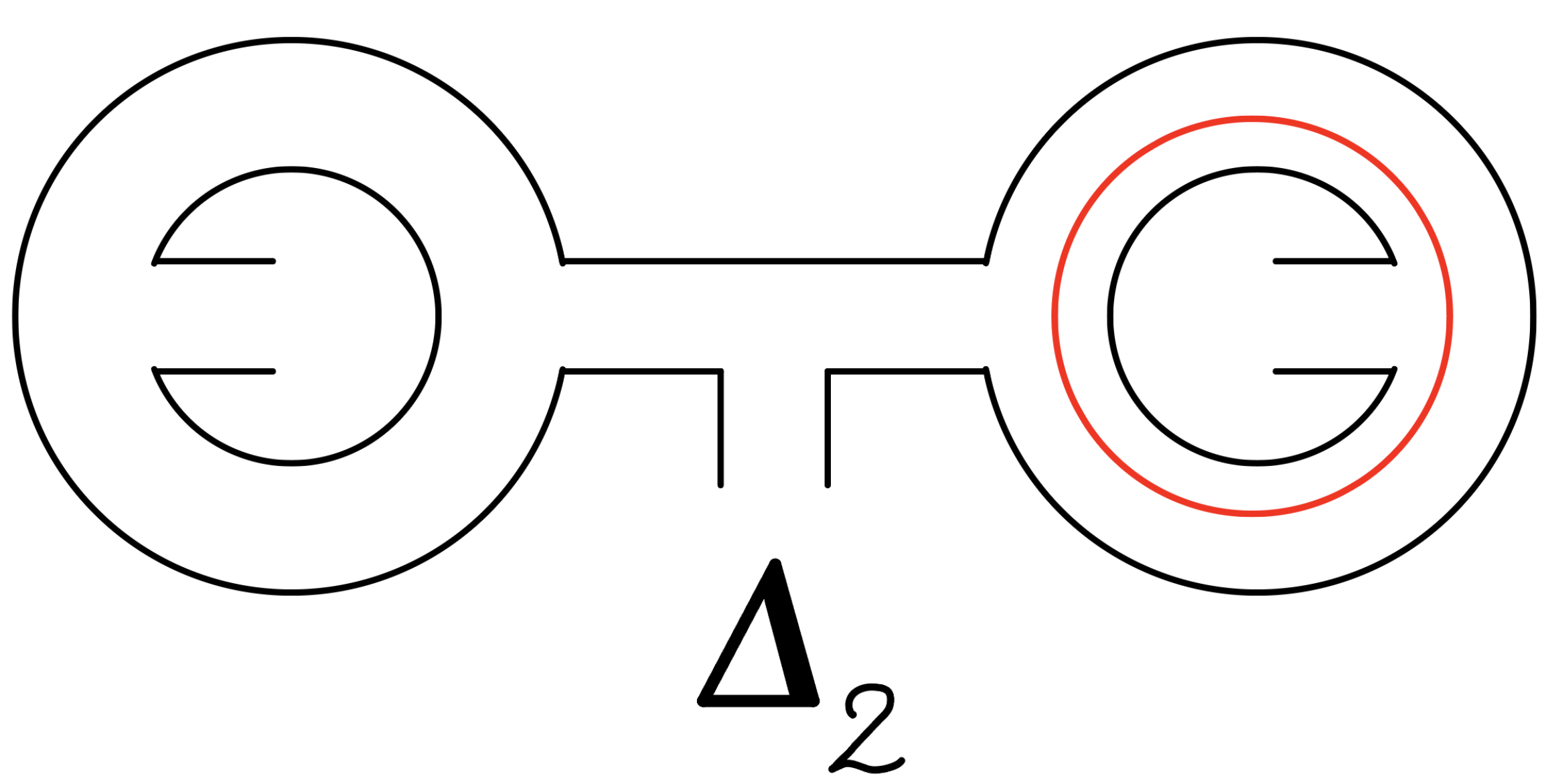}
       \caption{The three nontrivial closed curves, $\Delta_0,\Delta_1,\Delta_2$, on the 2-loop triple-trace tadpole graph.}
        \label{fig:A111delta}
        \end{center}
    \end{figure}

A triple-trace 2-loop tadpole fatgraph, $\Gamma_\ell$, is given in Figure \ref{fig:A111}. There are three nontrivial closed curves on $\Gamma_\ell$, shown in Figure \ref{fig:A111delta}. These closed curves have the paths:
\begin{gather}\label{eq:A111deltas1}
    \Delta_0 = w R y_2 R y_2 R w R z R x_2 R x_2 R z L,\\
    \Delta_1 = x_1 R x_2 L,\qquad \Delta_2 = y_1 L y_2 R. \label{eq:A111deltas2}
\end{gather}
Each of these closed curves surrounds one of the three trace-factors. Dehn twists around each of these $\Delta_i$, $\gamma_{\Delta_i}$, generate the $\MCG$. The $\Delta_i$ curves in Figure \ref{fig:A111delta} are non-intersecting. It follows that the Dehn twists around the $\Delta_i$ commute, and the $\MCG$ is isomorphic to $\mathbb{Z}^3$.

\begin{figure}
        \begin{center}
        \includegraphics[scale=0.55]{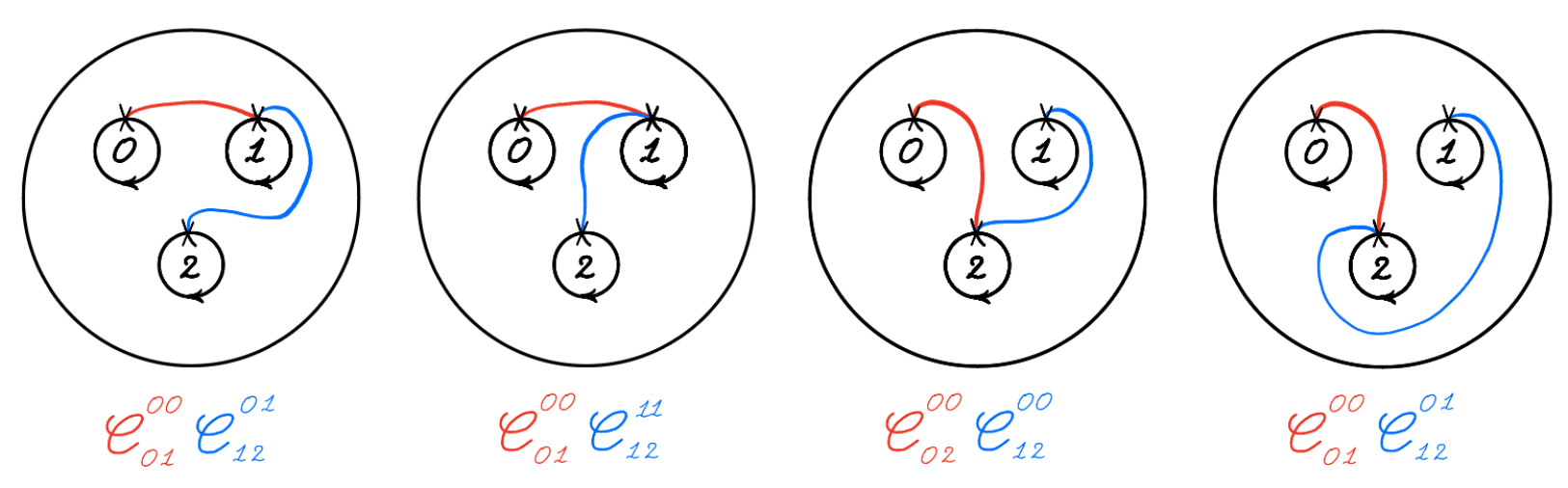}
       \caption{The surface obtained by fattening the tadpole fatgraph, and the four MCG-inequivalent cuts of the surface that give the Mirzakhani kernel.}
        \label{fig:A111surf}
        \end{center}
    \end{figure}

To find a Mirzakhani kernel, consider one trace-factor at a time. A curve, $C_{AB}$, connecting trace-factors $A$ and $B$, intersects $\Delta_A$ and $\Delta_B$, and is acted on by Dehn twists around these two curves. Consider the curves with one endpoint on trace-factor 0. From the surface diagram, Figure \ref{fig:A111surf}, these curves decompose into two cosets: the curves $C_{01}$ connecting 0 and 1, and the curves $C_{02}$ connecting 0 and 2.

Exploring paths on the fatgraph, $\Gamma_\ell$, we identify two possible coset representatives:
\begin{align}
    C_{01}^{00} = L z L x_2 R,\qquad C_{02}^{00} = R w R y_2 L.
\end{align}
These are the shortest paths in the two cosets. Other paths in these cosets can be obtained by acting with the Dehn twists: $\gamma_{\Delta_0}, \gamma_{\Delta_1},\gamma_{\Delta_2}$. Write
\begin{equation}
    \rho = \sum \alpha_{C_{01}} + \sum \alpha_{C_{02}}
\end{equation}
for the sum over curves with a single endpoint on trace-factor 0.

The stabilizer of $C_{01}^{00}$ is generated by $\gamma_{\Delta_2}$. To mod out by the stabilizer, consider the set of curves connecting 1 and 2. From the surface diagram, Figure \ref{fig:A111surf}, it can be seen that these curves form two cosets. Looking for curves compatible with $C_{01}^{00}$, we find
\begin{align}
    C_{12}^{01} = R x_1 L z L w L y_1 R, \qquad C_{12}^{11} = L x_2 R z L w L y_1 R,
\end{align}
which can be taken as coset representatives. Note that $C_{12}^{10}$ is obtained from $C_{12}^{00}$ by acting with $\gamma_{\Delta_1}$.

Likewise, the stabilizer of $C_{02}^{00}$ is generated by $\gamma_{\Delta_1}$. To mod out by this stabilizer, again consider the curves connecting 1 and 2. From the surface diagram, Figure \ref{fig:A111surf}, it can be seen that these curves form two cosets. Looking for curves compatible with $C_{02}^{00}$, we find
\begin{align}
    C_{12}^{00} = R x_1 L z L w R y_2 L, \qquad C_{12}^{01} = L x_1 L z L w L y_1 R,
\end{align}
which can be taken as coset representatives.

A Mirzakhani is then
\begin{equation}\label{eq:A111:K}
    {\cal K} = \frac{\alpha_{01}^{00}\alpha_{12}^{01}}{\rho \rho'} + \frac{\alpha_{01}^{00}\alpha_{12}^{11}}{\rho \rho'} + \frac{\alpha_{02}^{00}\alpha_{12}^{00}}{\rho \rho'} + \frac{\alpha_{02}^{00}\alpha_{12}^{01}}{\rho \rho'},
\end{equation}
where $\rho'$ is a sum
\begin{equation}
    \rho' = \sum \alpha_{C_{12}}
\end{equation}
over curves connecting 1 and 2. Only finitely many curves appear in $\rho$ and $\rho'$. Namely, only curves compatible with one of the pairs of curves appearing in the numerators of ${\cal K}$.

Let ${\cal S}$ be the set of \emph{all} curves on $\Gamma_\ell$ which are compatible with one of the pairs of curves appearing in the numerators of ${\cal K}$. 

We find that there are 33 curves in ${\cal S}$. 24 of these are curves that carry loop momentum, and so appear in ${\cal U}$, ${\cal K}$ and ${\cal F}_0$. Whereas 9 of these are curves do not carry loop momentum and only contribute to ${\cal Z}$. All 33 of these curves and their paths are listed in Appendix \ref{app:Apqr}.

\subsection{Tadpole Curves and Momenta}
The momentum assignment rule shows that the closed curves have momenta
\begin{equation}
    P_{\Delta_0} = p_0,~~P_{\Delta_1} = p_1,~~P_{\Delta_2} = p_2.
\end{equation}
A momentum routing can be chosen such that
\begin{equation}\label{eq:A111mom}
    P_{01}^{00}=\ell_1,\qquad P_{02}^{00}=\ell_2,\qquad P_{12}^{00} = \ell_1+\ell_2,
\end{equation}
for some loop momentum variables $\ell_1,\ell_2$. Acting with the $\MCG$, we obtain all other curves. A Dehn twist by $\Delta_i$ acting on $C_{ij}$ adds $\pm p_i$ to its momentum. So:
\begin{align}
    P_{01}^{ab} &=  \ell_1 + a p_0 + b p_1,\\
    P_{02}^{ab} &=  \ell_2 + a p_0 + b p_2,\\
    P_{12}^{ab} &=  \ell_1+\ell_2 + a p_1 + b p_2.
\end{align}

Finally, the curves that begin and end at the same trace-factor are homologous to the boundaries of the fatgraph. This means that the momentum of such a curve $C_{AA}$ ($A=0,1,2$) is a linear combination of the boundary momenta: $p_0,p_1,p_2$.

The $\Lambda$-matrix is constructed from $\alpha_C$ for curves $C$ that carry loop momentum. The only curves carrying loop momenta are those curves connecting two distinct trace-factors. Given \eqref{eq:A111mom}, the $\Lambda$-matrix is
\begin{equation}
    \Lambda = \begin{bmatrix} \alpha_{01}+\alpha_{12} & \alpha_{12} \\  \alpha_{12} & \alpha_{02}+\alpha_{12} \end{bmatrix},
\end{equation}
where
\begin{equation}
    \alpha_{AB} = \sum_{C_{AB}\in{\cal S}} \alpha_{C_{AB}}
\end{equation}
is the sum over those curves $C_{AB} \in {\cal S}$, connecting trace-factors $A$ and $B$, that are compatible with the Mirzakhani kernel, ${\cal K}$.  The surface Symanzik polynomial ${\cal U}$ is given by
\begin{equation}\label{eq:A111:U}
{\cal U} = \det \Lambda = \alpha_{01}\alpha_{02} + \alpha_{01}\alpha_{12}+\alpha_{02}\alpha_{12}.
\end{equation}

\subsection{The All-multiplicity Amplitudes}
To find the all-multiplicity amplitudes, attach tree comb graphs $\Gamma^0,\Gamma^1,\Gamma^2$ to $\Gamma_\ell$, to obtain the Tree-Loop graph, $\Gamma$. Let the number of external particles on each trace-factor be $n_0,n_1,n_2$, respectively, with $n=n_0+n_1+n_2$. From a curve $C_{AB}$ on $\Gamma_\ell$, we obtain curves of the form
\begin{equation}
    C_{iABj} = W_i^A C_{AB} \left(W_j^B\right)^T
\end{equation}
connecting particle $i$ on trace-factor $A$ to particle $j$ on trace-factor $B$. The momentum assigned to such a curve is (see Section \ref{sec:summary})
\begin{equation}
    P_{C_{iABj}} = P_{C_{AB}} + z_j^B - z_i^A.
\end{equation}
And the headlight function $\alpha_C$ is obtained by multiplying the matrix $M_{C_{AB}}$ with the tree matrices $W_i^A$ and $\left(W_j^B\right)^T$.

The curve integral for the amplitude is then (with $E=n$)
\begin{equation}
    {\cal A}_{n_1,n_2,n_3} = \int d^Et\, {\cal K}  \, \left(\frac{\pi^2}{\cal U}\right)^{\frac{D}{2}} \exp\left({-\frac{{\cal F}_0}{\cal U} - {\cal Z}}\right).
\end{equation}
Here, ${\cal K}$ and ${\cal U}$ are given as above by the tadpole calculation (equations \eqref{eq:A111:K} and \eqref{eq:A111:U}). 

The formulas for ${\cal F}_0$ and ${\cal Z}$ follow from the momentum assignments. For a curve $C_{iABj}$ connecting trace-factors $A$ and $B$, write
\begin{equation}
    P_C = K_C + \ell_{C_{AB}},
\end{equation}
where $\ell_{C_{AB}}$ is the part depending on the loop momentum variables. From \eqref{eq:A111mom},
\begin{equation}
    \ell_{C_{01}} = \ell_1,\qquad \ell_{C_{02}} = \ell_2, \qquad \ell_{C_{12}} = \ell_1+\ell_2.
\end{equation}
It follows that {\cal J} is
{
\everymath={\displaystyle}
\begin{equation}
    {\cal J} = \left[ \begin{array}{c} \sum_{\substack{C_{01}\in{\cal S}\\ C=W_i^0 C_{01} \left(W_j^1\right)^T}} K_C\, \alpha_C + \sum_{\substack{C_{12}\in{\cal S}\\ C=W_i^1 C_{12} \left(W_j^2\right)^T}} K_C\, \alpha_C \\ \sum_{\substack{C_{02}\in{\cal S}\\ C=W_i^0 C_{02} \left(W_j^2\right)^T}} K_C\, \alpha_C + \sum_{\substack{C_{12}\in{\cal S}\\ C=W_i^1 C_{12} \left(W_j^2\right)^T}} K_C\, \alpha_C. \end{array} \right],
\end{equation}
}
Then ${\cal F}_0$ is, as usual ${\cal F}_0 = {\cal J}^T \tilde{\Lambda} \cal{J}$, where the adjugate matrix of $\Lambda$ is
\begin{equation}
    \tilde{\Lambda} = \begin{bmatrix} \alpha_{02}+\alpha_{12} & -\alpha_{12} \\  -\alpha_{12} &  \alpha_{01}+\alpha_{12}\end{bmatrix}.
\end{equation}
Finally,
\begin{equation}
    {\cal Z} = \sum_{C_{AB}\in{\cal S}} \sum_{C} ((K_C)^2 + m^2)\alpha_{C},
\end{equation}
where the second sum is over curves $C=W_i^A C_{AB} \left(W_j^B\right)^T$ formed by extending $C_{AB}$ to a curve on $\Gamma$.

As in the previous examples, evaluating the curve integral for ${\cal A}_{n_1,n_2,n_3}$ is simplified by using the form of the Mirzakhani kernel ${\cal K}$, which has four terms. So the amplitude is a sum of four curve integrals. For example, the monomial $\alpha_{01}^{00}\alpha_{02}^{00}$ appears in ${\cal K}$. Restricting to the region $\alpha_{01}^{00}\alpha_{02}^{00}$, the curve integral simplifies. Cutting $\Gamma_\ell$ along $C_{01}^{00},C_{02}^{00}$ gives a 7-point tree graph, which has 14 possible curves. So this contribution to ${\cal A}_{n_1,n_2,n_3}$ can be computed using 14 matrices, valid for all multiplicities.

\section{The 2-Loop Amplitudes at Genus One}\label{sec:markov}
\begin{figure}
\begin{center}
\includegraphics[scale=0.2]{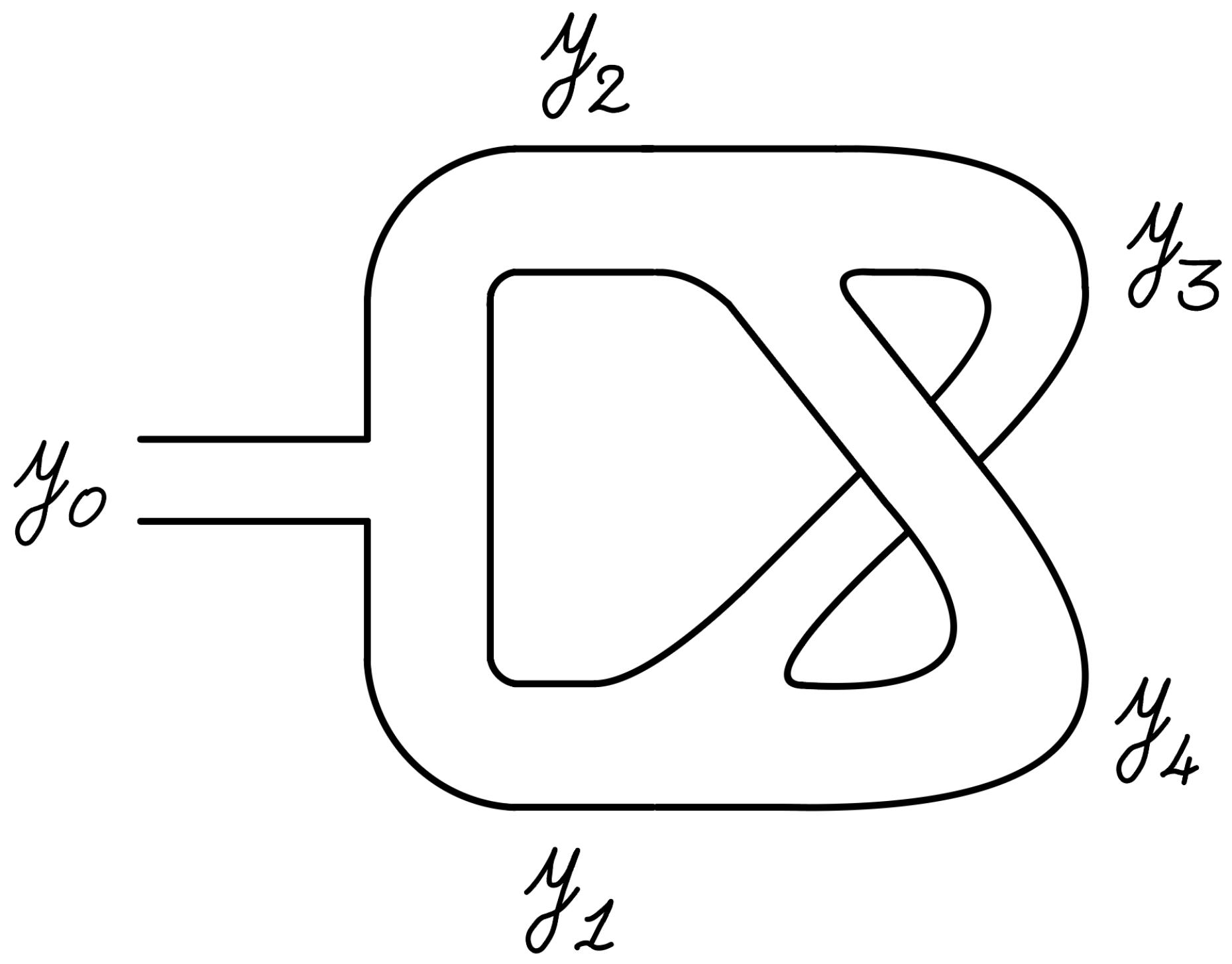}
\caption{The 2-loop genus-one tadpole fatgraph.}
\label{fig:markov}
\end{center}
\end{figure}

The first non-planar contribution to the single trace amplitude at two loops is given by genus one diagrams. These give a first example of a non-abelian Mapping Class Group, which is isomorphic in this case to ${\rm SL}_2\mathbb{Z}\times\mathbb{Z}$.

\subsection{MCG and Mirzakhani Kernel}
\begin{figure}
\centering
\includegraphics[width=.9\linewidth]{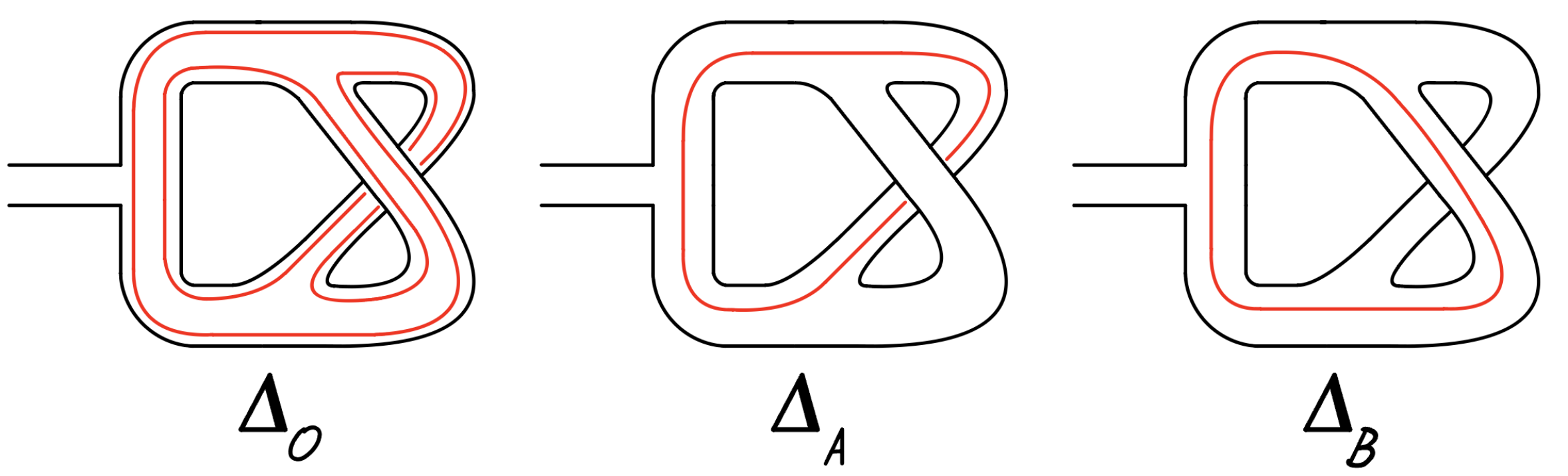}
  \caption{The three closed curves, $\Delta_0,\Delta_1,\Delta_2$, on the 2-loop genus-one tadpole graph. Dehn twists around these curves generate a $\mathbb{Z}\times\text{SL}_2\mathbb{Z}$ Mapping Class Group.}
\label{fig:mcgMarkov}
\end{figure}

The genus-one tadpole fatgraph, $\Gamma_\ell$, is given in Figure \ref{fig:markov}. There are three non-trivial closed curves on $\Gamma_\ell$. These closed curves are shown in Figure \ref{fig:mcgMarkov} and they have paths:
\begin{gather}
    \Delta_A = y_1Ry_2Ly_3R,\qquad \Delta_B = y_1Ry_2Ry_4L,\\
    \Delta_0 = y_1Ry_2Ry_4Ry_3Ry_2Ly_1Ry_4Ry_3R.
\end{gather}
$\Delta_0$ is the closed curve surrounding the trace-factor of the fatgraph. $\Delta_A$ and $\Delta_B$ are homologous to the A- and B- cycles of the torus (if $\Gamma_\ell$ is drawn embedded on the torus).

Dehn twists $\gamma_\Delta$ around these closed curves generate $\MCG$. $\Delta_0$ does not intersect either of $\Delta_A$ or $\Delta_B$, so $\gamma_{\Delta_0}$ commutes with the other two Dehn twists. Whereas $\Delta_A$ and $\Delta_B$ intersect, so that Dehn twists around these closed curves do \emph{not} commute. Since $\Delta_A$ and $\Delta_B$ are the A- and B- cycles of the torus, these Dehn twists generate the MCG of the torus, which gives an ${\rm SL}_2\mathbb{Z}$ subgroup of the full $\MCG$. The $\MCG$ is therefore isomorphic to $\mathbb{Z}\times{\rm SL}_2\mathbb{Z}$.

All open curves on $\Gamma_\ell$ begin and end on the external line, $0$.
Furthermore, all curves are equivalent up to MCG and therefore belong to the same coset. 

Exploring curves on the fatgraph, we find a short path 
\begin{equation}
    C_1 = Ly_2Ry_4Ly_1L
\end{equation}
which can be taken as coset representatives for the unique $\MCG$ coset. Note that $C_1$ intersects $\Delta_A$ and $\Delta_0$, but \emph{not} $\Delta_B$. So the stabilizer of this curve is the $\mathbb{Z}$ subgroup generated by $\gamma_{\Delta_B}$. Write
\begin{equation}
    \rho = \sum \alpha_{C}
\end{equation}
for the sum over all curves $C$ that are compatible with $C_1$. Then
\begin{equation}
    {\cal K}_A = \frac{\alpha_1}{\rho}
\end{equation}
is a Mirzakhani kernel that partially mods out by $\MCG$, but does not mod out by the stabilizer generated by $\gamma_{\Delta_B}$. 

Cutting the surface along $C_1$ gives an annulus with two marked points on one boundary, and one marked point on the other. $\Delta_B$ is the closed loop on the annulus and generates the $\MCG$ of the annulus, which is a $\mathbb{Z}$ subgroup of the full $\MCG$. So the curves compatible with $C_1$ decompose into two cosets under this $\mathbb{Z}$ subgroup: depending on their endpoints on the cut surface.

To mod out by the stabilizer, we need to find curves that intersect $\Delta_B$ and are compatible with $C_1$. We find, for example, the paths
\begin{equation}
C_2 = Ly_2Ly_3Ry_1L, \qquad     C_3 = Ry_1Ry_4Ry_3Ry_1Ry_2Ry_4Ly_1L.
\end{equation}
These are both compatible with $C_1$, intersect $\Delta_B$, but are not related to each other by $\gamma_{\Delta_B}$ (on the cut surface, the annulus, they connect distinct pairs of endpoints). So we can take $C_2$ and $C_3$ as our two coset representatives.

\begin{figure}
\centering
 \includegraphics[scale=0.55]{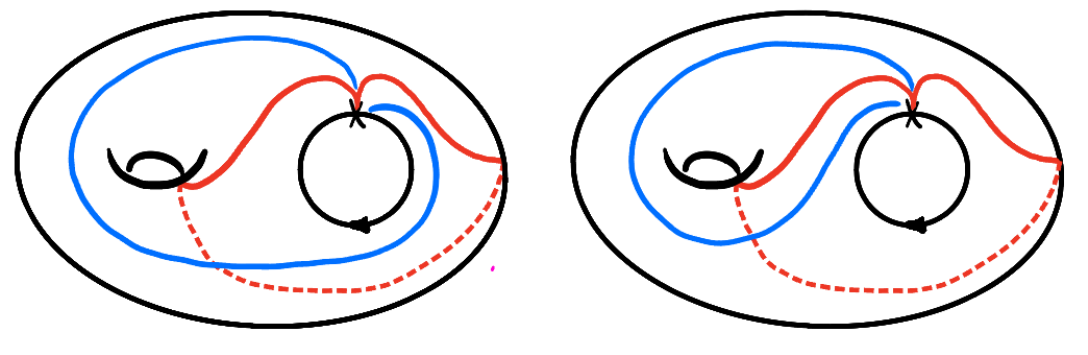}
  \caption{The four MCG-inequivalent pairs of curves that cut the torus to a disk. Each pair of curves corresponds to a distinct term in the Mirzakhani kernel, ${\cal K}$.}
\label{fig:markovsurface}
\end{figure}

Write
\begin{equation}
    \rho_B = \sum \alpha_{C}
\end{equation}
for the sum over all curves, $C$, which are compatible with $C_1$ and which intersect $\Delta_B$. Then the final Mirzakhani kernel for the full MCG is
\begin{equation}\label{eq:markov:K}
    {\cal K} = \frac{\alpha_1}{\rho}\frac{\alpha_2+\alpha_3}{\rho_B}.
\end{equation}
The two terms in ${\cal K}$ correspond to the two MCG-inequivalent cuts shown in Figure \ref{fig:markovsurface}. 

Let ${\cal S}$ be the set of all curves which are compatible with one of the two pairs of curves appearing in the numerators of ${\cal K}$. There are 10 curves in ${\cal S}$. Of these, 9 curves carry loop momentum, and contribute to the computation of ${\cal K}$ and ${\cal U}$. Only 1 curve in the set does not carry loop momentum: the curve $C_{00}$ that follows the boundary of $\Gamma_\ell$, and carries zero momentum:
\begin{equation}
    C_{00} = L y_2 L y_3 L y_4 L y_2 L y_1 L y_3 L y_4 L y_1 L.
\end{equation}
The paths for all 10 of the curves in ${\cal S}$ are listed in Appendix \ref{app:markov}.
    
\subsection{Momentum Assignments}
By momentum conservation, the external momentum entering the tadpole graph is zero. Assign a momentum routing to the graph by assigning edge 4 the loop momentum variable $\ell_1^\mu$, and edge 3 the loop momentum variable $\ell_2^\mu$. Then the momenta of the closed curves are
\begin{equation}
    P_{\Delta_0} =0,~~P_{\Delta_A} = \ell_1,~~P_{\Delta_B} = \ell_2.
\end{equation}
Moreover this momentum routing induces a momentum assignment to all open curves. For example, using the momentum assignment rule,
\begin{equation}
    P_{C_1} = \ell_2, \qquad P_{C_2} = \ell_1,\qquad P_{C_3} = \ell_1.
\end{equation}
Another way to understand this momentum assignment is that, on the surface (Figure \ref{fig:markovsurface}), the curve $C_1$ (red in the figure) is homologous to $\Delta_B$ (ignoring the boundary trace-factor), and so has the same momentum as $\Delta_B$. Likewise, the curves $C_2$ and $C_3$ (blue in the figure) are homologous to $\Delta_A$.\footnote{As explained in \cite{us_qft}, the momenta assigned to curves, $P_C^\mu$, satisfy the same additive relations as the classes of curves in homology.}

Now consider any curve $C$ in the set ${\cal S}$. The momentum assigned to $C$ is necessarily a linear combination of $\ell_1$ and $\ell_2$. It is convenient to write
\begin{equation}
C=C^{p/q;w}    
\end{equation}
for any such curve $C$ with momentum
\begin{equation}\label{eq:markov:mom}
    P_{C^{p/q;w}} = p\ell_1 + q \ell_2.
\end{equation}
Here, $w$ is an arbitrary index that distinguishes distinct curves that carry the same loop momentum. For example, we can write $C_1 = C^{0/1;0}$, $C_2 = C^{1/0;0}$, $C_3 = C^{1/0;1}$. 

The $\Lambda$-matrix follows from the momentum assignments, \eqref{eq:markov:mom}. It is given by
{\everymath={\displaystyle}
\begin{equation}\label{eq:markov:lambda}
    \Lambda = \left[ \begin{array}{cc}  \sum\limits_{\substack{C=C^{p/q;w} \\ C\in{\cal S}}} \,p^2\,\alpha_C  & \sum\limits_{\substack{C=C^{p/q;w} \\ C\in{\cal S}}} \,pq\,\alpha_C \\ \sum\limits_{\substack{C=C^{p/q;w} \\ C\in{\cal S}}} \,pq\,\alpha_C   & \sum\limits_{\substack{C=C^{p/q;w} \\ C\in{\cal S}}} \,q^2\,\alpha_C 
    \end{array} \right],
\end{equation}
}
\noindent
These sums exclude the one curve, $C_{00}$, that does not carry loop momentum. The surface Symanzik polynomial ${\cal U}$ is then the determinant, ${\cal U} = \det \Lambda$.

As an aside, note that this determinant can be expanded as the sum over pairs $C,D$ of curves in ${\cal S}$:
\begin{equation}
    \det \Lambda = \sum_{\substack{ C=C^{p/q;w}\\D=C^{r/s;w'}}} \left|\begin{matrix} p & r\\ q & s \end{matrix}\right|^2 \alpha_C\,\alpha_D.
\end{equation}
It can be checked that the determinant factors appearing in this sum, $(ps-qr)^2$, are equal to $1$ if and only if the two curves $C$ and $D$ cut $\Gamma_\ell$ to a tree. Otherwise the determinant factor is zero. So ${\cal U}$ is precisely a sum over maximal cuts of $\Gamma_\ell$ that can be formed from the set ${\cal S}$:
\begin{equation}
    {\cal U} = \sum_{\substack{C,D\in{\cal S}\\ \text{max. cut}}} \alpha_C\alpha_D.
\end{equation}

\subsection{The All-multiplicity Amplitudes}
To find the all-mulitplicity genus-one amplitudes, attach a tree comb graph, $\Gamma^0$, to the external leg of $\Gamma_\ell$, to obtain a Tree-Loop graph, $\Gamma$. Let $\Gamma$ have $n$ external legs. From any curve $C^{p/q;w}\in{\cal S}$ on $\Gamma_\ell$, we can obtain a curve
\begin{equation}
    C_{ij}^{p/q;w} = W_i C^{p/q;w} \left(W_j\right)^T,
\end{equation}
connecting particles $i$ and $j$, and such a curve has momentum (see Section \ref{sec:summary})
\begin{equation}
    P_{ij}^{p/q;w} = P_{C^{p/q;w}} + z_j - z_i.
\end{equation}
The headlight functions for these curves are obtained by matrix multiplication from the $n$ tree-level matrices $W_i$ ($i=1,\ldots,n$) and the 10 matrices $M_C$ for the curves $C^{p/q;w}$ in ${\cal S}$.

Note that there are also a set of curves that do not carry loop momentum and have tree-like propagators. These are (for $i<j-1$)
\begin{equation}
    C_{ij} = W_{ij},
\end{equation}
with momentum $P_{ij} = p_i + \cdots + p_{j-1}$, and also (for $i\leq j$)
\begin{equation}
    C_{ji} = W_i C_B \left(W_j\right)^T,
\end{equation}
with momentum $P_{ji} = p_{i+1}+\cdots+p_j$.

The curve integral for the amplitude is then (with $E=n+2$)
\begin{equation}
    {\cal A}_{n} = \int d^Et\, {\cal K}  \, \left(\frac{\pi^2}{\cal U}\right)^{\frac{D}{2}} \exp\left({-\frac{{\cal F}_0}{\cal U} - {\cal Z}}\right).
\end{equation}
Here, ${\cal K}$ (equation \eqref{eq:markov:K}) and ${\cal U}$ (equation \eqref{eq:markov:lambda}) are computed using the 13 curves $C^{p/q;w}$ in ${\cal S}$.

The formulas for ${\cal F}_0$ and ${\cal Z}$ follow from the momentum assignments. The vector ${\cal J}^\mu$ is
{
\everymath={\displaystyle}
\begin{equation}
    {\cal J}^\mu = \left[ \begin{array}{c} \sum\limits_{\substack{C^{p/q;w}\in{\cal S}\\ C=W_i C^{p/q;w} W_j^T}} \, p \,\alpha_C \,z_{ij}^\mu \\ \sum\limits_{\substack{C^{p/q;w}\in{\cal S}\\ C=W_i C^{p/q;w} W_j^T}} \, q\, \alpha_C \,z_{ij}^\mu \end{array} \right],
\end{equation}
}
where $z_{ij}^\mu = z_j^\mu - z_i^\mu$. Then ${\cal F}_0 = {\cal J}^T \tilde{\Lambda} {\cal J}$, where $\tilde{\Lambda}$ is the the adjugate of the $\Lambda$-matrix (equation \eqref{eq:markov:lambda}). Finally,
\begin{equation}
    {\cal Z} = \sum_{\substack{C\in{\cal S}\\D=W_i CW_j^T }} \left( z_{ij}^2+m^2\right) \alpha_D + \sum_{\substack{i=1\\ i<j-1}}^{j=n} \left( z_{ji}^2+m^2\right)\alpha_{C_{ij}}. 
\end{equation}


\section{The 2-Loop Double-trace Amplitudes}\label{sec:double}
\begin{figure}
\centering
\includegraphics[width=.3\linewidth]{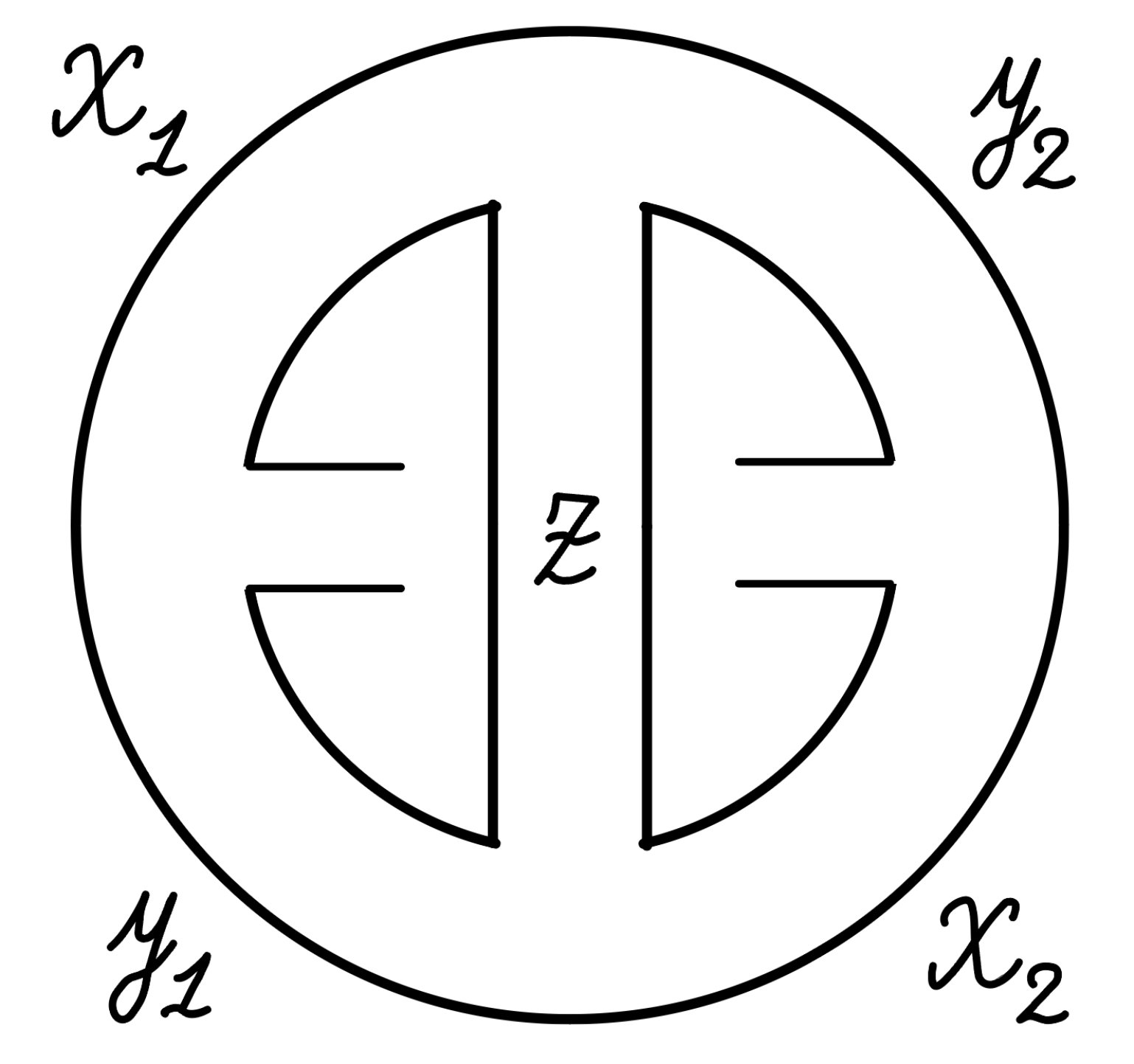}
  \caption{A tadpole graph for the 2-loop double-trace amplitudes.}
\label{fig:double:tadpole}
\end{figure}

To complete our list of 2-loop amplitude formulas, we give here a curve integral formula for the 2-loop double-trace amplitudes. The analysis is very similar to the computations in the foregoing sections.

\subsection{MCG and Mirzakhani}
\begin{figure}
\centering
\includegraphics[width=.95\linewidth]{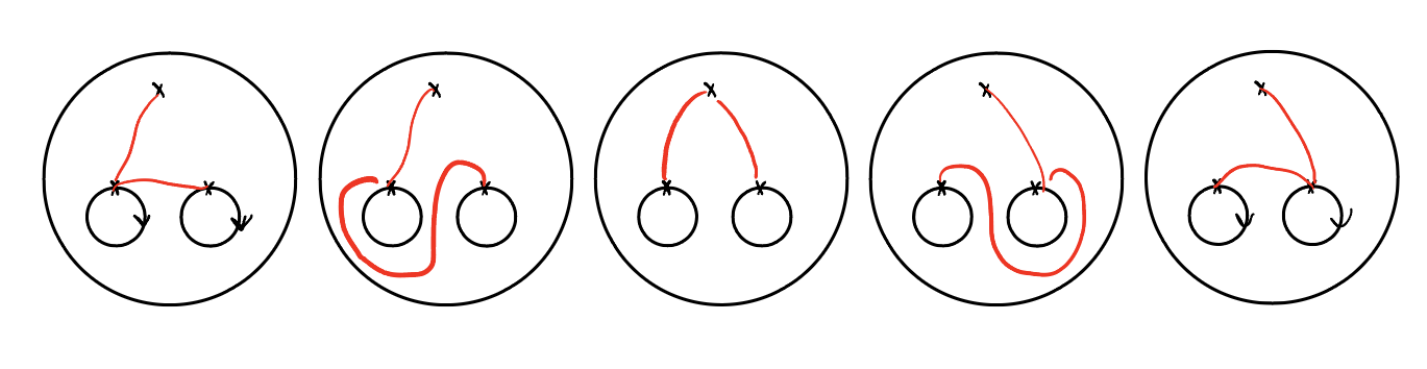}
  \caption{The surface obtained by fattening the double-trace 2-loop fatgraph is a sphere with one puncture and two trace-factors. There are five MCG-inequivalent cuts of this surface to a disk, which appear in the Mirzakhani kernel, ${\cal K}$. Note that the first two (and the last two) cuts are MCG-inequivalent due to the different order in which the curves are incident at the trace-factors.}
\label{fig:double:surface}
\end{figure}

The double-trace tadpole fatgraph, $\Gamma_\ell$, is given in Figure \ref{fig:double:tadpole}. The $\MCG$ is generated by Dehn twists around the closed curves, $\Delta_1$ and $\Delta_2$, that surround trace-factor 1 and trace-factor 2 of $\Gamma_\ell$, respectively. $\Delta_1$ and $\Delta_2$ are non-intersecting, and so these two Dehn twists commute, and the $\MCG$ is isomorphic to the free abelian group $\mathbb{Z}^2$.

Using the surface diagram, Figure \ref{fig:double:surface}, one can identify the MCG cosets. In fact, there are five MCG-inequivalent cuts of the surface to a disk (shown in the figure). Exploring paths on the tadpole fatgraph, $\Gamma_\ell$, we find that the following five pairs of curves can be taken as representatives for the five cuts:
\begin{align}\label{eq:double:cuts}
    (C_{1,0}^{0}, C_{1,2}^{1,0}), ~~(C_{1,0}^{0}, C_{1,2}^{0,0}), ~~(C_{1,0}^{0}, C_{2,0}^{0}), ~~(C_{2,0}^{0}, C_{1,2}^{0,0}), ~~( C_{2,0}^{0}, C_{1,2}^{0,1}),
\end{align}
where the explicit paths for the 5 curves appearing here are given in Figure \ref{fig:double:curves}. The paths for these curves are also listed in Appendix \ref{app:double}. We have chosen to label curves as $C_{AB}^w$, where $A$ and $B$ ($A,B = 0,1,2$) are trace-factors that the curve begins and ends on. $A=0$ is the trace-factor with no particles. $w$ is some decoration to distinguish distinct curves with the same endpoints. 

Let ${\cal S}$ be the finite set of curves compatible with at least one of the five cuts in equation \eqref{eq:double:cuts}. All curves in this set and their paths are listed in Appendix \ref{app:double}. There are 23 curves in total, of which 14 curves connect distinct trace-factors. Consider the sum, ${\cal U}$, over all possible cuts that can be formed from the set of curves ${\cal S}$:
\begin{equation}\label{eq:double:U}
    {\cal U} = \alpha_{01}\alpha_{12} + \alpha_{01}\alpha_{02} + \alpha_{02}\alpha_{12},
\end{equation}
where $\alpha_{AB}$ is a sum over curves in ${\cal S}$ connecting trace-factors $A$ and $B$,
\begin{equation}
    \alpha_{AB} = \sum\limits_{C_{AB}\in {\cal S}} \alpha_{C_{AB}}.
\end{equation}
Then a Mirzakhani kernel is given by
\begin{align}\label{eq:double:K}
    \mathcal{K} = \frac{
    \alpha_{10}^0 \alpha_{12}^{10}
    +\alpha_{10}^{0} \alpha_{12}^{00}
    +\alpha_{10}^{0} \alpha_{20}^{0}
    +\alpha_{20}^{0} \alpha_{12}^{00}
    +\alpha_{20}^{0} \alpha_{12}^{01}
    }{\mathcal{U}}.
\end{align}

\begin{figure}
\centering
\includegraphics[width=.95\linewidth]{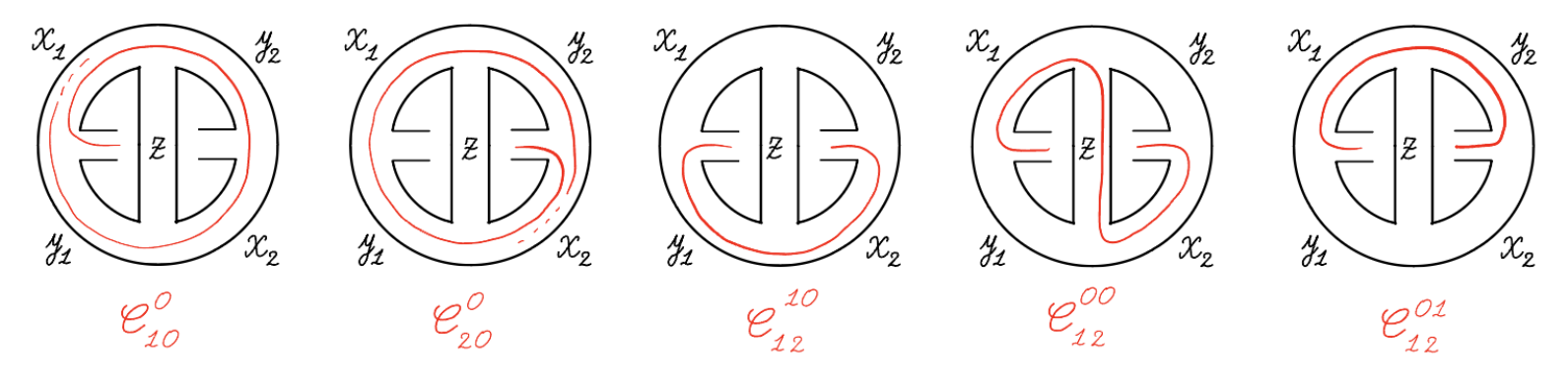}
  \caption{The five curves on $\Gamma_\ell$ which we use to construct all five MCG-inequivalent cuts.}
\label{fig:double:curves}
\end{figure}

\subsection{The All-multiplicity Amplitudes}
To assign momenta to the curves in the set ${\cal S}$, choose a routing of momenta on $\Gamma_\ell$ (Figure \ref{fig:double:tadpole}) such that edge $x_1$ has momentum $\ell_1^\mu$ and edge $x_2$ has momentum $\ell_2^\mu$, for some loop momentum variables, $\ell_1,\ell_2$. Let $p_1=p$ and $p_2=-p$ be the momenta of the two trace-factors. Then the momentum assignment rule gives, for example,
\begin{equation}
    P_{C_{01}^0} =  \ell_1,\qquad P_{C_{02}^0}=\ell_2.
\end{equation}
The only curves that carry a loop momentum are those connecting distinct trace-factors. In Appendix \ref{app:double}, these curves are labelled such that their momenta are given by
\begin{align}
    P_{C_{12}^{w_1,w_2}} &= \ell_1 - \ell_2 + (w_1-w_2-1)p, \\
    P_{C_{01}^{w}} &= \ell_1 + wp, \\
    P_{C_{02}^w} & = \ell_2 - wp.
\end{align}
In the notation of the summary, Section \ref{sec:summary}, these curves have momentum
\begin{equation}
    P_{C}^\mu = K_C^\mu + \ell_C^1\ell_1^\mu + \ell_C^2\ell_1^\mu,
\end{equation}
where $\ell_{C_{12}} = (1,-1)$, $\ell_{C_{01}} = (1,0)$ and $\ell_{C_{02}} = (0,1)$.

The all-multiplicity amplitude is then given by
\begin{equation}
    {\cal A} = \int d^E t\, {\cal K} \left(\frac{\pi^2}{\cal U}\right)^{\frac{D}{2}}\,\exp\left(-\frac{{\cal F}_0}{\cal U} - {\cal Z}\right),
\end{equation}
where ${\cal K}$ is given by \eqref{eq:double:K}, ${\cal U}$ is given by \eqref{eq:double:U}. ${\cal F}_0$ and ${\cal Z}$ are given by the formulas in Section \ref{sec:summary}, with the momenta assignments as above.

\section{Outlook}
This paper opens a systematic investigation of all-loop scattering amplitudes in the tr$\phi^3$ theory at all $n$. We have seen that certain especially simple fatgraph representatives for the curve integral formalism manifest a remarkable decoupling of $n$ and $L$, and we illustrated this at all $n$ for all amplitudes in the theory through to two loops.

In practice, the computations described in this paper are best done using a computer. The curve integral formalism is based on simple combinatorial rules, which are very amenable to implement in code. This will play an important role in applications of these formulas beyond two-loops. One possible application of these computer methods could lie in the numerical evaluation of amplitudes. Tropical methods have recently proved useful for the evaluation of individual Feynman integrals. \cite{Borinsky_2023,feyntrop}

The key result that underlies all the formulas in this paper is the existence of an explicit positive parametrization for Teichm\"uller spaces, which is most closely related to the point of view of \cite{fockgoncharov}. This parametrization is given by the definition of the so-called $u$-variables as rational functions in positive $y$-variables. \cite{us_qft}. However, to arrive at formulas for amplitudes, we have to take the tropicalization of this parametrization.\footnote{The headlight functions, $\alpha_C(t)$, are defined as the tropicalization of the $u$-variables, $u_C(y)$.} Tropical moduli spaces have a rich topological structure, which is an active area of research.\footnote{Tropical moduli spaces are central to recent calculations of previously unknown Betti numbers in the cohomology of ${\cal M}_{g,n}$, and in calculations of the cohomology of the Kontsevich graph complex, relevant to certain problems in number theory. \cite{Brown:2021umn,borinsky2020euler,andersen2021kontsevich}} It remains to be seen what importance these topological facts about tropical moduli spaces have for the new formulas in this paper, which compute amplitudes as integrals over tropical moduli spaces.

We close by discussing in greater detail some particularly interesting physics questions to explore using the results in this paper.

\subsection{Large $n$ limits}
Perhaps the most striking fact about the curve integral formulas is that an integral over a $\sim n$ dimensional space, built out of $\sim n^2$ simple \emph{headlight functions}, each with no more than $\sim n$ terms, magically produces the amplitude: which is naively the sum over $\sim 4^n$ Feynman diagrams. This suggests that the curve integral can be used to define a sensible large $n$ limit of amplitudes. Large $n$ limits have been studied for certain special cases, \cite{Voloshin:1992mz,Douglas:1989ve} and processes with multi-state emission of many particles may be important in phenomenomology.

Clearly, the external kinematics should be chosen carefully for the large $n$ limit to be well-defined. For instance, we can pick a smooth curve $C$ in some $D$ dimensional space, and at any $n$, define our canonical momentum polygons by picking a collection of ordered points on $C$; the density of points increases as $n \to \infty$, and one should expect the final amplitude to just be a function of $C$ itself. 
Another option is to think of the $X_{ij}$ as evaluating some smooth function $X(u,v)$ on the ``kinematic spacetime" on the discrete lattice associated with the $n$ point problem. We can begin with the simplest situation where $X(u,v) = X_0 = X_{ij}$ is just a constant. In this case, working for simplicity at tree-level, the sum over all diagrams is simply the total number of diagrams weighted by $1/X_0^{n-3}$. Since the total number of diagrams scales as $4^n$, we have that at large $n$ the amplitude scales as ${\cal M}_n \sim (4/X_0)^n$. Thus ${\rm lim}_{n \to \infty} (1/n) {\rm log}{\cal M}_n \to \log(4/X_0)$ has a good large $n$ behavior, reminiscent of the large $N$ behavior in the 't Hooft limit, where the action is enhanced by an overall power of $N$. It is then natural to understand ${\rm lim}_{n \to \infty} (1/n) {\rm log} {\cal M}_n(X)$ for smooth perturbations $X(u,v) = X_0 + \delta X(u,v)$. In both these scenarios for taking the large $n$ limit of kinematics, we expect that curve integral representation will also have a ``smooth" limit; with the integral of a piecewise linear action on an $(n-3)$ dimensional space asymptoting to a sort one-dimensional path integral with an action given by an integral over tropical functionals, and one can hope that critical points of this unusual ``tropical action" will determine the leading behavior of the amplitude in this limit. 

\subsection{Large $L$ limits} 
In this paper we have focused on understanding the large $n$ limit of amplitudes. This was made possible by choosing a particularly simple reference triangulation. This suggests a similar strategy for understanding large $L$ limits as well. For instance in the planar limit, we can work with an especially simple fat graph for a tadpole at $L$-loops, consisting of a string of $L$ bubbles. The enumeration of all the relevant curves that survive on the support on the Mirzakhani kernel, together with the computation of the corresponding matrices, can be carried out recursively in a straightforward way. 
It would be interesting to examine the behavior of the amplitudes at large $L$ in the 't Hooft limit. 

In a similar vein, we can simply count all planar vacuum diagrams; this is tantamount to simply setting all the $X$ variables to one, with no loop integrations. Diagram counting is the same thing as studying the tr$\phi^3$ \emph{matrix model}. This subject is usually studied using the techniques of random matrix theory, with the the famous and rich connections to two dimensional quantum gravity and string theory revealed in double-scaling limits. \cite{Gross:1989ni,Gross:1989vs,Saad:2019lba} These methods are however tailored to matrix integrals and do not straightforwardly generalize to field theory and non-trivial loop integrals. With a judicious choice of representative fat graphs for general vacuum diagrams, the curve integral formalism yields a different way of counting diagrams. This should not recover the results from the matrix model literature, but suggests a path to generalize those results to higher dimensions and to compute subleading terms in $1/L$.

\subsection{Renormalization}
One of the most important aspects of fundamental physics that has not yet been properly understood from the on-shell amplitude perspective is the exponentiation of UV divergences and the Wilsonian renormalization group. The reason appears obvious: the Wilsonian viewpoint is maximally off-shell, integrating out models of local quantum fields. Moreover, already when studying individual Feynman integrals, it can be intricate to identify and remove those divergences which need to be canceled. In this vein, tropical methods have recently proved useful. \cite{arkanimizerahillman,hillman2023subtraction}

The curve integral formalism offers a fresh perspective on this physics and is expected to manifest renormalization and renormalizability in a new way. Here again, the UV divergences depend on $n$ and $L$ in a way which suggests treating $n$ and $L$ independently, involving separate physics. The logarithmic UV divergences at $n$ points should be associated with specific $n$-dependent factors, reflecting the fact they can be reabsorbed into a redefinition of coupling constants. Whereas, the $L$-dependence must reflect the summation of leading and subleading UV log-divergences associated with the RG. It is plausible that the exponentiation of UV divergences is naturally reflected in the recursive structure of the fan itself.

A hint of how renormalization is encoded in the curve integral picture can be seen for the 1-loop planar amplitudes discussed in Sections \ref{sec:Dn} and \ref{sec:Dnrev}. In $D=2d-\epsilon$ dimensions, these amplitudes have $1/\epsilon$ UV divergences. The divergences are conventionally removed by redefinition of the propagators or vertices in the Lagrangian, with the exact correction required depending on the dimension. But the divergences can also be removed directly inside the curve integral, which has the form (after loop integration)
\begin{equation}
    {\cal A} = \int\limits_{\sum t_i \leq 0} d^nt\, F,
\end{equation}
for some $F(t)$. Every cone in the integration region of this curve integral corresponds to some fatgraph $\sigma$. The integral over a cone will give a UV divergence if (and only if) $\sigma$ contains a divergent subgraph, $\gamma$. By Euler's relation, if $\gamma$ has loop order $L_\gamma$ and $n_\gamma$ external lines, then the superficial divergence of $\gamma$ is
\begin{equation}
    d_\gamma = n_\gamma + 3(L_\gamma -1) - \frac{1}{2} LD,
\end{equation}
and it gives a divergence if $d_\gamma <0$. 
We can remove these divergences from the curve integral by defining tropical functions, $\mathcal{R}_D(y)$, such that
\begin{equation}
    \mathcal{R}_D|_\sigma = \left\{ \begin{matrix} 0 & \qquad \text{if a subgraph of $\sigma$ has $d_\gamma<0$}\\
    1 & \qquad \text{else}\end{matrix} \right.
\end{equation}
Then define a \emph{regularized amplitude} by
\begin{equation}
    {\cal A}^{\text{reg}}_{\text{1-loop}} = \int\limits_{\sum t_i \leq 0} d^n t\, {\cal R}_D\, F.
\end{equation}
The exact form of ${\cal R}_D$ depends on the dimension, $D$.

For instance, in $D=2+\epsilon$ dimensions, the divergent subgraphs are 1-loop tadpoles, which have $d_\gamma = - \epsilon /2 < 0$. In this case, one option for ${\cal R}_D$ turns out to be
\begin{equation}
\mathcal{R}_2 = \max\left(0,1- \sum_i g_{ii}^a \partial_{t_a} \alpha_{C_{ii}}(t)\right),
\end{equation}
where the sum is over all \emph{tadpole propagator} curves, $C_{ii}$, and ${\bf g}_{ii}$ is the $g$-vector of $C_{ii}$. Similar functions can be defined for higher dimensions. 

It will be fascinating to find generalizations of this to all loops, such that UV divergences are governed by tropical functions on the Feynman fan, to better understand renormalization and renormalizability from an on-shell perspective.


\section*{Acknowledgments}
\small{NAH is supported by the DOE under grant DE-SC0009988; further support was made possible by the Carl B. Feinberg cross-disciplinary program in innovation at the IAS. PGP is supported by ANR grant CHARMS (ANR-19-CE40-0017) and by the Institut Universitaire de France (IUF). HF is supported by Merton College, Oxford, and received additional support from ERC grant GALOP (ID: 724638). GS is supported by the European Union’s Horizon 2020 research and innovation programs {\it Novel structures in scattering amplitudes} (No. 725110) of Johannes Henn. HT is supported by NSERC Discovery Grant RGPIN-2022-03960 and the Canada Research Chairs program, grant number CRC-2021-00120.}

\appendix

\section{Curve Glossary}
\label{sec:Appendix}
For reference, and as a proof of principle, we record here all the curve paths needed to compute every amplitude up to 2-loops. In practice, these lists are most easily obtained by computer.

\subsection{1-loop Non-planar 2-point}\label{app:apq}
The Mirzakhani kernel in Section \ref{sec:apq} takes the form
\begin{equation}
    {\cal K} = \frac{\alpha_{12}^0}{\rho}.
\end{equation}
So the set of curves, ${\cal S}$, contributing to the curve integral for the amplitude are those which are compatible with $C_{12}^0$. There are five such curves on the tadpole graph. 

One way to enumerate these curves is to cut the tadpole graph along $C_{12}^0$, to produce a 5-point tree graph. The five curves we need are then the five curves on three 5-point tree graph.

Of the 5 curves in ${\cal S}$, 3 of them carry loop momentum and are given by
\begin{align}
    C_{12}^{-1} &= 1R e_2 L2,\\
    C_{12}^0 &= 1L e_1 R2,\\
    C_{12}^1 &= 1L e_1 L e_2 R e_1 R2.
\end{align}
In addition to these, there are the two curves that follow the boundaries of the fatgraph, and which contribute to ${\cal Z}$. These boundary curves have paths
\begin{align}
    C_{11} &= 1L e_1 L e_2 L1,\\
    C_{22} &= 2L e_1 L e_2 L2.
\end{align}

Note that it is straightforward to use these paths to obtain the curve matrices for these five curves by matrix multiplication. One finds
\begin{align}\label{eq:A11matrices}
   M_{11} = M_{22} &= \begin{bmatrix} 1 & 0\\ 1+y_1+y_1y_2 & y_1y_2\end{bmatrix}\\
    M^{-1}_{12} &=\begin{bmatrix} 1+y_2 & y_2 \\ y_2 & y_2 \end{bmatrix}\\
    M^0_{12} &=\begin{bmatrix} 1 & 1 \\ 1 & 1 + y_1 \end{bmatrix}\\
    M^1_{12} &= \begin{bmatrix}  1 & 1+y_1 \\
    1 + y_1 & 1 + 2y_1 +y_1^2+y_1^2 y_2 \end{bmatrix}
\end{align}
For the rest of this Appendix, we record only the paths of the curves.

\subsection{2-loop Planar Tadpole}\label{app:Dntilde}
The following is the finite set $\mathcal{S}$ of 16 non-boundary curves compatible with the Mirzakhani kernel ${\cal K}$ in Section \ref{sec:Dntilde}:
\begin{align*}
& 0 L w R (x L)^\infty \\
& 0 R z R (y L)^\infty \\ 
& 0 L w L x L w R 0 \\ 
& 0 R z L y L z L 0 \\ 
& 0 R z R y R z L 0 \\ 
& (R x)^\infty L w L z R (y L)^\infty \\ 
& (R y)^\infty L z R w R (x L)^\infty \\ 
& 0 L w L x L w L z R (y L)^\infty \\ 
& 0 R z R y R z R w R (x L)^\infty \\ 
& (R y)^\infty L z R w R x R w R 0 \\ 
& (R x)^\infty L w L z L y L z R w R (x L)^\infty \\ 
& (R y)^\infty L z R w L x L w L z R (y L)^\infty \\ 
& 0 L w L x L w L z L y L z R w R (x L)^\infty \\ 
& 0 R z R y R z R w R x R w L z R (y L)^\infty \\ 
& 0 L w L x L w L z L y L z R w R x R w R 0 \\ 
& 0 R z R y R z R w L x L w L z L y L z L 0
\end{align*}
The momentum of a curve $C_{AB}$ ($A,B=0,a,b$) connecting trace-factors $A$ and $B$ is given in dual variables by $x_B^\mu - x_A^\mu$. $x_a^\mu$ and $x_b^\mu$ are the loop momentum variables. So it follows that, of these 16 curves, 9 of them carry loop momentum and so contribute to $\mathcal{K}$, $\mathcal{U}$ and ${\cal F}_0$. The remaining 7 contribute only to ${\cal Z}$.

In addition, the tadpole graph has one boundary curve, $C_{00}$, which contributes to $\mathcal{Z}$, and has the path
\begin{align*}
C_{00} &= 0 R z R y R z R w R x R w R 0
\end{align*}
    
\subsection{2-loop Double Trace}\label{app:double}
The following is the finite set $\mathcal{S}$ of 21 non-boundary curves compatible with the Mirzakhani kernel ${\cal K}$ in Section \ref{sec:double}:
\begin{align*}
    C_{12}^{01} &= L y_2 R x_1 L \\
    C_{12}^{10} &= L y_1 R x_2 L \\
    C_{12}^{00} &= R x_2 R z L x_1 L \\
    C_{12}^{11} &= L y_1 L z R y_2 R \\
    C_{11}^{0} &= L y_1 R x_2 R y_2 R x_1 L \\
    C_{22}^{0} &= L y_2 R x_1 R y_1 R x_2 L \\
    C_{10}^{0} &= R (x_1 L y_2 L x_2 L y_1 L)^\infty \\
    C_{20}^{0} &= R (x_2 L y_1 L x_1 L y_2 L)^\infty\\
    C_{11}^{-1/2} &= L y_1 R x_2 R y_2 L z R y_1 R \\
    C_{11}^{1/2} &= R x_1 L y_2 L x_2 R z L x_1 L \\
    C_{22}^{-1/2} &= L y_2 R x_1 R y_1 L z R y_2 R \\
    C_{22}^{1/2}  &= R x_2 L y_1 L x_1 R z L x_2 L \\
    C_{12}^{1-1} &= L y_1 R x_2 R y_2 L z L x_2 L \\
    C_{12}^{-11} &= L y_2 R x_1 R y_1 L z L x_1 L \\
    C_{12}^{-10} &= R x_1 R z R y_1 L x_1 R z L x_2 L \\
    C_{12}^{0-1} &= R x_2 R z R y_2 L x_2 R z L x_1 L \\
    C_{10}^{-1} &= L y_1 L z R (y_2 L x_2 L y_1 L x_1 L)^\infty \\
    C_{10}^{1} &= R x_1 R z R (y_1 L x_1 L y_2 L x_2 L)^\infty \\
    C_{20}^{-1} &= L y_2 L z R (y_1 L x_1 L y_2 L x_2 L)^\infty \\
    C_{20}^{1} &= R x_2 R z R (y_2 L x_2 L y_1 L x_1 L)^\infty \\
    C_{00} &= (R x_2 R y_2 R x_1 R y_1)^\infty L z R (y_2 L x_2 L y_1 L x_1 L)^\infty
\end{align*}
The only curves that carry loop momentum are those curves $C_{AB}$ connecting distinct trace-factors, $A$ and $B$. 

It follows that, of these 21 curves, 14 of them carry loop momentum and so contribute to $\mathcal{K}$, $\mathcal{U}$ and ${\cal F}_0$. The remaining 7 contribute only to ${\cal Z}$.

In addition, the tadpole graph has two boundary curves, which contribute to $\mathcal{Z}$, and have paths
\begin{align*}
C_{11} &= 1 R x_1 R z R y_1 R 1\\
C_{22} &= 2 R x_2 R z R y_2 R 2
\end{align*}

\subsection{2-loop Triple Trace}\label{app:Apqr}
The following is the finite set $\mathcal{S}$ of 30 non-boundary curves compatible with the Mirzakhani kernel ${\cal K}$ in Section \ref{sec:Apqr}:
\begin{align*}
& 0 L z L x_2 R 1 \\ & 1 R x_1 L z R 0 \\ & 2 L y_1 R w L 0 \\ & 2 R y_2 L w L 0 \\ & 0 L z L x_2 L x_1 R x_2 R 1 \\ & 1 R x_1 L z L w L y_1 R 2 \\ & 2 L y_1 R w R z L x_2 R 1 \\ & 2 R y_2 L w R z L x_2 R 1 \\ & 2 R y_2 L w R z R x_1 L 1 \\ & 2 R y_2 R y_1 L y_2 L w L 0 \\ & 0 L z L x_2 L x_1 L z L w L y_1 R 2 \\ & 0 L z L x_2 L x_1 L z L w R y_2 L 2 \\ & 1 L x_2 R z L w L y_1 L y_2 L w L 0 \\ & 1 L x_2 R z L w L y_1 L y_2 R y_1 R 2 \\ & 1 R x_1 L z L w L y_1 L y_2 L w L 0 \\ & 1 R x_1 L z L w L y_1 L y_2 R y_1 R 2 \\ & 1 R x_1 R x_2 L x_1 L z L w L y_1 R 2 \\ & 1 R x_1 R x_2 L x_1 L z L w R y_2 L 2 \\ & 0 L z L x_2 L x_1 L z L w L y_1 L y_2 R y_1 R 2 \\ & 1 L x_2 R z L w L y_1 L y_2 L w R z L x_2 R 1 \\ & 1 R x_1 L z L w L y_1 L y_2 L w R z L x_2 R 1 \\ & 1 R x_1 L z L w L y_1 L y_2 L w R z R x_1 L 1 \\ & 1 R x_1 R x_2 L x_1 L z L w L y_1 L y_2 L w L 0 \\ & 2 L y_1 R w R z L x_2 L x_1 L z L w L y_1 R 2 \\ & 2 R y_2 L w R z L x_2 L x_1 L z L w L y_1 R 2 \\ & 2 R y_2 L w R z L x_2 L x_1 L z L w R y_2 L 2 \\ & 0 L z L x_2 L x_1 L z L w L y_1 L y_2 L w R z L x_2 R 1 \\ & 0 L z L x_2 L x_1 L z L w L y_1 L y_2 L w R z R x_1 L 1 \\ & 2 L y_1 R w R z L x_2 L x_1 L z L w L y_1 L y_2 L w L 0 \\ & 2 R y_2 L w R z L x_2 L x_1 L z L w L y_1 L y_2 L w L 0    
\end{align*}
The only curves that carry loop momentum are those curves $C_{AB}$ connecting distinct trace-factors, $A$ and $B$. 

It follows that, of these 30 curves, 24 of them carry loop momentum and so contribute to $\mathcal{K}$, $\mathcal{U}$ and ${\cal F}_0$. The remaining 6 contribute only to ${\cal Z}$.

In addition, the tadpole graph has three boundary curves, which contribute to $\mathcal{Z}$, and have paths
\begin{align*}
C_{11} &= 1 R x_1 R x_2 R 1\\
C_{22} &= 2 R y_2 R y_1 R 2\\
C_{00} &= 0 R w R y_2 R y_1 R w R z R x_1 R x_2 R z R 0
\end{align*}

\subsection{2-loop Genus One Tadpole}\label{app:markov}
The following is the finite set $\mathcal{S}$ of 9 non-boundary curves compatible with the Mirzakhani kernel ${\cal K}$ in Section \ref{sec:markov}:
\begin{align*}
C_1 = C^{0/1;0} &= 0Ly_2Ry_4Ly_1L 0\\
C_2 = C^{1/0;0} &= 0Ly_2Ly_3Ry_1L 0\\
C_3 = C^{1/0;1} &= 0Ry_1Ry_4Ry_3Ry_1Ry_2Ry_4Ly_1L 0\\
C_4 = C^{1/1;0} &= 0Ry_1Ry_4Ry_3Ry_1L 0\\
C_5 = C^{1/1;1} &= 0Ly_2Ly_3Ly_4Ly_2R0\\
C_6 = C^{-1/1;0} &= 0Ly_2Ly_3Ry_1Ry_2Ry_4Ly_1L0 \\
C_7 = C^{0/1; 1} &= 0Ly_2Ly_3Ry_1Ry_2Ry_4Ry_3Ry_2R0 \\
C_8 = C^{0/1;2} &=  0Ry_1Ry_4Ry_3Ry_1Ry_2Ry_4Ry_3Ly_4Ly_1L0 \\
C_9 = C^{-1/1;1} &= 0Ry_1Ry_4Ry_3Ry_1Ry_2Ry_4Ly_1Ry_2Ry_4Ly_1L0 \\
\end{align*}
All 9 of these curves carry loop momentum and so contribute to $\mathcal{K}$, $\mathcal{U}$ and ${\cal F}_0$.

In addition, the tadpole graph has one boundary curve, which contributes to $\mathcal{Z}$, and has path
\begin{align*}
C_{00} &= 0Ry_1Ry_4Ry_3Ry_1Ry_2Ry_4Ry_3Ry_2R0.   
\end{align*}

\bibliographystyle{unsrt}
\bibliography{refs}

\end{document}